\documentclass[11pt]{article}

\usepackage[margin=1.2in]{geometry}
\usepackage{amsmath, amssymb, mathtools}
\usepackage{amsthm}
\usepackage{graphicx}
\usepackage{color}
\usepackage{pifont, array}
\usepackage{xcolor}
\usepackage{booktabs}
\usepackage{gensymb}
\usepackage{enumitem}
\usepackage{setspace}
\usepackage{tikz}
\usetikzlibrary{arrows.meta,positioning,calc}
\usepackage{fancyhdr}

\usepackage{etoolbox}
\usepackage{authblk}

\makeatletter
\renewcommand\@fnsymbol[1]{\ifcase#1\or *\else\@arabic{#1}\fi}
\makeatother

\usepackage[colorlinks, citecolor=blue, linkcolor=blue, urlcolor=blue, linktoc=page]{hyperref}

\usepackage[
  backend=biber,
  natbib=true,
  style=authoryear,
  backref=true
]{biblatex}
\AtEveryBibitem{%
  \clearfield{urldate}%
  \clearfield{url}%
  \clearfield{urlyear}%
  \ifboolexpr{test{\ifentrytype{misc}} or test{\iffieldequalstr{note}{Accessed on 2023-07-12}}}{}{\clearfield{note}}%
  \clearfield{isbn}%
  \clearfield{issn}%
  \clearfield{language}%
}
\numberwithin{equation}{section}

\theoremstyle{plain}

\theoremstyle{remark}
\newtheorem{definition}{Definition}

\newtheorem{remark}{Remark}

\newcommand{\xmark}{\ding{55}}
\newcommand{\cmark}{\ding{51}}
\newcommand{\PA}{\mathrm{PA}}

\newcolumntype{P}[1]{>{\centering\arraybackslash}p{#1}}

\title{Identifying Elasticities in Autocorrelated Time Series \\ Using Causal Mechanism Graphs}

\author[1]{Lion Hirth}
\author[2]{Jonas Peters}
\author[3]{Oliver Ruhnau}
\author[1]{Jorge Sánchez Canales\thanks{Corresponding author (j.sanchez-canales@hertie-school.org). Replication code available at \url{https://github.com/jscanales/elasticities_with_graphs}}}
\author[2]{Felix Schur}
\author[1]{Raffaele Sgarlato}
\author[1]{Silvana Tiedemann}

\affil[1]{Centre for Sustainability, Hertie School}
\affil[2]{Department of Mathematics, ETH Zurich}
\affil[3]{Faculty of Management, Economics and Social Sciences, Institute of Energy Economics, University of Cologne}

\date{October 19th, 2026}

\begin{document}

\maketitle

\pagestyle{fancy}
\fancyhf{}
\fancyhead[C]{ELASTICITIES WITH GRAPHS}
\fancyfoot[C]{\thepage}
\renewcommand{\headrulewidth}{0pt}
\renewcommand{\footrulewidth}{0pt}
\setlength{\headheight}{14pt}

\thispagestyle{empty}

\begin{abstract}
Instrumental variable (IV) methods are widely used to estimate the price elasticity of demand from observational data. However, naive~IV estimators may be inconsistent in settings with autocorrelated time series.  We argue that causal graphs taking into account time structure can simplify IV identification and help select consistent estimators. More precisely, we propose to model equilibrium conditions using causal mechanism time graphs. These (acyclic) graphs allow us to maintain the assumption of a simultaneous determination of prices and quantities and come with an interpretable notion of intervention. They further allow us to exploit recent advances in graphical inference to derive valid IV estimators that are consistent under weak assumptions. Finally, we argue that observing significant differences between the estimates of presumably valid estimators can help to reject false model assumptions, thereby improving our understanding of underlying economic dynamics. We apply our approach to the German electricity market, estimating the price elasticity of demand from simulated and real-world data. The findings underscore the importance of accounting for structural autocorrelation in IV-based analysis.
\end{abstract}


\newpage
\tableofcontents
\thispagestyle{empty}

\section{Introduction}\label{sec:intro}

\begin{figure}
    \centering
    \includegraphics[width=\linewidth]{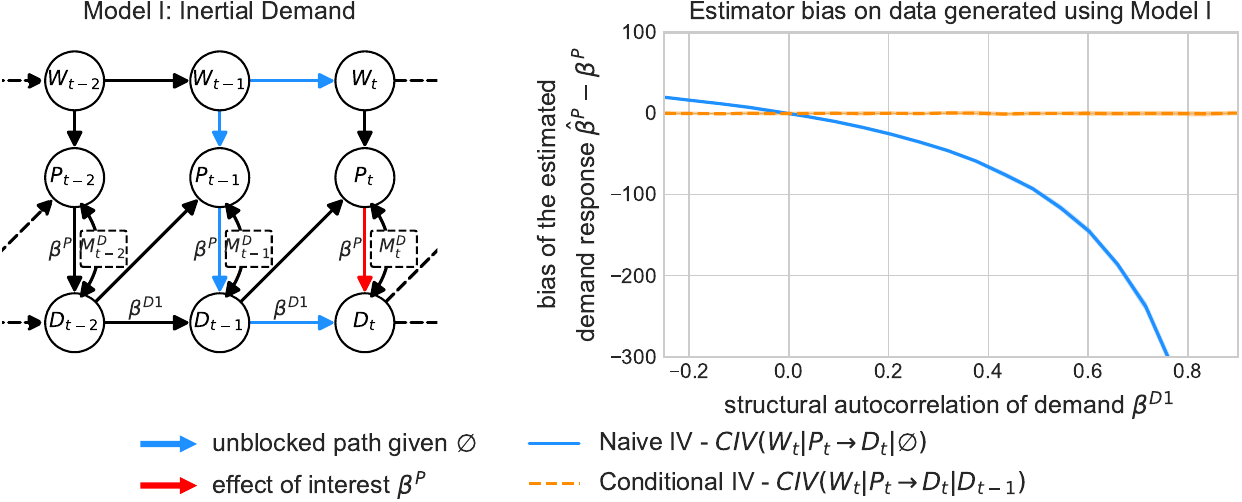}
    \caption{
    The effect of time dependence
    on the empirical bias of IV estimators.
    Left: a marginalized version of the causal mechanism time graph for Model~I (see \eqref{eq:demand_autocorrelated}); demand is inertial, as $D_t$ structurally depends on $D_{t-1}$ (with coefficient $\beta^{D1}$), and the causal effect of interest is $P_t \to D_t$ (red edge, with coefficient $\beta^P$).
    The nodes $M^D$ are mechanism nodes: unobservable variables 
    that help distinguish different types of interventions (see Section~\ref{sec:demand_autocorrelated}).
    Our framework provides a simple, graphical condition for evaluating validity of estimators. Using
    $W_t$ as an instrument for $P_t$, the 
    blue path is unblocked for the naive~IV estimator (so, it is invalid) and blocked when conditioning on $D_{t-1}$ (see Section~\ref{sec:conditionsIV}).
    Right: the empirical biases 
    with respect to the true value ($\beta^P=-100$)
    of the estimators for varying strength of the structural autocorrelation in demand:
    while the naive estimator indeed suffers from severe bias, conditioning on the right variables does not.
    }
    \label{fig:intro}
\end{figure}

Estimating the price elasticity of demand from observational time series is a fundamental problem in econometrics.
Observable prices and quantities are determined simultaneously through the equilibrium of the underlying economic relationships, the demand and supply equations.
Identifying the price elasticity of demand from data, therefore, requires non-trivial methodology, such as using credible supply-shifting instrumental variables (IVs) \citep{wright_tariff_1928, tinbergen_econometric_1940, angrist_interpretation_2000}. 
Time dependencies in the underlying relationships pose an additional challenge to identification.
IV estimators that do not account for this are, in general, inconsistent and
the resulting bias can be large: in 
our simulation, an estimator that neglects the time dynamics is off by a large margin (see, e.g., Figure~\ref{fig:intro}).
 
The statistics and computer science communities have developed a framework based on structural and graphical causal models \citep{pearl_causality_2009, spirtes_causation_2000}
that
can help to find consistent estimators for causal effects. 
Starting from a set of assumptions encoded in a structural causal model (SCM) and its associated graph, graphical criteria based on d-separation (see Definition~\ref{def:dsep}) can determine which estimators are valid;
this approach applies, for instance, to covariate adjustment or conditional instrumental variable regression \citep{Brito2002generalized,henckel_graphical_2023}.
Recently, \citet{thams_identifying_2022} have extended the graphical criteria for the validity of conditional IV (CIV) estimators to infinite causal time series graphs \citep{peters_causal_2013}.
The application of the 
graphical framework to economic problems has, however, remained limited.
The reason is that graphical criteria require acyclic causal graphs, and it is not obvious how a system in which prices and quantities are determined jointly should be represented as an acyclic graph \citep{imbens_potential_2020}.
 
In this paper, we extend the causal graphical framework to such economic settings with market equilibrium.
Specifically, we introduce mechanism-augmented SCMs and causal mechanism graphs, and extend them to the time series setting.
These constructions 
include the demand and supply functions as additional mechanism nodes. This  yields an acyclic graph of the system while maintaining the assumption that prices and quantities are determined 
in equilibrium.
We discuss that the global Markov property continues to hold in the linear setting we consider (Appendix~\ref{sec:constrained_graph}), and
that the graphical criteria apply unchanged.
Including mechanism nodes explicitly 
makes interventions in the system easier to interpret.
The idea of representing mechanisms as nodes has been discussed before -- although, to the best of our knowledge, without SCMs, in different graphs, in non-time-series settings and without formal introduction \citep[see, e.g.,][Figure~12]{imbens_potential_2020}. 
We mostly study linear models but extend our framework to nonlinear models in Appendix~\ref{sec:non-linear-ext}.

Figure~\ref{fig:intro} illustrates the proposed approach.
We start from an equilibrium system in which demand is inertial, that is, $D_t$ structurally depends on $D_{t-1}$ (see Section~\ref{sec:demand_autocorrelated} and Appendix~\ref{app:data_generation} for details), and construct the corresponding causal mechanism graph; the left panel shows a marginalized version in which some nodes are omitted.
We show 
in Section~\ref{sec:CIV-estimation}
that the graphical criteria for CIV validity \citep{Brito2002generalized, henckel_graphical_2023, thams_identifying_2022} 
hold in equilibrium systems; thus, an IV estimator is, in general, inconsistent if the instrument and the dependent variable are connected by a path that remains unblocked once the edge of interest $P_t \to D_t$ is removed
(see Section~\ref{sec:conditionsIV} for details).
Using $W_t$ as an instrument for $P_t$, the naive~IV estimator (no additional controls) is therefore inconsistent -- the blue path is unblocked -- whereas the CIV estimator (conditioning on $D_{t-1}$) is valid (and consistent under weak conditions).%
The right panel confirms this in simulation with known ground truth: the naive estimator overestimates the demand response by a large margin, while the conditional IV estimator remains consistent across the range of structural autocorrelation.
 
Applying the same graphical criteria, it is generally possible to derive several valid CIV estimators for a given model. 
This can be done by
blocking paths via conditioning (this
recovers several known estimators, such as the lag-augmented local projection IV estimator \citep{stock_identification_2018, montiel_olea_local_2021, montiel_olea_double_2024})
or, see 
\citet{thams_identifying_2022},
via estimating nuisance effects, 
which 
simultaneously estimates and later disregards effects that are not of primary interest.
We argue that deriving several valid estimators allows us to reject false assumptions. 
If the researcher's assumptions encoded in the structural model are correct, two valid estimators should yield comparable results (up to finite-sample errors).
Conversely, a statistically significant difference (such as non-overlapping confidence intervals) between estimates of two estimators that were considered valid should lead the researcher to reject her model assumptions, thereby improving our understanding of economic dynamics.
 
We apply this methodology to the price elasticity of demand in the electricity market, using wind power generation as an instrument, both on simulated data (Section~\ref{sec:Simulation}) and on real-world data (Section~\ref{sec:application}).
Electricity is a suitable case because demand is inherently dynamic at high frequencies.
Many electricity-consuming activities involve technological, physical, or behavioral constraints that create intertemporal dependence. Industrial production processes, household appliances, and other electricity uses cannot generally adjust instantaneously to price changes (`demand inertia') and may require balancing upward and downward adjustments over time (`load shifting').
As a result, current demand may depend not only on current prices but also on past consumption decisions and other state variables.
Moreover, the case is empirically relevant: estimates of this elasticity are widely used in energy policy and market design, yet the identification strategies on which they rest rarely account for these dynamics (see the related literature below).
Which form these dynamics take -- inertia, load shifting, or both -- is not known a priori, and we therefore study three models:
Model~I represents demand inertia; 
Model~II captures heterogeneous inertia and exposure to real-time prices;
and Model~III represents load shifting through a stock of energy services.
Each model implies a different set of valid estimators.
 
Our analysis on real data
shows that disregarding the time dynamics matters: the naive~IV estimator understates the demand response by a factor of two relative to our benchmark estimator, with non-overlapping confidence intervals.
The estimators that are valid across the largest model class agree on a linear demand response of approximately $-200$ MW/(EUR/MWh), corresponding to a log-log elasticity of $-0.1$.
The pattern of 
overlapping intervals
allows us to reject Models~I and~II, whereas Model~III cannot be rejected.
By construction, our method can only falsify assumptions, so 
the results are in line with the existence of
load shifting in electricity markets but do not provide strong evidence.

\paragraph*{Related literature.}
There is an ongoing discussion about whether (and if so, how) causal graphs can help solve problems in economics.
Proponents argue that causal graphs representing structural causal models allow researchers to derive valid non-parametric estimators transparently  \citep{spirtes_causation_2000, pearl_causality_2009, pearl_book_2020, hunermund_causal_2023}, as exemplified by \citet{wright_tariff_1928}, who derived the instrumental variable estimator for the price elasticity of demand for flaxseed oil via graph-based path analysis (see \citet{cunningham_causal_2021} for a historical account).
Opponents point out that the utility of causal graphs is limited to problems that can be expressed in terms of acyclic graphs and argue that this precludes a broad class of econometric problems represented by simultaneous equations, such as textbook examples of supply and demand.
They further argue \citep[see, e.g.,][]{imbens_potential_2020} that, so far, using causal graphs does not solve problems that could not be solved with alternative frameworks such as the potential outcomes framework \citep{neyman_application_1923,holland_statistics_1986, angrist_interpretation_2000, rubin_causal_2005,angrist_mostly_2009}.
Despite the debate, graphs representing structural equations are not alien to econometrics 
\citep[see][for an early example]{griliches_education_1972}, especially not to identification in time series: \citet{tinbergen_econometric_1940} and \citet{wold_econometric_1964} graphically represented causal dependencies between variables in economic models over time, using what they call an `arrow scheme'.
Arguably, these representations were not widely adopted because they were tied to what \citet{wold_econometric_1964} termed the `causal chain model', which posits that economic processes are recursive, and thus conflicted with the prevailing notion that economic systems are generally in equilibrium and should be modeled using simultaneous equations \citep[see, e.g.,][]{christ_cowles_1994}.
This paper engages directly with this debate by proposing a method to represent equilibrium in acyclic graphs, specifically by including the demand and supply functions as additional mechanism nodes.
We believe that this paper demonstrates the value of using graphs in econometrics.

Several papers study causal representations of equilibrium or constrained systems. \citet{blom2020beyond} retain equilibrium relations as causal constraints, whereas we assume a uniquely solvable equilibrium and encode its solution as an acyclic structural assignment. \citet{blom2021conditional} derive causal and conditional-independence graphs directly from systems of equations; in our construction, the demand and supply functions instead appear explicitly as function-valued mechanism nodes. \citet{beckers2023causal} introduce constraints and a separate disconnection intervention, while we use standard interventions on either realized variables or mechanisms.
Our specialized construction yields graphs from which one can directly  derive valid conditional IV estimators for autocorrelated time series.

Causal time graphs also speak to the literature on identifying dynamic causal effects in macroeconomic time series. 
Macroeconomic applications often focus on estimating causal effects at different horizons $h$, known as impulse response functions: the response
of a (macro)economic system at time $t+h$ to a shock at time $t$. 
One can use frameworks such as local projections and SVARs \citep{jorda_betting_2015, ramey_government_2018, stock_identification_2018}. 
We provide a detailed comparison to our approach in Appendix~\ref{sec:SVAR} 
and summarize the most important points here.
The two frameworks target different causal estimands: our estimand is the demand response to a hypothetical intervention on the price, whereas SVAR analysis targets the response of equilibrium quantities to shocks, which generally depends on the supply slope as well.
We show that the former problem is strictly easier and 
comes with different identifiability conditions (there are settings in which our estimand is identifiable but the SVAR estimand is not).
Arguably closest in spirit is the Proxy-VAR and local projection IV literature.
The lag-augmented local projection IV estimator \citep{stock_identification_2018, montiel_olea_local_2021, montiel_olea_double_2024} 
can be represented using the conditional IV (CIV) notation developed in this paper
(see estimator~$\#4$ in Section~\ref{sec:CIV-estimation}). 
More precisely, lag-augmented local projection IV can be viewed as a sequence of CIV estimators for the effects $P_t \rightarrow D_t$, $P_t \rightarrow D_{t+1}$, $\ldots$, $P_t \rightarrow D_{t+h}$ at successive horizons. 
Lag-augmented local projection IV estimators are not always valid; 
our framework provides a means to assess their validity and to construct alternative valid estimators
(see Section~\ref{sec:consistent_estimators}).
Furthermore, graphical analysis provides a straightforward explanation for recent findings in this literature, such as the importance of lead--lag exogeneity of the instrument \citep{stock_identification_2018} and the robustness of lag-augmented local projections to model misspecification \citep{montiel_olea_local_2021, montiel_olea_double_2024}. 

Historically, electricity demand was considered inelastic, and power plants adjusted supply 
to whatever demand occurred \citep{borenstein_trouble_2002,cramton_capacity_2013}. Liberalization, digitalization, and decarbonization 
have since increased the scope of demand response, making the short-run price elasticity of electricity demand an important object to study. A substantial empirical literature estimates this elasticity using high-frequency electricity market data 
\citep[e.g.,][]{lijesen_real-time_2007, bonte_price_2015, knaut_when_2017, fabra_estimating_2021, hirth_how_2024, herrmann_wind_2026}.
Academics and practitioners routinely use the resulting estimates for model-based assessments
of system reliability \citep[e.g.,][]{bushnell_electricity_2005, hogan_energy_2005, acer_decision_2020}, market power \citep[e.g.,][]{borenstein_market_1999}, renewable energy policies \citep[e.g.,][]{abrell_economics_2019, reguant_efficiency_2019}, and geographic market configurations \citep[e.g.,][]{ENTSOE_zone_2025}.
As electricity prices become more volatile with growing market shares
of intermittent wind and solar power, the price elasticity of electricity demand may become even more important \citep{joskow_challenges_2019, ciarreta_renewable_2020, maniatis_impact_2022, hosius_impact_2023, ruhnau_flexible_2022}.

\section{
Causal Representation of
Autocorrelated Equilibrium Systems
}\label{sec:graphical_representation}
This section links the economic equilibrium model of autocorrelated price and demand with structural causal models \citep{pearl_causality_2009,bongers_foundations_2021} and (causal) graphical models \citep{Lauritzen1996, spirtes_causation_2000, pearl_causality_2009} in the context of time series data  \citep[e.g.,][]{peters_causal_2022}.%
We propose 
causal mechanism graphs, which can be used to represent an extension of the SCMs in which%
constraints,
such as the requirement that demand equals supply for the market to clear, form part of the structural assignments defining the model.
Specifically,
for our example,
we 
propose to
include the structural equations for demand (and supply) as explicit yet unobserved
mechanism nodes that jointly determine the market-clearing price and, evaluated at that price, the realized quantities. 
The
corresponding causal mechanism graph is 
a 
directed acyclic graph (DAG) representation of the system at equilibrium, in which the simultaneity of price and demand appears as confounding by the demand mechanism. 
The SCM serves both as an observational model and as an interventional model \citep{peters_causal_2016}, and allows us to distinguish two types of interventions: interventions on realized variables and interventions on the mechanisms generating them (see Appendix~\ref{sec:constrained_graph} for details).

We introduce these concepts in the context of 
three stylized dynamic demand models.
Each 
model takes the form of a classic econometric
single-equation dynamic model used to approximate economic time series \citep[see][chapter 7]{hendry_dynamic_1995}.
Each of these models captures a qualitatively distinct pattern of dynamic behavior
plausible in electricity markets: 
Model~I captures physical inertia in electricity consumption; Model~II captures heterogeneous inertia and exposure to real-time prices; 
and Model~III captures intertemporal substitution through storage or load shifting. 
Although all three models generate autocorrelated demand series (see Section~\ref{sec:Simulation}), they imply different causal structures and 
induce different sets of valid 
conditional IV estimators (see Section~\ref{sec:CIV-estimation}).
Appendix~\ref{sec:microfoundations} relates these demand models to the behavior of a forward-looking, utility-maximizing consumer.
We introduce our models assuming linearity. Appendix~\ref{sec:non-linear-ext} extends our results to nonlinear settings.

\subsection{Equilibrium System with Inertial Demand (Model~I)}\label{sec:demand_autocorrelated}
Our starting point is an adaptation of SCMs \citep{pearl_causality_2009, peters_elements_2017,bongers_foundations_2021}
which allows for time series models \citep{peters_causal_2013, peters_causal_2022}.
In short, here, an SCM is a set of structural equations
$X^j_t \coloneqq f^j(\mathrm{PA}(t,j), \epsilon^j_t)$, $j \in \{1, \ldots, d\}$,
where $\mathrm{PA}(t,j)$ are the 
\emph{parents} 
of $X^j_t$ and 
$\epsilon^j_t$, $t \in \mathbb{Z}, j \in \{1, \ldots, d\}$, are jointly independent random error variables. We provide a formal mathematical introduction to structural causal models for time series in Appendix~\ref{sec:appendix_scm}.

We model demand $D_t$ 
as depending
structurally on price $P_t$, past demand $D_{t-1}$
(we refer to this dependence as
\emph{(direct) structural autocorrelation}),
and an i.i.d.\
exogenous shock
$U_t^D$.
We introduce 
this type of
structural autocorrelation to model demand inertia: electricity consumption often involves multi-hour processes, so that feasible consumption in hour $t$ depends on past operating decisions.
We further assume linearity, so that the demand equation reads 
\begin{equation}\label{eq:demand_autocorrelated}
   \textit{Model~I:} \qquad 
   D_t \coloneqq
   M_t^D(P_t, D_{t-1}) \coloneqq
   D^0 + \beta^P P_t + \beta^{D1} D_{t-1} + U^D_t;
\end{equation}
here,  
$M_t^D(\cdot, \cdot)$
denotes the 
mechanism of demand, which is a random function
due to the noise term;
$\beta^P$ is the slope of the demand equation. 
Equation~\eqref{eq:demand_autocorrelated} is structural, so it models how $D_t$ reacts to an intervention on $P_t$ and $D_{t-1}$ (formally defined in Appendix~\ref{sec:constrained_graph}); 
for example, $\beta^P$ corresponds to the effect of a hypothetical change
 in the price $P_t$ on the quantity demanded $D_t$,
understood as willingness to consume at that price
(see also Section~\ref{sec:application}).
We take $\beta^P$ as the causal (or structural) parameter of interest throughout the paper.
We return to the identification of $\beta^P$
 in Section~\ref{sec:CIV-estimation}.

Regarding supply $S_t$, for all $t \in \mathbb{Z}$ we assume:
\begin{equation}\label{eq:supply}
    S_t  \coloneqq 
    M_t^S(P_t, W_t) \coloneqq
    S^0 +\gamma^P P_{t} + \gamma^W W_t +  U^S_t,
\end{equation}
where $W_t$ is exogenous, $U_t^S$ is an i.i.d.\ exogenous shock,
and $\gamma^P$ is the slope of the supply equation.
The structural equation of $W_t$ does not depend on $P_t$ and follows an AR(1) process:
$W_t \coloneqq \beta^W W_{t-1} + U_t^W$,
where 
$U_t^W$ is i.i.d.\ and $(U_t^W)_t$, $(U_t^S)_t$, and $(U_t^D)_t$ are jointly independent.\footnote{\label{fn:shock-independence}
None of the identification results in this paper requires 
the 
shocks $U_t^D$, $U_t^S$, and $U_t^W$
to be Gaussian; consistent estimation only requires finite second
moments (see Section~\ref{sec:CIV-estimation}).}

At equilibrium, demand equals supply and the equilibrium price clears the market. 
This yields a structural constraint:
for all $t$ we have
\begin{equation}\label{eq:equilibrium}
    S_t = D_t.
\end{equation}
From the structural equations for demand~\eqref{eq:demand_autocorrelated} and supply~\eqref{eq:supply}, and the equilibrium constraint~\eqref{eq:equilibrium}, we obtain the following expression for prices at equilibrium
\begin{equation}\label{eq:price_autocorrelated}
\begin{split}
   \textit{Model~I:} \quad 
    P_t 
    &\coloneqq 
    \Phi(M_t^D, M_t^S; D_{t-1}, W_t) \\
    &\coloneqq 
    \frac{S^0-D^0}{\beta^P - \gamma^P}+ 
    \frac{\gamma^W}{\beta^P- \gamma^P}W_t - 
    \frac{\beta^{D1} }{\beta^P - \gamma^P}D_{t-1} + 
    \frac{U^S_t - U^D_t}{\beta^P- \gamma^P},  
\end{split}
\end{equation} 
where we have assumed that $\beta^P-\gamma^P \neq 0$ (a prerequisite of market clearing);
$\Phi$ is the solution operator of \eqref{eq:equilibrium}
mapping
 two functions (the demand and supply equations) and two real numbers (observed lagged demand and wind) to the
price that solves
\eqref{eq:equilibrium}.

Constraints like \eqref{eq:equilibrium} are not part of the standard SCM framework
\citep{pearl_causality_2009, bongers_foundations_2021}.
As the notation above suggests, we propose
to write the 
structural assignment for $P_t$ using the
solution operator $\Phi$ (which itself depends on the structural equations for $D_t$ and $S_t$), 
that is,
$    P_t 
    \coloneqq 
    \Phi(M_t^D, M_t^S; D_{t-1}, W_t)$. 
We refer to this  as a \emph{mechanism-augmented} SCM, which we introduce formally in Appendix~\ref{sec:constrained_graph}.
There, we also show how
mechanism-augmented SCMs 
model 
observational
and intervention distributions.

The structure of
standard SCMs 
can be represented by
directed acyclic graphs (DAGs) in which each variable in the SCM is a node
and directed edges point from causal parents to their children.
In the case of time series, these are graphs with infinitely many nodes.\footnote{
Appendix~\ref{sec:appendix_graphs} formally introduces directed acyclic graphs for time series.} 
Mechanism-augmented SCMs 
can be extended to time series analogously.
We call the corresponding graphical representations
\emph{causal mechanism 
graphs} 
and
\emph{causal mechanism (time)
graphs}, respectively.
Both of these constructions are detailed
in Appendix~\ref{sec:constrained_graph}.
In contrast to graphs of standard SCMs, where the mechanisms are part of the background model, 
causal mechanism graphs
contain explicit mechanism nodes.
Figure~\ref{fig:augmented-equilibrium-dag} 
shows an example of a causal mechanism graph for Model~I if there is no time dependence ($\beta^W = \beta^{D1} = 0$). 
\begin{figure}[t]
    \centering
    \begin{tikzpicture}[
        node distance=0.5cm and 2.0cm,
        every node/.style={font=\small},
        var/.style={circle, draw, minimum size=8mm, inner sep=1pt},
        dashedvar/.style={circle, draw, dashed, minimum size=8mm, inner sep=1pt},
        mech/.style={rectangle, draw, dashed, rounded corners,
                     minimum height=7mm, minimum width=11mm, inner sep=2pt},
        >=Latex
        ]   
        \node[var] (W) {\(W_t\)};
        \node[var, below=of W] (P) {\(P_t\)};
        \node[var, below=of P] (D) {\(D_t\)};
        \node[mech, left=1.85cm of P] (MS) {\(M_t^S\)};
        \node[dashedvar, left=2.0cm of D] (S) {\(S_t\)};
        \coordinate (midPD) at ($(P)!0.58!(D)$);
        \node[mech, right=2.0cm of midPD] (MD) {\(M_t^D\)};
        \draw[->] (W) -- (P);
        \draw[->] (W) -- (S);
        \draw[->] (MS) -- (P);
        \draw[->] (MS) -- (S);
        \draw[->] (P) -- (D);
        \draw[->] (P) -- (S);
        \draw[->] (MD) -- (P);
        \draw[->] (MD) -- (D);
    \end{tikzpicture}
    \caption{Causal mechanism graph for an equilibrium system without time dependence.
    The mechanism nodes \(M_t^D\) and \(M_t^S\) determine the equilibrium variable
    \(P_t\) through the solution operator \(\Phi\). Demand and supply are then
    realized by evaluating the corresponding mechanisms at \(P_t\).}
    \label{fig:augmented-equilibrium-dag}
\end{figure}
As expected from Equations~\eqref{eq:demand_autocorrelated}
and~\eqref{eq:price_autocorrelated}, 
$M_t^D$ appears as a parent of both $D_t$ and $P_t$ --
the equilibrium constraint~\eqref{eq:equilibrium} has made the demand mechanism $M_t^D$ a confounder of $P_t$ and $D_t$.
Figure~\ref{fig:intro} (left) shows 
a marginalized version 
of the causal mechanism time graph for Model~I:
following the marginalization procedure of latent projections
\citep{Verma1991}, see Appendix~\ref{sec:marginalization},
we have marginalized out
supply $S_t$ and its mechanism $M_t^S$, which are irrelevant for identifying $\beta^P$.

The distribution induced by a standard (time series) SCM is Markov with respect to the induced graph \citep{pearl_causality_2009, Lauritzen1990, thams_identifying_2022}, \
which lets us read off conditional independences among the variables from the graph: if two sets of nodes are d-separated in the graph (see Definition~\ref{def:dsep}), they are conditionally independent.
The same holds for all (marginalized) causal mechanism graphs considered in this paper:
we can also read off conditional independence statements over the retained random variables
by interpreting the mechanism nodes
as hidden confounders and then using the d-separation criterion (see
Appendix~\ref{sec:constrained_graph} for a formal argument).

We argue that displaying the mechanism node explicitly, as in Figure~\ref{fig:intro}, has the additional benefit of clarifying which interventions the model supports.
We 
discuss three examples of
interventions. 
First, 
arguably most naturally,
an intervention on the mechanism,
 $\mathrm{do}(M_t^D = m)$, changes the demand function entering the market-clearing condition, thus affecting the equilibrium price
 (Appendix~\ref{sec:constrained_graph}). 
An example is direct provision, which replaces the function by a price-inelastic one. 
Second, an intervention on the realized variable $D_t$, 
$\mathrm{do}(D_t = d)$, overwrites the demand outcome after the market has cleared; it does not change $M_t^D$ and does not affect the equilibrium price $P_t$.
An example of such an intervention in the electricity market is involuntary load shedding due to some equipment failure or measures taken by the grid operator.
(Some such interventions could require additional measures, not discussed here, to re-balance demand and supply for physical system stability, see Section~\ref{sec:application}.)
Either type of intervention (on the variable or its mechanism node)
can also be applied
to the lagged state $D_{t-1}$.
Third, interventions on $P_t$ and $W_t$ are standard: $\mathrm{do}(P_t = p)$ removes the edges into $P_t$,
and the demand equation~\eqref{eq:demand_autocorrelated}, being structural in $P_t$, determines the response of $D_t$. 
It is this interventional effect, $\beta^P$, that 
we 
aim to
identify in Section~\ref{sec:CIV-estimation} using $W_t$ as an instrument. 

A similar construction 
to the one for mechanism-augmented SCMs
can be done using the notation of potential outcomes \citep{neyman_application_1923, rubin_causal_2005, imbens_causal_2015, hernan_causal_2020}, where, again, we need to make use of the mechanisms directly. For example, for fixed mechanisms and fixed realizations of $U_t^D$ and $U_t^S$, $D_t(M_t^D, p, d) \coloneqq M_t^D(p, d) = D^0 + \beta^P p + \beta^{D1} d + U_t^D$ and $S_t(M_t^S, p, w) \coloneqq M_t^S(p, w)$ denote the potential outcomes of demand and supply at time $t$.
Such an extended potential outcome framework, however, does not lend itself to a straightforward graphical 
representation or to a methodology for determining the validity of conditional IV estimators.

\subsection{Alternative Demand Equations}\label{sec:demand_alternatives}
\begin{figure}
    \centering
    \includegraphics[width=0.9\linewidth]{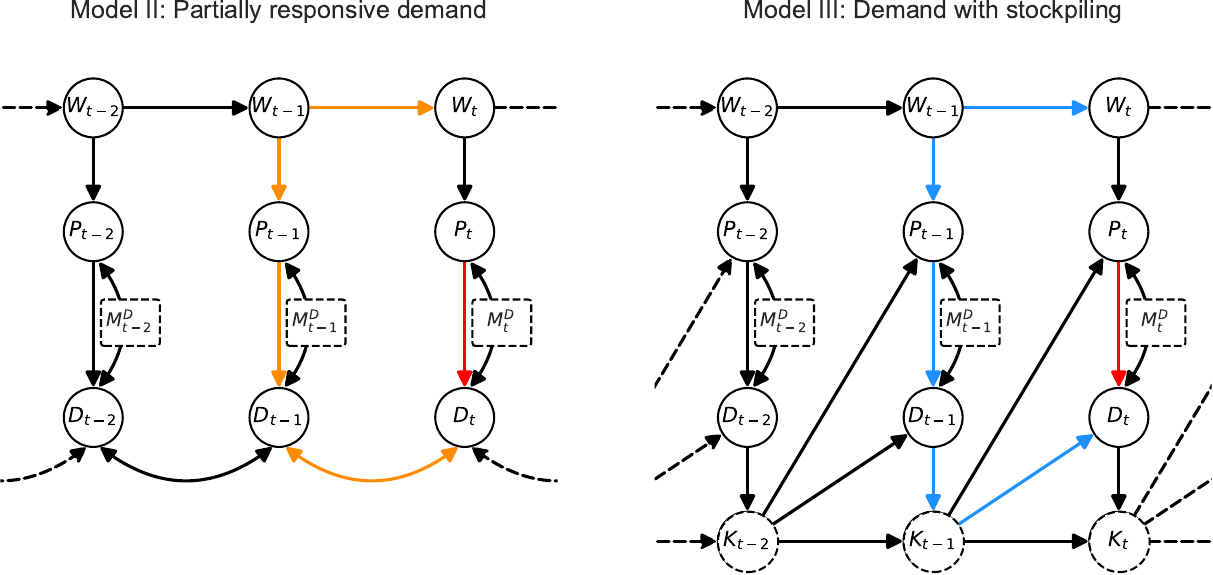}
    \caption{
        (Marginalized) causal mechanism time graphs for
        alternative demand equations. 
        A bidirected edge $X \leftrightarrow Y$ indicates that $X$ and $Y$ share an unobserved common cause.
     Left: Model~II~(Partially responsive demand), where aggregate demand can be divided into two components as in
    ~\eqref{eq:quantities_partially_unexposed}, one price-exposed and not autocorrelated, one non-price-exposed and with inertia.
    Right: Model~III~(Demand with stockpiling), where demand depends on the last state of an inventory, as in~\eqref{eq:demand_inventory}. 
    The blue path highlights why a naive~IV estimator fails to consistently estimate the causal effect of price on demand in the presence of stockpiling; by contrast, the orange path does not yield a similar violation.}
\label{fig:DAG_demand_variants}
\end{figure}

We now present two alternative models that also generate autocorrelated time series
 but with different structural dependencies in time.
First, Model~II (`Partially responsive demand') 
assumes that aggregate demand consists of two components: the first, $A_t$, is 
price-exposed but not directly structurally dependent on its own past, and the second, $B_t$, is non-price-exposed and exhibits direct structural autocorrelation.
In electricity markets, part of demand, such as retail consumers without real-time pricing, is not exposed to high-frequency price fluctuations.
The corresponding structural equations read
\begin{equation}\label{eq:quantities_partially_unexposed}
\begin{split} 
      \textit{Model~II:} \qquad D_t 
      & \coloneqq M_t^D(P_t, B_{t-1}) \\
       & \coloneqq A_{t} + B_{t}
        = (A^0 + \beta^P P_t + U^{A}_t) +  (B^0+ \beta^{B1} B_{t-1} + U^B_t)\\
       &= D^0 + \beta^P P_t +  \beta^{B1} B_{t-1} + U_t^{D}
\end{split}
\end{equation}
and
the price function becomes 
\begin{equation}\label{eq:price_partially_responsive}
\begin{split} 
    \textit{Model~II:} \qquad P_t 
    & \coloneqq \Phi(M_t^D, M_t^S; B_{t-1}, W_t) \\
    & \coloneqq \frac
       {S^0 - D^0}
       { \beta^P - \gamma^P}  
   +\frac
    {\gamma^W}
    {\beta^P - \gamma^P}
   W_t  
   - \frac{\beta^{B1} }
    {\beta^P- \gamma^P}B_{t-1}
   + \frac{ U^S_t - U^D_{t} }
    { \beta^P - \gamma^P}.
\end{split}
\end{equation}
Figure~\ref{fig:DAG_demand_variants} (left) depicts the corresponding graph.

Second, Model~III (`Demand with stockpiling') includes demand optimization over time: consumption shortly before or after $t$ may substitute for consumption at time $t$. In electricity markets, this is often called load shifting.
More precisely,
we introduce a parsimonious stockpiling mechanism in the spirit of \citet{hendel_measuring_2006}. 
 We let $K_t$ denote
 an inventory (of energy services) that accumulates past consumption and depletes over time, representing, for example, accumulated industrial production or the charge level of electric vehicles. Low prices at time $t$ encourage higher demand and thus a higher inventory, which in turn depresses future demand; 
 high prices work in reverse.
 We capture this by specifying demand and prices as
\begin{equation}\label{eq:demand_inventory}
         \textit{Model~III:} \qquad 
         D_t \coloneqq
         M_t^D(P_t, K_{t-1})\coloneqq
         D^0 + \beta^P P_t + \beta^{K} K_{t-1} + U^D_t,
\end{equation}
with $\beta^K<0$ and 
\begin{equation}
\label{eq:price_inventory}
\begin{split}
    \textit{Model~III:} \qquad P_t 
    &\coloneqq
    \Phi(M_t^D, M_t^S; W_t, K_{t-1}) \\
    &\coloneqq
    \frac
        {S^0 - D^0}
        { \beta^P - \gamma^P}  
   +\frac
    {\gamma^W}
    {\beta^P - \gamma^P}
    W_t  
   - \frac{\beta^{K}}{\beta^P- \gamma^P}K_{t-1}
   + \frac{ U^S_t - U^{D}_{t} }
    { \beta^P - \gamma^P}.
\end{split}
\end{equation}
The inventory state $K_t$ evolves according to
\begin{equation}
\label{eq:inventory}
    K_t \coloneqq 
        (1-\delta^K)K_{t-1}+\delta^DD_t-K^0,
\end{equation}
where $0<\delta^K<1$ represents depreciation of energy services, $\delta^D>0$ is the rate at which consumption is turned into energy services, and $K^0$ is the steady
consumption from the inventory.
Figure~\ref{fig:DAG_demand_variants} (right) depicts the corresponding graph.

\section{Graph-based IV Estimation 
in Equilibrium Models
} \label{sec:CIV-estimation}
We now 
review
sufficient conditions for valid conditional instrumental variable (CIV)
estimation in time series SCMs
 (Section~\ref{sec:conditionsIV}).
Their main component is a set of
graphical criteria
which have 
recently been extended from the
i.i.d.\ 
setting \citep{Brito2002generalized,henckel_graphical_2023} to time series settings \citep{thams_identifying_2022}.
In Section~\ref{sec:consistent_estimators}, we  
apply these CIV criteria
 to the three models introduced in Section~\ref{sec:graphical_representation} and present several valid CIV estimators. We argue that 
 significant differences
 between the estimates
 can inform researchers
 about the validity of the underlying models and assumptions.

Both of the following 
sections%
rely on
the demand equation being
additively separable and linear. 
In case of nonlinear models, we can either apply the same strategies 
(which yields a weighted average of local causal effects, in the spirit of \citet{angrist_interpretation_2000})
or extend our methodology; the latter extensions are discussed in
Appendix~\ref{sec:non-linear-ext}.

\subsection{Conditional IV Estimation in Linear Time Series}\label{sec:conditionsIV}

In the remainder of this paper,
we make repeated
use of
d-separation
\citep[short for directed separation;][]{pearl_causality_2009}. 
D-separation is a purely graphical criterion connected to conditional independences of random variables via the Markov condition \citep{Lauritzen1996}.
The original formulation of \citet{pearl_causality_2009} considers finite graphs; here, we use a version adapted
to (marginalized) infinite time series graphs.
\begin{definition}[Pearl's d-separation \citep{pearl_causality_2009}] \label{def:dsep}
Let $\mathcal{G}$ be the marginalized
version of an acyclic directed full time graph \citep{peters_causal_2013}, 
which contains a node for each variable and time point, 
with vertices\footnote{Here the $\mathbb{Z}$ component encodes time, e.g., $D_t$ is a vertex denoted as $(t,1)$.}
 $V = \mathbb{Z} \times \{1,\dots,d\}$,
and let $i_1, \dots, i_m$ be a path between vertices $i_1$ and $ i_m $ in $\mathcal{G}$. We say that the path $i_1, \dots, i_m$ is \textit{blocked by the set of variables} $ S \subseteq V \setminus \{i_1, i_m\}$ if there exists $k \in \{2,\dots, m-1\}$
such that either
\begin{enumerate}
    \item[(i)] $i_k \in S$ and $ i_k $ is not a collider\footnote{Let $\mathcal{G}$ be a marginalized version of an acyclic directed full time graph with vertices $V$.  For such graphs we define $v$ to be a \emph{collider} on a path whenever two consecutive edges have arrowheads at $v$ (for example, $u_1 \leftrightarrow v \leftarrow u_2$).} on the path $i_1, \dots, i_m$, or
    \item[(ii)] $ i_k $ is a collider on the path and 
    $(\{i_k\} \cup \mathrm{DE}(i_k)) \cap S = \emptyset$, where $\mathrm{DE}(i_k)$ are the descendants of $i_k$;
\end{enumerate}
if a path is not blocked by $S$, we say it is \emph{unblocked, given $S$}. 
For three mutually disjoint sets of vertices $ A, B, S \subset V$, we say that $A$ and $B$ are \emph{d-separated by $ S $} if all paths between $ A $ and $ B $ are blocked by $ S $. If this is the case, we write 
 $   A \perp\!\!\!\perp_{\mathcal{G}} B \mid S$. 
\end{definition}
D-separation helps us construct valid IV estimators
 in time series. The following statement is an adapted version of Theorems~5~and~8 of \cite{thams_identifying_2022}.
Let $d,d' \in \mathbb{N}$, $d \leq d'$ and let $V' = \mathbb{Z}\times\{1,\dots,d'\}$. Consider a linear SCM for time series 
(see Appendix~\ref{sec:appendix_scm})
over variables $X_{V'}$ and let $\mathcal{G}$ be the induced marginalized version of an acyclic directed full time graph with vertices $V = \mathbb{Z} \times \{1,\dots,d\}$.
Let\footnote{Depending on the context, and slightly overloading notation, $X_S$ for $S \subseteq V$ is either the set of random variables $\{X_s \ \vert \ s \in S\}$ or the random vector $(X_{s_1}, \dots, X_{s_n})$, where $s_i$ are in lexicographic order.} $\mathcal{I}, \mathcal{X}, \mathcal{B}, \{Y\} \subseteq V$
be finite and pairwise disjoint,
assume that $X_\mathcal{I}, X_\mathcal{X}, X_\mathcal{B}$ and $X_Y$ have zero means and finite second moments, and let $\beta \in \mathbb{R}^{|\mathcal{X}|}$ be the causal coefficient with which $X_\mathcal{X}$ enters the structural equation for $X_Y$, that is,
\begin{equation}
    X_Y = \beta^T X_\mathcal{X} + \gamma^T X_K + \epsilon^Y,
\end{equation}
for some variables
$X_K \subseteq X_V \setminus X_\mathcal{X}$ 
(some of the components of $\beta$ can be zero, so not all variables in $X_\mathcal{X}$ have to be parents of $X_Y$). 

The following three requirements on $\mathcal{I}, \mathcal{X}, {\mathcal{B}}$ and $Y$
are the CIV criteria:
\begin{itemize}
    \item[(CIV1)] $\mathcal{I}$ and $Y$ are d-separated given $\mathcal{B}$ in the graph $\mathcal{G}_{\mathcal{X} \not\rightarrow Y}$, which denotes the graph obtained by removing all direct edges from $\mathcal{X}$ to $Y$ from $\mathcal{G}$.
    \item[(CIV2)] $\mathcal{B}$ does not contain any descendant of $\{Y\}$ in $\mathcal{G}$ \footnote{We use a slightly weaker version of (CIV2) than \cite[Proposition 20]{thams_identifying_2022}. For identification of direct structural coefficients,
their proof of Proposition 20 only requires the conditioning set to exclude descendants
of $Y$, rather than descendants of $\mathcal X\cup{Y}$.}.
    \item[(CIV3)] the matrix $\mathbb{E}[\text{cov}(X_{\mathcal{X}}, X_\mathcal{I}|X_\mathcal{B})]$ has rank $|\mathcal{X}|$, that is, full row rank. 
\end{itemize}
(CIV1) and (CIV2) are properties of the underlying causal graph; (CIV3) is a property of the induced distribution. (CIV1) is a graphical exclusion restriction: once the direct effect of $X_{\mathcal X}$ on $Y$ is removed, the instruments $X_{\mathcal I}$ must not 
be connected to
$Y$ through any remaining unblocked 
path. (CIV2) ensures that the conditioning variables $X_{\mathcal B}$ are not themselves consequences of the outcome, avoiding collider bias. Finally, (CIV3) is an instrument relevance condition: after conditioning on $X_{\mathcal B}$, the instruments must remain sufficiently correlated with $X_{\mathcal X}$. 

Assume we have observations at time points $1, \ldots, T$.
If
an estimator is based on 
$\mathcal{I}, \mathcal{X}, {\mathcal{B}}$
such that
requirements (CIV1), (CIV2), and (CIV3) are met, 
we call it a \emph{valid} CIV estimator. 
In this case,
\begin{equation}
\label{eq:civ}
    \hat{\beta} \coloneqq \arg\min_{b} \|\hat{\text{cov}}(X_Y - b^TX_\mathcal{X}, X_\mathcal{I}|X_\mathcal{B})\|_{\hat\Omega}, \quad \hat\Omega \coloneqq \hat{\text{cov}}(X_\mathcal{I} \mid X_\mathcal{B})^{-1},
    \end{equation}
is a 
consistent estimator for $\beta$ as $T \to \infty$ (assuming the system has a unique, strongly stationary and ergodic solution, Appendix~\ref{sec:appendix_scm}); for a detailed definition of $\hat{\beta}$ and its closed-form solution, see Appendix~\ref{app:tsls}.\footnote{
The estimator in \eqref{eq:civ} is 
similar
to a standard IV (or GMM) estimator 
(e.g., \cite{hansen1982large, chamberlain1987asymptotic})
based on the conditional moment restriction $\mathbb E[(X_Y-\beta^T X_{\mathcal X})X_{\mathcal I}\mid X_{\mathcal B}] = 0$,
where $X_Y$ is the dependent variable, $X_{\mathcal X}$ the set of endogenous variables, $X_{\mathcal I}$ the instrument set, and $X_{\mathcal B}$ the set of conditioning variables;
here, we are considering time series. The norm is $\|v\|_{\hat\Omega}^2 = v^\top \hat\Omega v$. The chosen weighting yields 2SLS and affects the point estimate only in
overidentified specifications among estimators \#5--\#7.%
}
We call $\hat{\beta}$ the \emph{conditional instrumental variable (CIV)} estimator and denote it by $\mathrm{CIV}(\mathcal{I} \mid \mathcal{X} \to Y \mid \mathcal{B})$. 
If at least one%
condition is not met, 
we
call the estimator \emph{invalid}
 and, in general, it may be inconsistent \citep{henckel_graphical_2023}.
If (CIV1) and (CIV2) are satisfied
but (CIV3) is not, any estimation technique 
with
valid inference
 yields unbounded confidence sets containing the true causal effect with large probability. 
\subsection{
Consistent Estimation of Price Elasticities
}\label{sec:consistent_estimators}
We now apply the CIV criteria to derive valid estimators for Models~I--III. The theory of 
Section~\ref{sec:conditionsIV}
applies because, for 
all three examples, we can create a corresponding surrogate 
linear SCM for time series.
To do so, we start with the equations 
defining the model
but then, for the assignment of the price,
we use the explicit form of the solution operator.
In all three cases, this results in an SCM with a hidden confounder between price and demand
($U^D_t$ in the case of Model~I, see Equation~\eqref{eq:price_autocorrelated});
by construction, the graph 
induced by 
the surrogate 
SCM
equals
the 
marginalized version of the causal mechanism graph.
We can thus apply the above theory to identify $\beta^P$. 
As an aside, the 
two causal models are not interventionally equivalent \citep{peters_elements_2017}:
while, for example, interventions on $P_t$ are modeled correctly,  
interventions on $D_t$ are not 
-- 
the reason is that we have used Equation~\eqref{eq:demand_autocorrelated} in deriving Equation~\eqref{eq:price_autocorrelated}.\footnote{
For example, if we were to consider an intervention on $D_t$ and set it to $x$, we would obtain the price
equation $P_t = (x - S^0 - \gamma^W W_t - U^S_t)/\gamma^P$.
}
Using an SCM to model interventions
on some but not all variables is similar in spirit to the decision-theoretic framework by 
\citet{Dawid2021}. 

For example, consider Model~I, 
  described
   by
   Equations~\eqref{eq:demand_autocorrelated}~and~\eqref{eq:price_autocorrelated}, and the corresponding causal mechanism time graph in Figure~\ref{fig:intro}. The naive~IV estimator $\#1$ that neglects time dependencies equals 
   $\mathrm{CIV}(W_t \mid P_t \to D_t \mid \emptyset)$, that is, we set $X_{\mathcal{X}} = (P_t)$, $X_{Y} = \{D_t\}$, $X_{\mathcal{B}} = \emptyset$ and $X_{\mathcal{I}} = (W_t)$. We see that (CIV1) is not satisfied: After removing the edge $P_t \to D_t$ in Figure~\ref{fig:intro}, we obtain the graph $ \mathcal{G}_{P_t \not\rightarrow D_t}$. In this graph, there are several paths  through $W_{t-1}$ that connect
 the instrument $W_t$ to the dependent variable $D_t$
  which are unblocked given $X_{\mathcal{B}} = \emptyset$.
   An example is the blue path in Figure~\ref{fig:intro}, $W_t \leftarrow W_{t-1} \to P_{t-1} \to D_{t-1} \to D_t$. $W_t$ is therefore not d-separated from $D_t$.%
     
We provide an overview of some available estimators and their validity in Table~\ref{tab:estimators}. 
Each entry is written 
using the $\mathrm{CIV}(\cdot \mid \cdot \mid \cdot)$ notation 
introduced above; 
the corresponding estimator is given by Equation~\eqref{eq:civ} when replacing
$\mathcal{I}$, 
$\mathcal{X}$, 
$Y$,
and $\mathcal{B}$
with the variables indicated by the arguments
of $\mathrm{CIV}$.

The CIV estimators 
fall into two categories.
The first category uses a selection of path-blocking variables as the conditioning set. Such a set can contain past realizations of the instrument $W_t$, the endogenous variable $P_t$, or the outcome variable $D_t$. For example, the estimator $\#3$, $\mathrm{CIV}(W_t \mid P_t \to D_t \mid D_{t-1})$, contains only one lag of the outcome variable; it is valid for Model~I. The estimator $\#2$, $\mathrm{CIV}(W_t \mid P_t \to D_t \mid \{W_{t-s}\}_{s=1}^L)$, contains $L$ lags of the instrument and is valid for Models~I--III.%
Intuitively, by including the past realizations of the instrument in the conditioning set, we block any path starting with
 $W_t \leftarrow W_{t-k}$ for some $k\geq 1$.
Estimator $\#2$ is valid for a strictly more general model class: any model yielding a (marginalized) graph in which
 $P_t \rightarrow D_t$ is the only directed path between $P_t$ and $D_t$, $W_t$ has no other directed path to $D_t$, $W_t$ and $D_t$ are unconfounded, and the set of parents of $W_t$ is a subset of $\{W_{t-1}, \ldots, W_{t-L}\}$. This model class contains Models~I,~II, and~III. We therefore consider $\#2$ our benchmark estimator. 

The second category
is called Nuisance IV \citep{thams_identifying_2022}. It estimates additional (nuisance) coefficients alongside the coefficient of interest. This%
can 
yield identifiability
in settings where a conventional IV specification with only the treatment regressor fails. As an example, estimator~$\#7$ regresses $D_t$ on the 
regressor vector $X_t = (P_t, P_{t-1}, D_{t-1})^\top$ using the instrument vector $I_t = (W_t, W_{t-1}, \ldots, W_{t-L})^\top$, solving the moment condition
\begin{equation}
\mathbb{E}\left[I_t\left(D_t - \beta^{P} P_t - \beta^{P1} P_{t-1} - \beta^{D1} D_{t-1}\right)\right] = 0,
\end{equation}
jointly for $(\hat\beta^P, \hat\beta^{P1}, \hat\beta^{D1})$. In practice, this can be implemented as an ordinary multivariate two-stage least-squares regression: each component of $X_t$ is projected onto $I_t$ in a first stage, and $D_t$ is regressed on the fitted values in a second stage. Only $\hat\beta^P$ is of interest; the coefficients on $P_{t-1}$ and $D_{t-1}$ are estimated but ignored
and need not be valid themselves.
Jointly estimating these nuisance coefficients makes estimator~$\#7$ valid for Models~I and~II.
\begin{table}[ht]
\centering
\begin{tabular}{ p{0.2cm} p{7.1cm}||P{1.5cm}|P{1.5cm}|P{1.5cm} }
    & Estimator & 
        Model~I
        Figure~\ref{fig:intro} (left)
        &
        Model~II
        Figure~\ref{fig:DAG_demand_variants} (left)%
        &
        Model~III
Figure~\ref{fig:DAG_demand_variants} (right)
\\
    \hline
    & Naive~IV &&& \\
   $\#1$& $\mathrm{CIV}(W_t \mid P_t \to D_t \mid \emptyset)$
   &\xmark&\cmark&\xmark\\
\hline
     &Conditional IV  & &&\\
     $\#2$& $\mathrm{CIV}(W_t \mid P_t \to D_t \mid \{W_{t-s}\}_{s=1}^L)$
   & \cmark  & \cmark  & \cmark\\
    $\#3$&$\mathrm{CIV}(W_t \mid P_t \to D_t \mid D_{t-1})$
   & \cmark  & \xmark & \xmark \\
    $\#4$&$\mathrm{CIV}(W_t\mid P_t \to D_t \mid \{W_{t-s}, P_{t-s}, D_{t-s}\}_{s=1}^{L}$)%
    & \cmark  & \cmark & \cmark \\

    \hline

    & Nuisance IV &&&\\

    $\#5$&$\mathrm{CIV}(\{W_{t-s}\}_{s=0}^L \mid (P_t, P_{t-1}) \to D_t \mid \emptyset)$%
    & \xmark  &\cmark &\xmark \\

    $\#6$&$\mathrm{CIV}(\{W_{t-s}\}_{s=0}^L \mid (P_t, P_{t-1}) \to D_t \mid D_{t-1})$%
    & \cmark  & \xmark 
    &\xmark \\

     $\#7$&$\mathrm{CIV}(\{W_{t-s}\}_{s=0}^L\mid (P_t, P_{t-1}, D_{t-1}) \to D_t \mid \emptyset)$%
    & \cmark  & \cmark   &\xmark\\

      $\#8$&$\mathrm{CIV}((W_{t}, W_{t-1})\mid (P_t, D_{t-1}) \to D_t \mid \{W_{t-s}\}_{s=2}^{L+1})$%
    & \cmark  & \cmark   &\cmark\\
\end{tabular}
\caption{
Estimators and their validity for the causal effect $\beta^P$ from $P_t \to D_t$ for Models~I--III. 
}
\label{tab:estimators}
\end{table}

\begin{figure}
    \centering
    \includegraphics[width=1\linewidth]{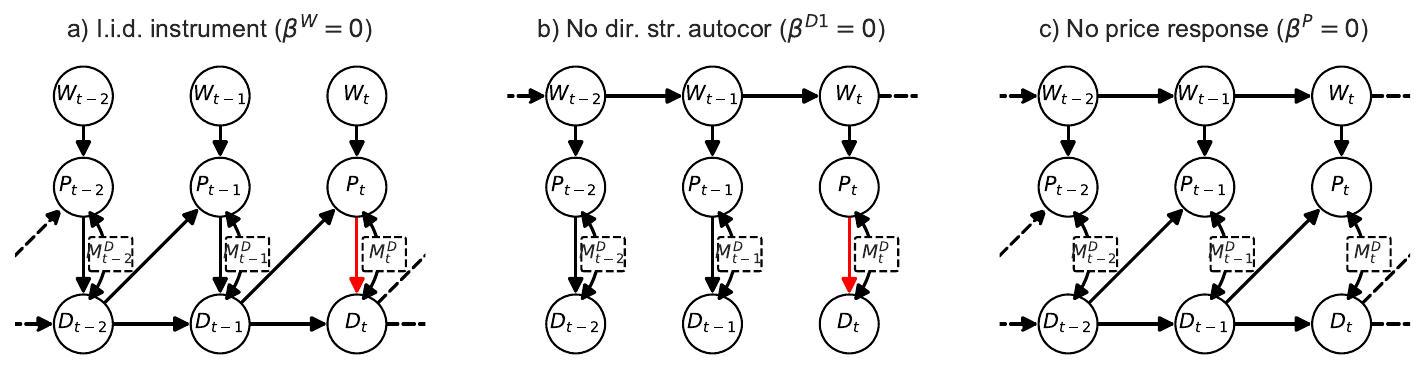}
    \caption{
    Graphical representation of special cases of Model~I, for which naive~IV is a valid estimator. Left: The instrument is i.i.d., i.e., not autocorrelated.
 Middle: The demand function is not autocorrelated. Right: Demand is not responsive to prices.
}
    \label{fig:DAG_special_cases}
\end{figure}

\paragraph*{Special cases for a valid naive~IV estimator.}
Even though the naive~IV estimator does not necessarily satisfy (CIV1),  there are special cases in which the naive estimator is valid, as visualized in Figure~\ref{fig:DAG_special_cases}. 
These occur (i) when the instrumental time series is i.i.d., 
(ii) when demand is not directly affected by its past realizations, or (iii) when there is no response to prices, i.e., $\beta^P = 0$.
In all three cases, no unblocked path exists (given $\emptyset$) from $W_t$ to $D_t$ through the past.

\paragraph*{Local projection IV estimator.}\label{par:lp}
Estimator $\#4$ in Table~\ref{tab:estimators} is the lag-augmented local projection IV estimator for $P_t \to D_t$ at impact (time horizon $h=0$).
This estimator is a standard tool for identifying impulse response functions in macroeconomics \citep{stock_identification_2018, montiel_olea_local_2021}\footnote{The general formulation of the local projection IV estimator includes lags up to infinity \citep[see e.g.,][Section 3.3]{montiel_olea_local_2021}, which in practice translates into a `large number of lags', which we represent by $L$.}.
It is a CIV estimator with lags of the instrumental, endogenous, and dependent variables in the conditioning set.
Our framework allows one to determine its validity given a graph.
For Models~I--III,
conditioning only on the lags of the instrument, as done by estimator $\#2$, suffices.
It is straightforward to see that 
$\#2$ and $\#4$ are valid well beyond Models~I--III. 
Importantly,
the local projection IV estimator $\#4$ is not valid in 
more model classes than the other estimators: 
if in Model~I there is  additional hidden
confounding between $D_{t-1}$ and $W_t$ and between $D_{t-1}$ and $D_{t}$, 
then $\#4$ and $\#2$ are both invalid while the nuisance IV estimator $\#8$ remains valid. 

\paragraph*{Model validation.} The existence of several valid IV estimators
allows us to
test 
our assumptions
about the dynamics of 
demand response. 
More precisely, we test the null hypothesis that the data follow the assumed model. Under that null, any two estimators that are valid for the model are consistent for the same $\beta^P$, and the probability that their confidence intervals overlap tends to one and is at least one minus twice the significance level.
If, for example, Model~II describes the true time dependencies, the intervals of all estimators but $\#3$ and $\#6$ in Table~\ref{tab:estimators} should overlap, and if they do not, we should reject Model~II. 

\section{Simulating Electricity Market Data
}\label{sec:Simulation}
\begin{figure}[t]
    \centering
    \includegraphics[width=.9\linewidth]{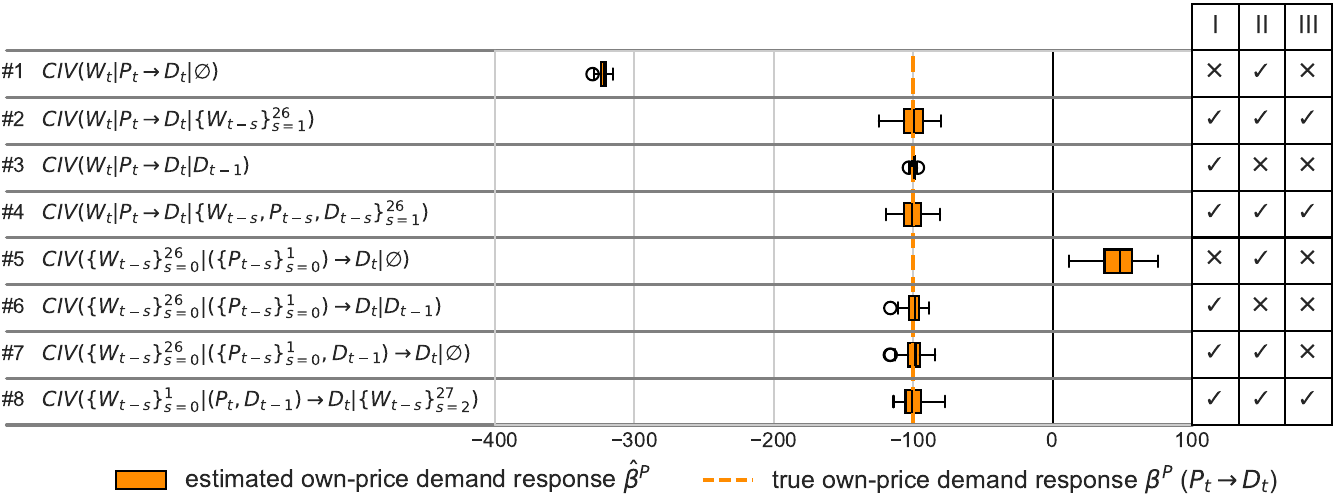}
    \caption{
    CIV estimates of the true demand response $\beta^P = -100$ MW/(EUR/MWh) when data are simulated from Model~I (Figure \ref{fig:intro} left). The structural autocorrelation of demand is $\beta^{D1} = 0.7$. The columns on the right indicate the validity according to the CIV criteria (see Table~\ref{tab:estimators}).}
    \label{fig:simulation_autocorrelated}
\end{figure}

\paragraph*{Data generation and estimation.}
We now evaluate the CIV estimators of Section~\ref{sec:consistent_estimators} in a controlled environment motivated by the German electricity market (see Section~\ref{sec:application}).  
We generate multiple datasets of three hourly series (wind generation, equilibrium demand\footnote{In the electricity market, equilibrium demand is also referred to as load or electricity consumption.} and the equilibrium price) under the linear specification of Models~I--III; Appendix~\ref{app:data_generation} gives the full data-generating process; Appendix~\ref{sec:non-linear-ext} extends the experiment to a quadratic demand curve.
All CIV estimators are calculated with the Python package linearmodels \citep{sheppard_linearmodels_2024}, using the kernel correction for heteroskedasticity- and autocorrelation-consistent (HAC) standard errors.%
\paragraph*{Validity of estimators.}
We validate the theoretical findings of Section~\ref{sec:CIV-estimation}. In Figure~\ref{fig:simulation_autocorrelated}, we apply the CIV estimators specified in Table~\ref{tab:estimators} to the data simulated from Model~I.\footnote{Appendix~\ref{app:divergence} provides simulation results for Model~II in Figure~\ref{fig:simulation_two_demand_types} and for Model~III in Figure~\ref{fig:simulation_cross}.}
We find that the estimate ranges of all valid estimators overlap (compare the overlap with the check marks for Model~I on the right of Figure~\ref{fig:simulation_autocorrelated}). By contrast, if one had assumed Model~II, one would be forced to reject the associated structural assumptions since estimators $\#1$ and $\#5$ deviate from other valid estimators such as $\#2$ and $\#8$.  
The simulation thus underscores our argument that comparing the overlap of confidence intervals 
from several presumably valid estimators provides a means of rejecting incorrect structural assumptions. 

\paragraph*{Magnitude of bias.}
Under Model~I, the bias of the naive~IV estimator $\#1$ grows with the structural autocorrelation of both the instrument ($\beta^W$) and demand ($\beta^{D1}$); \citet[][Prop.~17, for a special case]{thams_identifying_2022} compute it in closed form. When the instrument is strongly autocorrelated, even a moderate structural autocorrelation of demand produces a large bias (Figure~\ref{fig:heatmap} left), and the lag-one autocorrelation of wind generation exceeds $0.9$ in our application (Figure~\ref{fig:autocorrelation}). 
Furthermore,
this bias cannot be inferred from 
observed autocorrelations of wind and demand,
since the same 
observed autocorrelations 
may correspond to a
bias ranging from zero to over $100\%$ (Figure~\ref{fig:observed_ar}). 
Obtaining a consistent estimate requires knowing the underlying structure; thus the appeal of deriving and comparing several (supposedly) valid estimators to falsify models.

\begin{figure}
    \centering
    \includegraphics[width=1\linewidth]{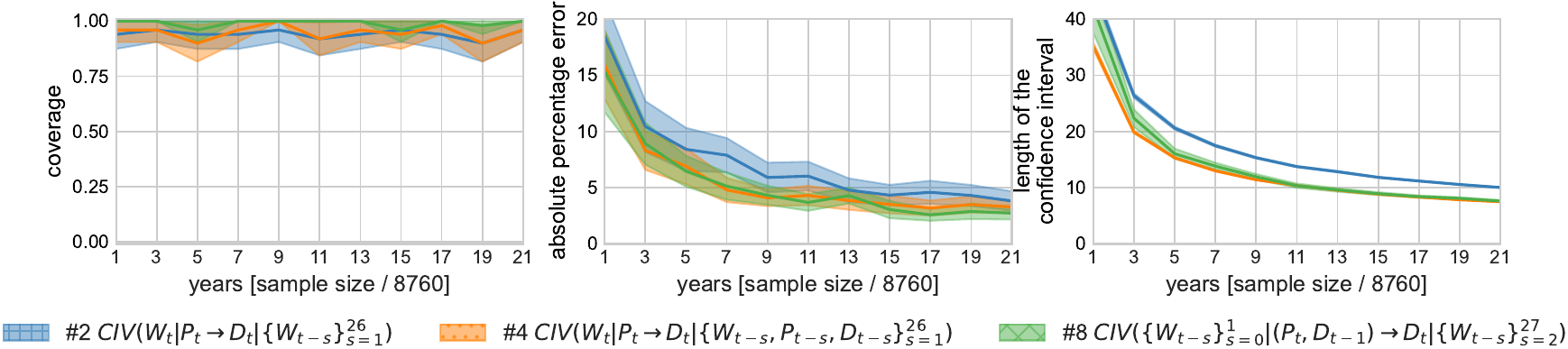}
    \caption{
    Evaluation of the statistical performance of conditional vs. nuisance IV estimators for Model~I based on the performance indicators: (left) coverage, (middle) the average absolute percentage error, and (right) the length of the confidence interval. $(\beta^{D1} = 0.9$.) The shaded areas represent the 95\% confidence intervals. For each sample size, we run $ 50$ simulations.
    }
    \label{fig:indicators_model_1}
\end{figure}

\paragraph*{Empirical performance of different CIV estimators.}
On data generated from Model~I,
we compare the three estimators valid across Models~I--III, $\#2$, $\#4$, and $\#8$, 
with respect to
 \emph{coverage},  
 \emph{average absolute percentage error}, 
 and \emph{length of the confidence interval}.
Figure~\ref{fig:indicators_model_1} reports these indicators
as a function of the sample size;
Appendix~\ref{app:indicator_plots} shows results for Models~II and III. 
As suggested by our validity analysis,
all estimators attain nominal coverage (left) across all three models.
The lag-augmented local projection IV estimator $\#4$ performs at least as well as the other two under every model, and clearly best under Model~II (see Figure~\ref{fig:indicators_model_2}). The benchmark $\#2$ is the least efficient under Model~I.

\section{Estimation of the Price Elasticity of Hourly German Electricity Demand}\label{sec:application}
We estimate the own-price elasticity of the aggregate German electricity demand using the CIV estimators derived in Section~\ref{sec:CIV-estimation}.
We also
discuss what the disagreement between the estimators can reveal
 about the structure of the dynamic response. 

\paragraph*{The electricity market.} The electricity system must always be in balance for physical reasons. The primary mechanism to equalize demand and supply is the day-ahead auction of the wholesale market. This is where generators, retail suppliers, and large industrial consumers submit their price-volume bids for each hour of the following day.\footnote{The bid volume does not correspond to aggregate demand, and the bid curves do not correspond to the demand equation. This is because market participants can submit net bids (pool-based bidding) and have the option of trading on other marketplaces. Hence, the demand elasticity is not observable but must be estimated. However, the clearing price is informative about the equilibrium price \citep{knaut_when_2017}.
} Because electricity storage is expensive and limited in capacity, wholesale electricity prices are much more volatile than those of other commodities \citep{ciarreta_renewable_2020, maniatis_impact_2022, hosius_impact_2023}.%
Our Models~I--III abstract from the day-ahead auction and model separate market clearings for each hour. 

\paragraph*{Data.} We estimate the own-price elasticity on electricity market data between November 1, 2017, and December 31, 2020 (see Table~\ref{tab:summary_stats} for summary statistics). Electricity demand 
is defined as consumption in the joint German-Luxembourg market. The electricity price in euro per megawatt-hour (EUR/MWh) corresponds to the clearing price of the day-ahead auction on EPEX Spot, the largest marketplace for the market zone. 
A detailed description of the data and sources can be found in Appendix~\ref{app:data}. When we split the sample, as for the on-peak and off-peak analysis of Appendix~\ref{app:results_on_off_peak}, we determine the relevant lags before cutting the dataset.
\begin{table}[htbp]
    \centering
    \footnotesize
    \setlength{\tabcolsep}{4pt}
    \begin{tabular}{lccccccc}
        \toprule
        & \textbf{Mean} & \textbf{Std. Dev.} & \textbf{Min} & \textbf{Median} & \textbf{Max} & \textbf{Skewness} & \textbf{Excess kurtosis} \\
        \midrule
        \textbf{Consumption [GWh/h]} & 59.9 & 11.3 & 34.0 & 59.6 & 92.0 & 0.17 & -0.69 \\
        \textbf{Price [EUR/MWh]} & 38.0 & 17.6 & -90.0 & 38.0 & 200.0 & -0.35 & 3.8 \\
        \textbf{Wind Generation [GWh/h]} & 14.1 & 10.1 & 0.1 & 11.6 & 46.1 & 0.87 & -0.06 \\
        \bottomrule
    \end{tabular}
    \caption{
    Summary statistics; $T = 27{,}072$ after removing national holidays.}
    \label{tab:summary_stats}
\end{table}

\paragraph*{Identification.} To overcome the endogeneity problem induced by the equilibrium condition, we use wind power generation as an exogenous time series.
 Wind speed and wind power generation are commonly used as supply-shifting instruments because they are relevant and credibly exogenous \citep{bonte_price_2015, knaut_when_2017, fabra_estimating_2021, hirth_how_2024}. 
The assumption that the structural equation of wind power generation does not depend on the price of electricity is based on the argument that renewable generators have marginal costs close to zero. Additionally, most German wind power generators receive a subsidy
per unit of electricity produced, which induces an opportunity cost of not producing, even if prices are negative.\footnote{Negative prices represent approximately $2.5\%$ of the observations. If wind generators stopped producing at negative prices, this would indicate that the instrument is not valid for that price range. We observe wind generation at negative prices, which does not rule out partial curtailment.}

We follow \citet{hirth_how_2024} in proposing that wind power is independent of the noise terms of demand and supply after conditioning on suitable covariates.
While it is reasonable to assume that wind power does not affect demand in any way other than through the price of electricity (see Appendix~\ref{sec:microfoundations} for a discussion on consumers using wind to forecast prices), 
we expect the existence of common causes such as seasonality and weather events. 
We therefore condition on the following covariates: seasonal dummies (hour of the week, month, and year) and weather controls (sunlight, and heating and cooling degree hours). Furthermore, we include other covariates to increase the precision of the estimate, namely commodity prices (natural gas, coal, and emission allowances), solar PV generation, and other calendar controls, see details in Appendix~\ref{app:data}. 
In Models~I--III, we allow $W_t$ to have a structural autocorrelation of one lag, and in that case, 
some estimators (such as $\#2$) that have a single time lag in the conditioning (or nuisance) set are valid.
However, observed wind power exhibits a higher order of autocorrelation (see Figure~\ref{fig:autocorrelation}), which needs to be reflected in the construction of the different estimators (the argument remains the same). 
Given the observed autocorrelation pattern, we include up to $50$ lags of wind power generation in the conditioning set of the CIV estimators $\#2$ and $\#4$ and of the nuisance IV estimator $\#8$, and as instruments for the nuisance IV estimators $\#5$ to $\#7$. We also observe a significant autocorrelation pattern in the demand time series (this is expected if there is indeed a causal effect from wind power generation on demand, but it may have other sources, too).%

\paragraph*{Models of electricity demand.}
The literature identifies two main dynamic mechanisms: demand inertia and load shifting (for a comprehensive classification of demand response in the electricity market, see \cite{albadi_summary_2008}).
First,
processes that run for extended periods, such as multi-hour industrial or appliance cycles, make demand at $t$ depend on past load, as in Model~I.
Second, electricity-consuming processes can be shifted in time to exploit (high-frequency) power price differentials. 
Load shifting may arise whenever electricity is used to build up an inventory of energy services, such as charge in an electric vehicle or thermal mass in a building, as captured in
Model~III.
Our alternative demand Model~II assumes that demand can be divided into two components: one that is price-exposed and reacts instantaneously, and one that is inertial
 and not exposed to price variations, motivated by retail consumers who lack real-time pricing.

\begin{figure}[t] 
    \centering
    \includegraphics[width=1\linewidth]{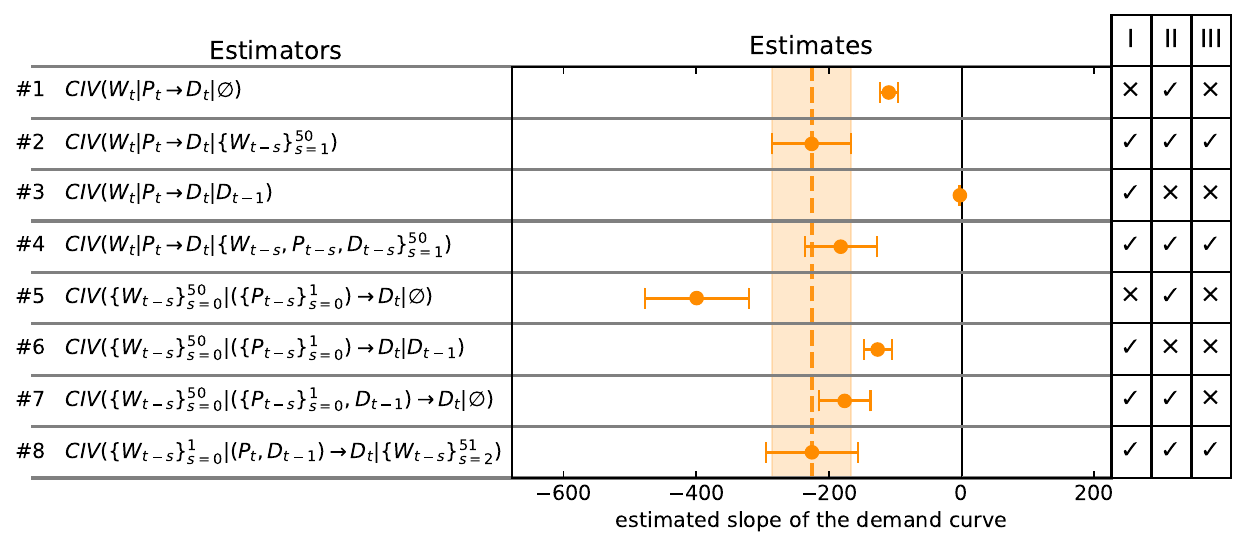}
    \includegraphics[width=1\linewidth]{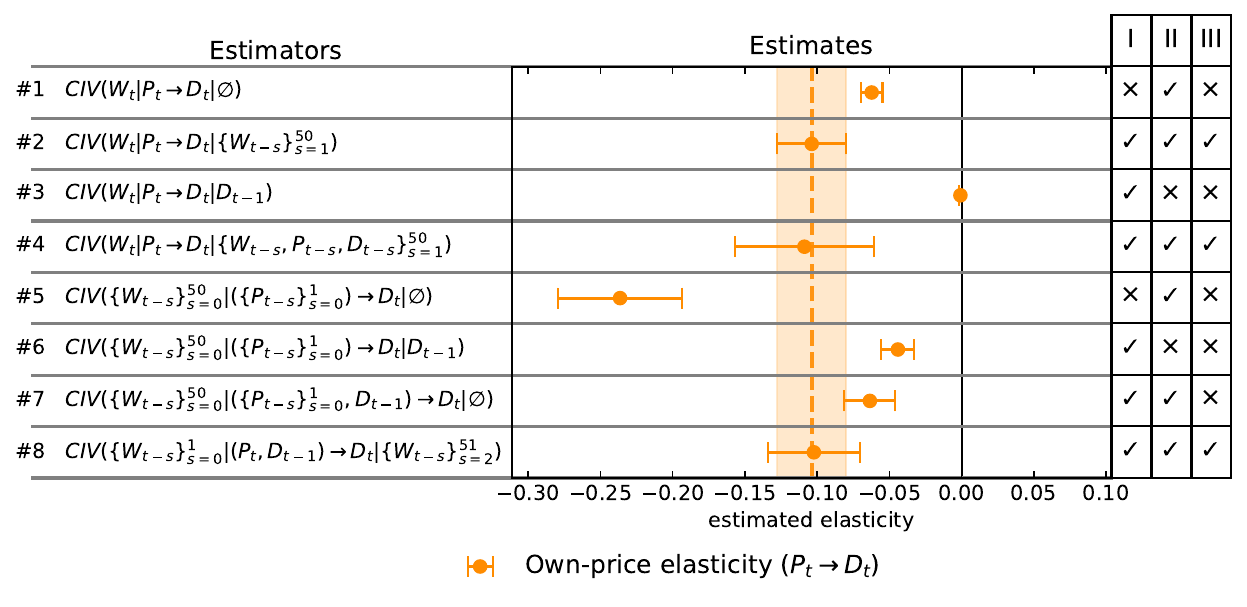}
 \caption{
 The own-price elasticity of aggregate German electricity demand, obtained by CIV estimators (top: linear specification; bottom: log-log specification). The highlighted area corresponds to the $95$\% confidence interval of estimator $\#2$. The tables on the right are reproductions of Table~\ref{tab:estimators}, showing the validity of each strategy for the structural causal models described in Section~\ref{sec:graphical_representation}.}
    \label{fig:sensitivity_aggregated}
\end{figure}

\paragraph*{Alternative assumptions on the demand equation.}
The models we consider make additional assumptions about the demand response. First, they posit that 
demand depends
linearly on price. 
To show that dynamics matter irrespective of the functional form, we include results for%
a log-log (constant-elasticity) specification
in Figure~\ref{fig:sensitivity_aggregated};\footnote{For the log-log transformation, we exclude prices of zero and below. 
} Appendix~\ref{sec:non-linear-ext} discusses how the same type of estimator could be implemented under more general forms. Second, the models assume that the elasticity is constant across hours and seasons. This implies that consumers who are active during the day have, on average, the same elasticity as consumers who are active at night. We investigate the relevance of this assumption in more detail in a robustness analysis in Appendix~\ref{app:results_on_off_peak}.

\paragraph*{Quantitative results.}
Figure~\ref{fig:sensitivity_aggregated} reports the quantitative results.
The CIV estimators $\#2$ and $\#4$, and the nuisance IV estimator $\#8$, which are valid across a large model class, have overlapping confidence intervals,
all estimating
a linear demand response $\hat{\beta}^P$ of approximately $-200$ MW/(EUR/MWh)\footnote{$\#2$ estimates a value of $-226$, $\#4$ of $-182$, and $\#8$ of $-226$. All estimates in MW/(EUR/MWh).}
and a log-log elasticity of $-0.1$\footnote{$\#2$ estimates a value of $-0.104$, $\#4$ of $-0.109$, and $\#8$ of $-0.102$. Estimates are unitless.}.
To gauge magnitude, a supply shift of one standard deviation of the instrument ($816$ MW) lowers the equilibrium price by about $-0.78$ EUR/MWh and raises the equilibrium quantity by $174$ MW.\footnote{
The standard deviation is taken from the wind residuals after conditioning on estimator $\#2$'s controls; the price change is this deviation times the first-stage coefficient $\hat{\pi}=-0.00095$ (EUR/MWh)/MW.}

\paragraph*{Overlap pattern.}
We can also analyze the pattern of the different CIV estimators to learn about the real-world dynamics of electricity demand.
We can reject a linear model without relevant structural autocorrelation. 
If the real-world dynamics were such that the naive estimator were consistent, then estimator $\#2$ would be consistent as well.
 But the pattern in Figure~\ref{fig:sensitivity_aggregated} shows that this is unlikely to be the case:
the naive~IV estimator that ignores time dynamics yields $-110$ MW/(EUR/MWh) 
against $-226$ 
MW/(EUR/MWh)
for the benchmark estimator $\#2$, understating the response by a factor of two, with non-overlapping confidence intervals.
Turning to our structural assumptions, 
first, we reject Model~I: estimators $\#3$ and $\#6$, both valid under Model~I, are far from $\#2$, $\#4$, and $\#8$, with non-overlapping confidence intervals.
Second, we reject Model~II: estimators $\#1$ and $\#5$, both valid under Model~II, similarly disagree with $\#2$, $\#4$, and $\#8$.
Third, we cannot reject Model~III. The three estimators valid under Model~III ($\#2$, $\#4$, and $\#8$) have overlapping confidence intervals, and in the log-log specification their point estimates are almost identical. 
By construction,
our approach only allows us to falsify structural assumptions but not to confirm them. 
Thus, not rejecting
Model~III should not be read as strong evidence for stockpiling.

\paragraph*{Robustness.}
To relax the constant-elasticity assumption, Appendix~\ref{app:results_on_off_peak} repeats the analysis separately for on-peak and off-peak hours \citep[as in][]{knaut_when_2017, hirth_how_2024}.
The linear on-peak 
response is roughly twice the off-peak response; 
a factor larger than one%
is to be expected given the diurnal nature of most economic activity. The rejection pattern is the same in both and in the aggregate.

\begin{figure}
    \centering
    \includegraphics[width=\linewidth]{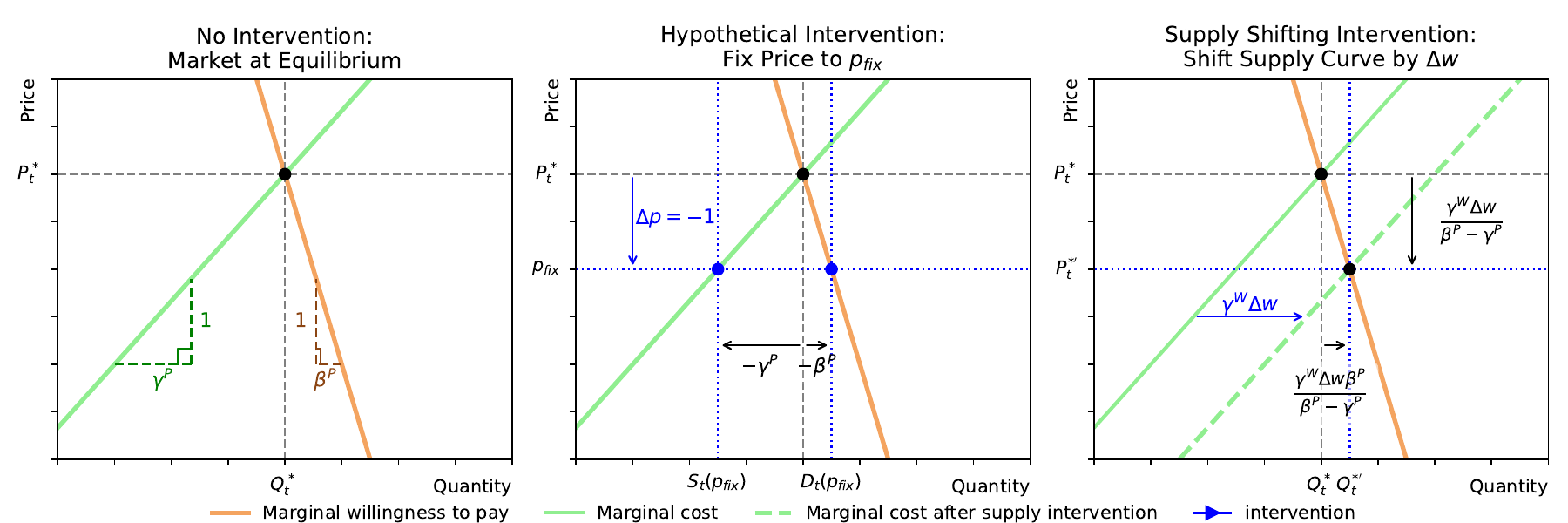}
    \caption{Illustrations of different interventions in a market at equilibrium.
    Left: the market equilibrium without intervention, with demand slope $\beta^P$ and supply slope $\gamma^P$. 
    Middle: an intervention that fixes the equilibrium price at $p_\mathrm{fix}$ ($\mathrm{do}(P_t=p_\mathrm{fix})$); the demand elasticity $\beta^P$ determines the change in quantity demanded, but the market does not clear, $S_t(p_\mathrm{fix}) \neq D_t(p_\mathrm{fix})$. 
    Right: 
    a supply-shifting intervention (e.g., $\mathrm{do}(W_t=W_t + \Delta w)$) that shifts supply outward to a new equilibrium; the change in equilibrium quantity, $\beta^P/(\beta^P-\gamma^P)$ per unit outward shift of the supply curve ($\gamma^W\Delta w$), depends on both slopes; choosing
    $\Delta w = 1/\gamma^W$ yields the target of inference of an SVAR; choosing
    $\Delta w = (\gamma^P-\beta^P)/\gamma^W$ yields the same change in demand as in the middle figure.
    }
    \label{fig:demand-supply-interventions}
\end{figure}

\paragraph*{Interventions.} Causal inference is concerned with predicting the effect of interventions on a system. $\hat{\beta}^P$ estimates the structural coefficient $\beta^P$ of the demand equation and can be interpreted as the instantaneous demand response to a hypothetical intervention on the 
price (Appendix~\ref{sec:microfoundations} distinguishes this from the long-run response and from the response to a change in the price process). 
If $\beta^P$ were indeed $-200$ MW/(EUR/MWh) and it were possible to reduce the price by $1$ EUR/MWh, demand would increase by $200$ MWh/h 
(see Figure~\ref{fig:demand-supply-interventions} (middle)).
We call it hypothetical because fixing the price at $p_\mathrm{fix}$ leaves the market out of balance $(S_t(p_\mathrm{fix}) \neq D_t(p_\mathrm{fix}))$, 
and additional measures 
not provided
by our model, 
such as rationing or subsidies, are needed to re-balance demand and supply.
Figure~\ref{fig:demand-supply-interventions} (right) shows 
alternative interventions. 
A supply shock, for instance, a change in wind generation, $\mathrm{do}(W_t = w)$, or a change in the supply function itself, $\mathrm{do}(M_t^S = m)$, 
such as a change in the asset fleet (see Section~\ref{sec:demand_autocorrelated}), shifts the supply curve, 
and the market clears at a new equilibrium. Here, price and quantity are endogenous, and the change in the equilibrium quantity
(as targeted by SVAR impulse-response analysis)
depends 
on the supply slope $\gamma^P$ as well as on $\beta^P$. Therefore, predicting realized market outcomes under policy interventions might require more than the demand elasticity.

\section{Summary, Conclusion and Future Work}\label{sec:conclusion}
This paper introduces 
causal mechanism graphs and
demonstrates the advantages of using 
them to estimate the price elasticity of demand in a setting with autocorrelated instruments.
These acyclic graphs
are able to represent equilibrium constraints (whereby prices and quantities are determined simultaneously) and
include demand and supply functions as mechanism nodes, which allows for a straightforward definition of interventions, both on quantities and on mechanisms.
We provide a reduction of
causal mechanism graphs to a known class of graphs,
which allows us to exploit existing graphical CIV criteria. 
Causal mechanism
graphs allow us to express structural assumptions transparently, 
understand biasing dynamics,
derive multiple valid estimators, and  test the validity of assumptions about structural dynamics.

One version of a lag-augmented local projection IV estimator appears as a special case, estimator $\#4$.
In the Proxy-VAR and linear local-projection settings,
it is used as part of a procedure for
impulse response analysis (see Appendix~\ref{sec:SVAR-IV}).%
Furthermore, our framework allows us to 
determine when this estimator is valid (and consistent) 
and 
provides a large family of related estimators that may be valid when lag-augmented IV is not. 

We have applied the above framework to estimate the own-price elasticity of electricity demand under three competing structural assumptions: a model in which demand exhibits inertia, a model in which consumption accumulates into an inventory of energy services, and a model of heterogeneous response.
We have shown through simulations that each model can produce identical levels of observed demand autocorrelation, illustrating that the observed correlation alone cannot predict potential bias. 
Several estimates disagree when conditional IV estimators are applied to German electricity demand. First, the widely used naive~IV estimator disagrees with all other estimators, and thus seems unable to identify the own-price elasticity in this setting. Second, the comparison lets us reject
Models~I and~II.
 All estimators that are valid for Model~III agree with each other.
The procedure can be applied to other models too by going through the same four steps: structural equations, mechanism graph, valid estimators, falsification.

In this paper, we have introduced causal mechanism graphs in a linear setting.
 However, the graphical criteria apply to more general settings. Under nonlinearity, the IV strategies identify a weighted-average estimand \citep{angrist_interpretation_2000}, which we validate in the basis-expansion case
 (see Appendix~\ref{sec:non-linear-ext}).

Looking ahead, 
besides linearity, our analysis assumes that the structural coefficients are constant over time, a restrictive assumption in economics: preferences, technology, and beliefs may change, and these determine $\beta^P$ (see Appendix~\ref{sec:microfoundations}).
The graphical conditions (CIV1)--(CIV2), however, concern causal structure alone and can be combined with recent developments on nonstationary causal effects to define the relevant estimands.
Where coefficients vary smoothly, local kernel-weighted IV estimation may offer a route forward under additional smoothness and local-relevance assumptions \citep{dahlhaus1997fitting, giraitis2021timevarying};
where the effect is genuinely period-specific, the time-series potential-outcome frameworks of \citet{bojinov2019time} and \citet{rambachan2019common} define estimands accordingly.

Causal mechanism graphs can help strengthen the intuition behind recent results in this literature. 
Local projection IV estimators identify the same impulse response function as a vector autoregressive model with the instrument ordered first~\citep{plagborg-moller_local_2021}, which our CIV estimators recover at $h=0$ (Appendix~\ref{sec:SVAR-IV}).
The robustness of local projection IV to misspecification \citep{montiel_olea_double_2024} 
might admit similar graphical explanations.
Finally, it would be interesting to understand whether non-Gaussianity-based identification strategies \citep[e.g.,][]{shimizu_linear_2006, Hoyer2008b} can be applied to mechanism-augmented SCMs.

We consider
the graphical approach a powerful addition to the toolkit of every economist working with data structured as a time series, with many applications yet to be explored.

\section*{Funding and Competing Interests}
Researchers from the Hertie School gratefully acknowledge financial support from the German Federal Ministry of Education and Research through the ARIADNE Project (FKZ 03SFK5K0). Lion Hirth and Silvana Tiedemann are owners and directors of Neon Neue Energieökonomik GmbH, an energy economics consulting firm. Oliver Ruhnau is owner and director of tournesol energy GmbH, an energy consulting and software firm. 

\printbibliography

\begin{appendix}
\section{Structural Causal Models for Equilibrium Time Series}
\label{app:appendix_scm_general}
First, we give a formal definition of structural causal models (SCMs), followed by an introduction to directed acyclic graphs (DAGs) for time series. We then formalize the mechanism-augmented SCMs we propose in this article. Finally, we introduce the concept of marginalization, which is a way to simplify the representation of graphs.  

\subsection{Structural Causal Models for Time Series}
\label{sec:appendix_scm}
Let $\{U_{t}^j\}_{t \in \mathbb{Z}, j \in \{1, \dots, d\}}$ 
be a family of jointly independent
random error variables. 
The index $t$ encodes time, and the index $j$ is the time series component. 
For all $j \in \{1, \dots, d\}$ 
define the \emph{parents} of $X^j_{t}$ as $\mathrm{PA}(t,j) \coloneqq (X^{s_1}_{t-k_1}, \dots, X^{s_{p_j}}_{t-k_{p_j}})$
for some $p_j$
and 
let 
$f^j: \mathbb{R}^{p_j} \times \mathbb{R} \to \mathbb{R}$ be a measurable function. 
For all $j \in \{1, \dots, d\}$, $t \in \mathbb{Z}$ we consider the structural equations
\begin{equation}
\label{eq:scm}
    X_t^j \coloneqq f^j\left(\mathrm{PA}(t,j), U_t^j\right).
\end{equation}
Intervening in the system corresponds to 
replacing one (or more) of these structural equations. 
This gives rise to a new SCM and the induced distribution is called an \emph{intervention distribution}. 
Two SCMs are called \emph{observationally (interventionally) equivalent} if they induce the same observational (interventional) distributions
\citep{peters_elements_2017}.
SCMs are not unique in that there are different SCMs that are interventionally equivalent (an analogous statement holds for mechanism-augmented SCMs introduced below). 
However, all procedures described in this work for constructing SCMs or graphs are unique.

Throughout our work, we assume there is a unique solution\footnote{More precisely, we assume that 
all solutions induce the same distribution.} to \eqref{eq:scm}
(and the intervened systems)
that is 
strongly stationary and ergodic.
We further assume that the induced distribution 
is Markov with respect to the induced graph (see Section~\ref{sec:appendix_graphs}). 
For linear functions $f^1, \dots, f^d$ these properties are well-studied 
\citep[see, e.g.,][Thm.~1]{thams_identifying_2022}. 

\subsection{Directed Acyclic Graphs for Time Series}\label{sec:appendix_graphs}

We now introduce some standard graph terminology \citep[e.g.,][]{Lauritzen1996, FriedmanKoller:MLJ03, pearl_causality_2009}, adapted to the case of infinitely many nodes.
Let $V \subseteq
\mathbb{Z} \times \{1,\dots,d\}$. A \emph{directed graph} $ \mathcal{G} = (V, E) $ is composed of nodes $V$ (also called vertices) and edges $ E \subseteq V \times V $ with the restriction that
for all $u,v \in V$ either $ (u, v) \not\in E $ or $ (v, u) \not\in E $ or both. We sometimes use the notation $v \to u$, meaning that $(v, u) \in E$.
A \emph{path} in
$\mathcal{G}$ is a sequence of distinct vertices $ v_1, \ldots, v_m $,
such that an edge connects each consecutive pair of vertices.
If $v_k \to v_{k+1}$ for all $k \in \{1,\dots, m-1\}$,
we say that the path is \emph{directed from $v_1 $ to $ v_m $}. A directed path $v_1 \to ... \to v_m$ such that $(v_m, v_1) \in E$ is called a \emph{cycle}.
A directed graph $\mathcal{G}$ is classified as a \emph{directed acyclic graph (DAG)} if it has no cycles.

We also consider marginalized graphs, which are acyclic directed mixed graphs (ADMGs), containing both directed $(\rightarrow)$ and bidirected $(\leftrightarrow)$ edges. A path in an ADMG is a sequence of distinct vertices in which each consecutive pair is joined by an edge of either type, and we define $v$ to be a collider on a path whenever two consecutive edges have arrowheads at $v$ (e.g., $u_1 \leftrightarrow v \leftarrow u_2)$.
Because we deal with time series, the structural equations induce a graph with infinitely many nodes, i.e., a \emph{full time graph} \citep{peters_causal_2013} $ \mathcal{G}_{\mathrm{FT}}$ over nodes $V = \{(t,s) \ \vert \ t \in \mathbb{Z}, s \in \{1, \dots, d\}\}$, where we draw an edge from $(t_1,s_1)$ to $(t_2,s_2)$ if $(t_1,s_1) \in \mathrm{PA}(t_2,s_2)$.
We assume that the parent sets are such that $\mathcal{G}_{\mathrm{FT}}$ is directed and acyclic. 
We say $v \in V$ is a \emph{descendant} of $u \in V$ if $v \neq u$, and there exists a directed path from $u$ to $v$. We denote the set of descendants of $u$ by $\mathrm{DE}(u) \coloneqq \{v \in V \ \vert \ \text{$v$ is a descendant of $u$}\}$.

\subsection{Mechanism-augmented Structural Causal Models}
\label{sec:constrained_graph}

We use an augmented version of structural causal models in which the structural mechanisms are represented explicitly as nodes, which we call mechanism-augmented SCM. In the SCM of Section~\ref{sec:appendix_scm}, a variable $X_t^j$ is generated by the structural assignment~\eqref{eq:scm}, where the function $f^j$ is part of the background model.
In the mechanism-augmented representation, we instead introduce a mechanism node $M_t^j$, a random function defined by $M_t^j(\cdot) \coloneqq f^j(\cdot, U_t^j)$, and generate the realized variable by evaluating it: $X_t^j \coloneqq \operatorname{ev}(M_t^j, \mathrm{PA}(t,j)) = M_t^j(\mathrm{PA}(t,j))$, where $\operatorname{ev}\colon (f,x) \mapsto f(x)$ is the evaluation map. We therefore extend the parent set of $X_t^j$ to include $M_t^j$ and change the structural equation $f^j$ of $X_t^j$ to $\mathrm{ev}$.
The mechanism-augmented SCM induces a graph in the sense of Section~\ref{sec:appendix_graphs} with an extended node and edge set: in addition to the realized variables, each mechanism is a node, $V^{\mathrm{aug}} \coloneqq \{(t,s) \ \vert \ t \in \mathbb{Z}, s \in \{1, \dots, d\}\} \cup \{(t, M^s) \ \vert \ t \in \mathbb{Z}, s \in \{1, \dots, d\}\}$ and we extend the edge set to include edges $(t, M^s) \to (t,s)$ for all $t \in \mathbb{Z}$ and $s \in \{1, \dots, d\}$.
For equilibrium systems, we additionally specify one scalar constraint
per time point,
$
    C(p,\mathbf M_t,Z_t)=0,
$
where $\mathbf M_t$ collects the mechanisms entering the constraint and $Z_t$ collects the additional inputs to the mechanisms, such as
$(D_{t-1},W_t)$ in Model~I. We assume that, for admissible mechanisms and inputs, this equation
has a unique solution and that the solution operator $\Phi$ is measurable.
We then define the structural assignment
\[
    P_t:=\Phi(\mathbf M_t,Z_t).
\]
The graph includes an edge into $P_t$ from each argument on which $\Phi$
depends.
For market clearing, the constraint is
$
    C(p,\mathbf M_t,Z_t)
    :=M_t^D(p,Z_{D,t})-M_t^S(p,Z_{S,t})=0.
$
Importantly, interventions on the original parents $\mathrm{PA}(t,j)$ induce the same interventional distributions in the original and mechanism-augmented SCM formulations. The mechanism-augmented SCM additionally allows interventions that set a mechanism node to a given function. Replacing a structural equation is the standard notion of intervention in an SCM (Section~\ref{sec:appendix_scm}); what the mechanism nodes add is that such an intervention becomes an intervention on a node of the graph, so that the graphical criteria apply to it, and that the equilibrium constraint is re-solved with the modified mechanism.
Thus, for interventions on the original variables, the augmentation leaves the
induced interventional distributions unchanged. Its additional role is to represent
the structural mechanisms explicitly as nodes, which can themselves be targeted
by interventions.
Although we present the construction for time series to match our setting, the same augmentation applies to static, cross-sectional SCMs.

This representation of an SCM as a mechanism-augmented SCM is useful for equilibrium systems. Consider realized demand
and supply variables \(D_t\) and \(S_t\), with mechanisms
\[
    M_t^D(p) \coloneqq  f^D(p,U^D_t),
    \qquad
    M_t^S(p) \coloneqq  f^S(p,U^S_t),
\]
i.e., here, there are no inputs other than price,
and suppose that an equilibrium variable \(P_t\) is determined by the market-clearing condition \(D_t=S_t\). We can write
\[
    P_t \coloneqq  \Phi(M_t^D,M_t^S),
\]
where \(\Phi\) returns the unique \(p\) satisfying\footnote{Throughout, we assume that the equilibrium exists and is unique, so that $\Phi$ is well defined.}
$
    M_t^D(p)=M_t^S(p).
$
The realized variables are then
\[
    D_t \coloneqq \operatorname{ev}(M_t^D, P_t) = M_t^D(P_t),
    \qquad
    S_t \coloneqq \operatorname{ev}(M_t^S, P_t) = M_t^S(P_t).
\]
The corresponding causal mechanism graph is acyclic: the mechanism nodes \(M_t^D\) and
\(M_t^S\) are upstream of \(P_t\), while \(D_t\) and \(S_t\) are realized only after
\(P_t\) has been determined, see Figure~\ref{fig:augmented-equilibrium-dag} but without wind.
This distinction is important for interventions. An intervention on realized
demand,
$
    \mathrm{do}(D_t=d),
$
overwrites the realized outcome after the equilibrium mechanism has operated.
It does not change the demand function \(M^D_t\), and therefore does not make the
equilibrium price \(P_t\) adjust. By contrast, an intervention on the mechanism,
$
    \mathrm{do}(M_t^D=m),
$
changes the demand function that enters the equilibrium problem. The equilibrium
price is then recomputed using the modified mechanism. In particular, an
equilibrium-preserving intervention that fixes demand at \(d\) can be written as
$
    \mathrm{do}^{\mathrm{eq}}(D_t=d)
    \coloneqq 
    \mathrm{do}(M^D_t=c_d),
$
where \(c_d\) is the constant function \(c_d(p)=d\) for all \(p\). Thus,
\(\mathrm{do}(D_t=d)\) is an after-the-fact outcome intervention, whereas
\(\mathrm{do}^{\mathrm{eq}}(D_t=d)\) changes the mechanism generating demand
before the system re-equilibrates.
The construction extends directly to the dynamic models of the main text: there, the mechanisms take additional state arguments, for instance $M_t^D(p, D_{t-1})$ and $M_t^S(p, W_t)$ in Model~I.
In the linear setting considered in the main text, this representation can be
reduced to an ordinary finite-dimensional linear SCM. In particular, write
$
f^j(x,u) = a_j^\top x + c_j + b_j u
$
and define $m_t^j \coloneqq c_j + b_j U_t^j$. Then
$
M_t^j(x) = a_j^\top x + m_t^j.
$
Since $a_j$ is fixed, $M_t^j$ and $m_t^j$ are in one-to-one
correspondence. Under this identification, the evaluation map
$\mathrm{ev}(M,x)=M(x)$ is represented by
$
\widetilde{\mathrm{ev}}_j(m,x) = a_j^\top x + m,
$
which is linear in $(m,x)$. Hence the structural assignment
$X_t^j=\mathrm{ev}(M_t^j,\PA(t,j))$ becomes
$
X_t^j = a_j^\top \PA(t,j) + m_t^j.
$
Moreover, the linear equilibrium constraints considered in this article are
assumed to have a unique solution, and solving them yields linear structural
assignments for the equilibrium variables as functions of their parents.
Consequently, after replacing each $M_t^j$ by its coordinate $m_t^j$ and
solving the equilibrium constraints, all structural assignments of the
mechanism-augmented SCM are linear. The resulting finite-dimensional
representation therefore forms a linear structural time-series SCM with
jointly independent exogenous variables and an acyclic full time graph,
possibly containing additional directed edges between variables at the same
time point.

Importantly, replacing $M_t^j$ by $m_t^j$ is only a node-wise one-to-one
reparametrization. In particular,
$\sigma(M_t^j) = \sigma(m_t^j)$,
so the function-valued and finite-dimensional representations induce the same
conditional-independence relations and the same graph up to relabeling of the
mechanism nodes.
Theorem~1 of \cite{thams_identifying_2022} establishes the global Markov property for the corresponding infinite full time graphs without instantaneous effects. The same argument extends to the present setting: acyclicity allows the nodes within each time point to be arranged in a topological order, after which the finite-window argument of \cite{thams_identifying_2022} applies unchanged. Hence, for finite disjoint sets of nodes $A$, $B$, and~$C$,
\begin{equation*}
A\perp\!\!\!\perp_{\mathcal{G}}C\mid B
\quad\Longrightarrow\quad
A\perp\!\!\!\perp C\mid B.
\end{equation*}
Finally, marginalizing the mechanism nodes produces the corresponding latent-confounding edges; for example, $M_t^D\to P_t$ and $M_t^D\to D_t$ induce $P_t\leftrightarrow D_t$. The resulting marginalized graph is therefore of the type considered in \cite{thams_identifying_2022}, see Section~\ref{sec:consistent_estimators}.

\subsubsection{Relation to Imbens (2020)}
\label{sec:relationimbens}
The causal mechanism representation above can be understood as a response to the concern raised
by \citet[Section~4.3]{imbens_potential_2020} that standard directed acyclic graphs do not
naturally represent equilibrium systems. In a textbook supply--demand model, the primitive
economic objects are not only the realized equilibrium price and quantity. Rather, they are
the demand and supply functions. For each possible price $p$, the functions $q^D(p)$ and
$q^S(p)$ describe the quantities that would be demanded and supplied at that price. The
observed price is then determined by the market-clearing condition $q^D(p)=q^S(p)$.
Imbens points out that this creates a difficulty for ordinary DAGs: the natural nodes in the
graph would be the functions $q^D(\cdot)$ and $q^S(\cdot)$, but these are not ordinary realized
variables.

The mechanism-augmented SCM formalizes precisely this idea. 
The mechanism nodes $M^D$ and $M^S$ are
upstream of the equilibrium price $P$, while realized demand and supply are downstream of
$P$. In this sense, the representation keeps the economic structure emphasized by Imbens
but embeds it in a graphical SCM language.
 The 
 causal mechanism
 graphs that we propose deviate from the graph suggested by \citet[][Fig.~12(b)]{imbens_potential_2020}. We still use the variables supply and demand instead of a single quantity variable. This way we can allow, for example, interventions on demand (not the mechanism) that would yield an inequality between 
 demand and supply.
 Furthermore, 
 unlike \citet{imbens_potential_2020},
 we 
 propose to add an edge from price to demand, for example, 
 which is important for 
 intervention effects and 
 identification after marginalization.

The mechanism-augmented SCM therefore answers Imbens's representational question by making the
function-valued economic primitives explicit. Once these primitives are represented as
mechanism nodes, the equilibrium system can be written as an acyclic graph, and graphical
reasoning can be applied without abandoning the standard economic view that prices and
quantities are jointly determined in equilibrium.

\subsection{Graph Marginalization} 
\label{sec:marginalization}
We make use of the graph marginalization 
technique called \emph{latent projections}
\citep[see, e.g.,][]{Verma1991}. The main idea 
is to marginalize out nodes
while keeping the same d-separation statements over the retained nodes. 
In the context of this paper, we, for example, marginalize out the supply variable $S_t$ and mechanism $M_t^S$ (see Remark~\ref{rem:identification_device}).
A formal definition of a marginalized time graph is also given in \citet{thams_identifying_2022}, which, too, considers graphs with infinitely many nodes.  
\begin{definition}\label{def:marginalzed_graph}
    Consider a
    full time graph
    $\mathcal{G}_{\mathrm{FT}}$ over nodes 
    $V = \{(t,s) \ \vert \ t \in \mathbb{Z}, s \in \{1, \dots, d\}\}$.
    Let $\mathcal{R} \subseteq V$ be a set of nodes, which we call the \emph{retained} nodes.
    The \emph{marginalized time graph}, $\mathcal{G}_\mathcal{R}$, is the graph over nodes $\mathcal{R}$ where for all $i, j \in \mathcal{R}$ there is:
    \begin{enumerate}
        \item a directed edge $i \rightarrow j$ if and only if $i \rightarrow j$ in $\mathcal{G}_{\mathrm{FT}}$ or 
        there exist $k_1 \in \mathbb{N}$, $v_1, \ldots, v_{k_1} \notin \mathcal{R}$ and a directed path $i\rightarrow v_1 \rightarrow \cdots \rightarrow v_{k_1}\rightarrow j$ in $\mathcal{G}_{\mathrm{FT}}$, and
       \item a bidirected edge $i \leftrightarrow j$ if and only if there exist $k_1, k_2 \in \mathbb{N}_0$, $v_1, \ldots, v_{k_1}$, $w_1, \ldots, w_{k_2}$, $U\notin \mathcal{R}$ in $\mathcal{G}_{\mathrm{FT}}$ such that there exist directed paths $U\to v_1\to \cdots \to v_{k_1}\to i$ and $U\rightarrow w_1\rightarrow \cdots \to w_{k_2} \rightarrow j$, where an empty sequence ($k=0$) means the edge $U \to i$ respectively $U \to j$.
    \end{enumerate}
\end{definition}
Latent projections are purely graphical operations. We use them for identifying parameters in the original graph. (We do not perform interventions in marginalized graphs.)

\begin{remark}[Exchanging the role of supply and demand]
\label{rem:identification_device}
The mechanism-augmented SCM treats demand and supply in a parallel way: it represents both mechanisms and models interventions on either. The marginalized graphs used in the main text do not: they are purpose-built for one target parameter, the price elasticity of electricity demand. Marginalizing out the supply side (as in Figure~\ref{fig:intro}) yields a graph that supports identification of the demand slope $\beta^P$ from a supply-side instrument $W_t$; 
similarly,
marginalizing out the demand side yields a graph that, given a demand-side instrument, supports identification of the supply slope $\gamma^P$. 
\end{remark}

\section{Details on the Simulation and the Application}
\subsection{Data Generation}\label{app:data_generation}
Our datasets contain three series: wind generation, the equilibrium price, and equilibrium demand.
We model wind generation as an AR$(L)$ process with $L=26$, autocorrelation coefficients and contemporaneous shock $U_t^W$
derived from observed wind generation data in Germany in 2019 (see Figure~\ref{fig:autocorrelation}, bottom left).

The supply equation corresponds to Equation~\eqref{eq:supply} with $S^0 = 25{,}000$ MW, $\gamma^P = +500$ MW/(EUR/MWh) and $\gamma^W = 1$, so that every unit of generated wind electricity is immediately offered to the market. The positive constant $S^0$ implies significant supply at negative prices, a known feature of the German market. The supply shock $U_t^S \sim \mathcal{N}(0,1^2)$ is negligible relative to market quantities of approximately $60{,}000$ MW and only serves to ensure smooth market clearing. In other words, we take wind generation as the sole source of stochastic variation in supply, the remainder consisting of must-run and subsidized generation $(S^0 > 0)$ and of flexible generation that adjusts deterministically to prices $(\gamma^P = +500)$.

On the demand side, the own-price response is $\beta^P = -100$ MW/(EUR/MWh), close to the estimate obtained from the naive~IV specification. To fix the scale of the shocks, we regress observed demand on a constant, the contemporaneous price and lagged demand, which yields residuals with a standard deviation of approximately $2{,}500$ MW. We set the demand shocks $(U_t^D, U_t^A, U_t^B)$ to i.i.d.\ $\mathcal{N}(0,2000^2)$, somewhat below that value because the regression omits the controls of Section~\ref{sec:application} and therefore overstates the shock variance. The constant $D^0$ is chosen in each case so that mean demand matches German electricity load (approximately $60{,}000$ MW).

Equating supply with the model-specific demand equation then yields the hourly equilibrium prices and quantities.

\subsection{Data for the Application}\label{app:data}
The main variables are electricity load (MW), electricity price (EUR/MWh), and
wind generation (MW, including both onshore and offshore generation), all obtained from \citet{entsoe_data}, the European Network of Transmission System Operators for Electricity. ENTSO-E's measurement and reporting of variables is bidding-zone-specific. Since October 1, 2018, the bidding zone includes the territory of Germany and Luxembourg. Previously, it also included the territory of Austria.
The data used in our analysis spans from November 1, 2017, to December 31, 2020.%

In addition 
to the main variables, we use several control series.
If the data exist only at a daily frequency, we assume the same value for every hour of the day. 
    (A) At hourly frequency: 
        solar PV power generation (from \citet{entsoe_data});
        surface temperature data for Germany from NASA's MERRA-2 \citep{global_modeling_and_assimilation_office_merra-2_2015}, transformed using Eurostat's methodology \citep{eurostat_heating_nodate};
        sunlight, indicating whether it is day (1) or night (0) at the geographic center of Germany (at 51\degree09' N, 10\degree26' E) using Python's `suntime' package \citep{stopa_suntime_2019};
    (B)
    At daily frequency, all from \citet{investing_commodities_2023}: 
    price of emissions allowances (EUA yearly futures);
    coal prices (API 2 CIF ARA ARGUS-McCloskey futures);
    natural gas price (Dutch TTF futures).
Furthermore, we include hour of the week (one-hot encoding, 167 variables), month of the year (one-hot encoding, 11 variables), and year (one-hot encoding, 3 variables).

Public holidays and school holidays are two daily variables describing the fraction of the German population affected by a holiday on a given day, using data from the Federal Statistical Office of Germany 
for the population by state \citep{destatistis_population_2023}, public holiday dates from Python's `holidays' package \citep{montel_holidays_2023}, and school holiday dates from \citet{feiertagskalender_school_2023}. Instances where holiday$_t = 1$ (e.g., Easter, Christmas) are dropped. The next largest observed value equals~$0.64$. 

\begin{figure}[t]
    \centering
    \includegraphics[width=0.7\linewidth]{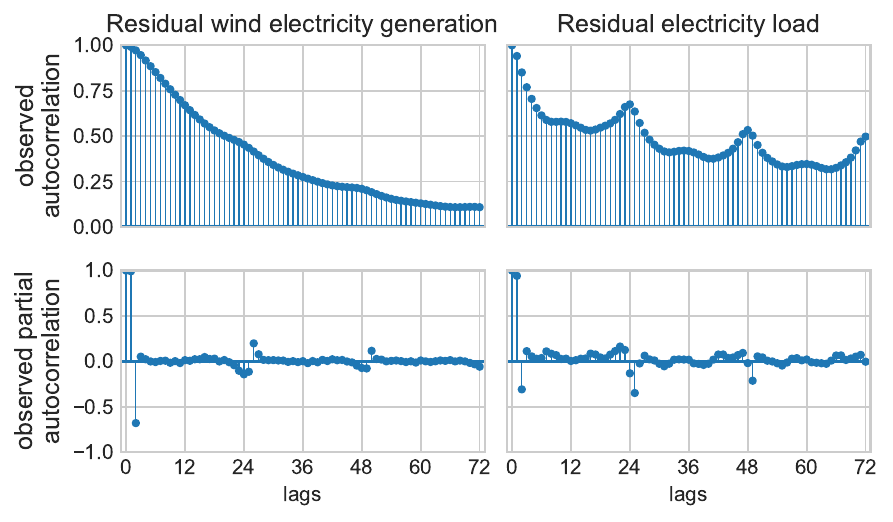}
    \caption{
    Autocorrelation function (above) and partial autocorrelation function ({below}) of the residuals of the instrumental time series wind power generation (left), and the residuals of the dependent variable German electricity demand (right). In all four cases, we first regress the original data on the covariates and then obtain the autocorrelation of the residual time series.}
    \label{fig:autocorrelation}
\end{figure}

Figure~\ref{fig:autocorrelation} shows the autocorrelation and partial autocorrelation plots of the instrument and the electricity load after controlling for the control variables. We observe a significant autocorrelation, which is most pronounced for lags 1 and 2 but conditionally extends to up to 50 lags, corresponding to about two days.   

\subsection{Sensitivity Analysis for the Price Elasticity of German Electricity Demand}\label{app:results_on_off_peak}

\begin{figure}[t]
    \centering
    \includegraphics[width=1\linewidth]{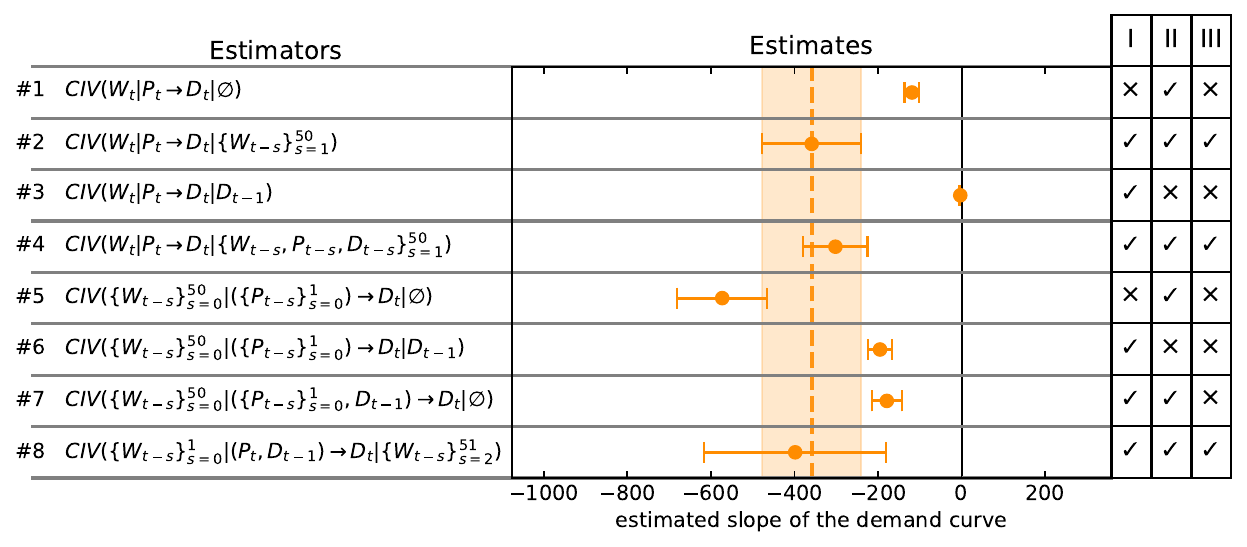}
    \includegraphics[width=1\linewidth]{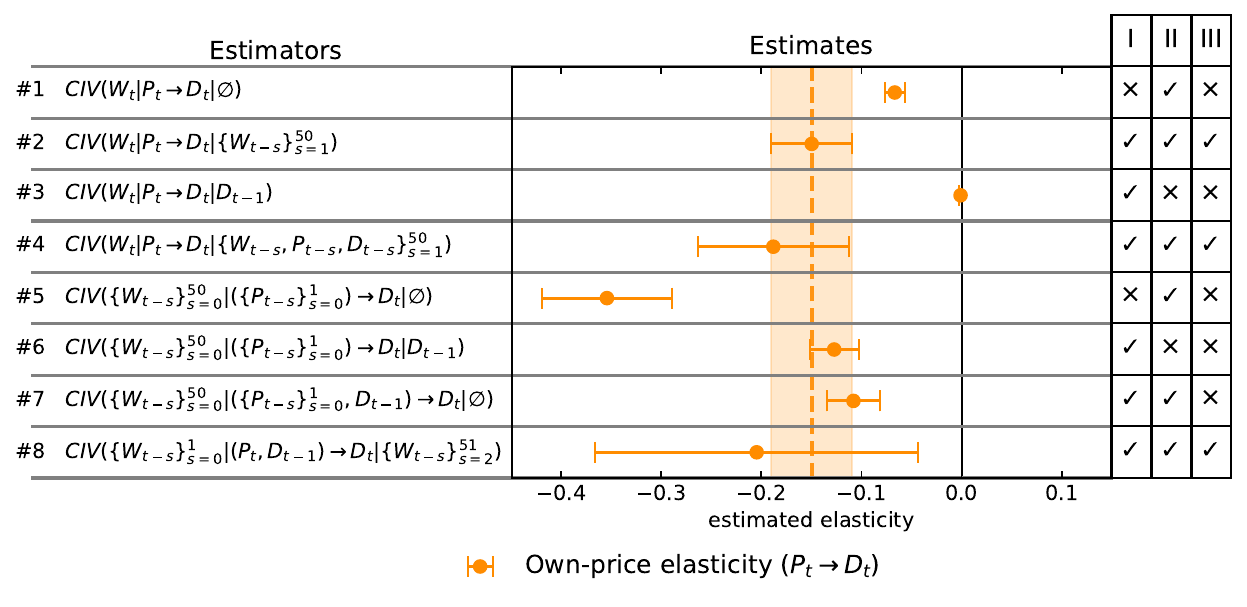}
    \caption{Own-price elasticity of on-peak German electricity demand, obtained from the CIV estimators given in Table~\ref{tab:estimators} in linear specification (top) and log-log specification (bottom). The highlighted area corresponds to the confidence interval of estimator $\#2$. 
    The tables on the right are reproductions of Table~\ref{tab:estimators}, showing the validity of each strategy under different structural model assumptions. 
    On-peak hours refer to the time between 8:00 and 19:59.}
    \label{fig:application_t-1}
\end{figure}

\begin{figure}[t] 
    \centering
    \includegraphics[width=1\linewidth]{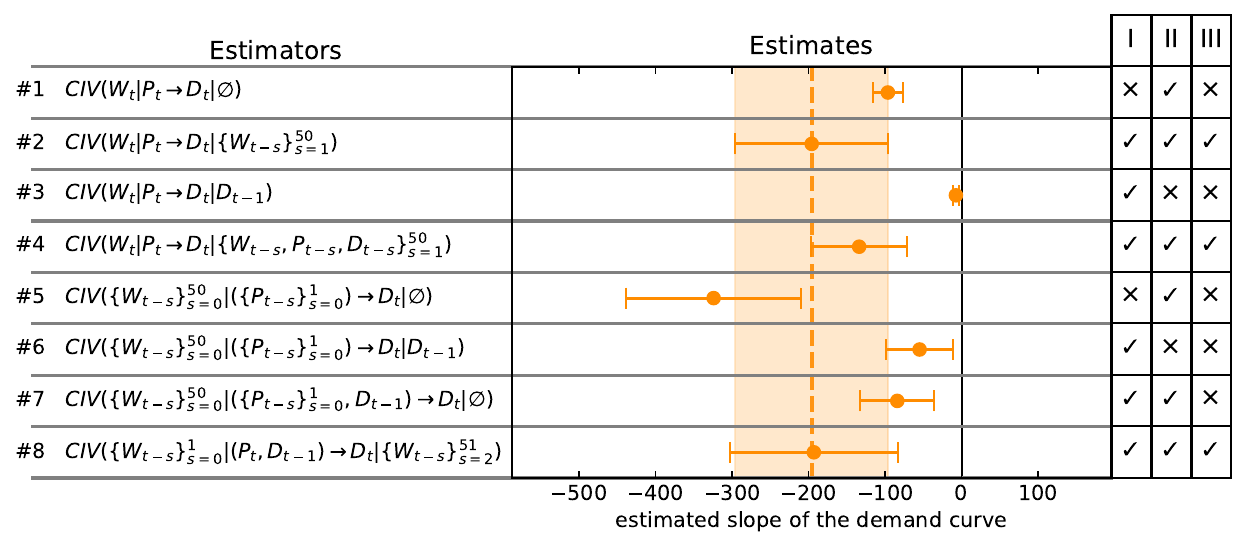}
    \includegraphics[width=1\linewidth]{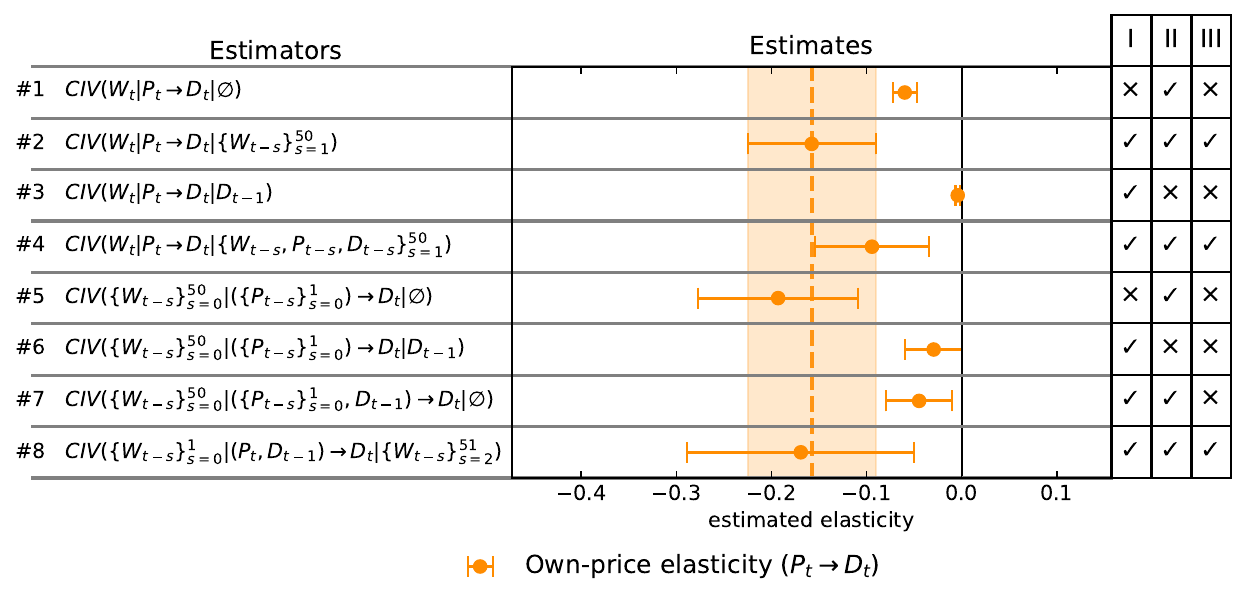}
 \caption{Own-price elasticity of off-peak German electricity demand, obtained from the CIV estimators given in Table~\ref{tab:estimators} in the linear specification (top) and log-log specification (bottom). The highlighted area corresponds to the $95\%$ confidence interval of estimator $\#2$. The tables on the right are reproductions of Table~\ref{tab:estimators}, showing the validity of each strategy under different structural model assumptions. 
 Off-peak hours refer to the time between 20:00 and 7:59.
 }
    \label{fig:sensitivity_20_to_8}
\end{figure}

In Section~\ref{sec:application}, we assume a homogeneous elasticity across hours. However, structurally different consumers may be active across different hours of the day. In particular, one may expect different consumers during peak and off-peak hours. They could either have different elasticities or different dynamic behavior.
We split the data into two subsets and perform the same analysis as in Section~\ref{sec:application} on each of the splits: on-peak in Figure~\ref{fig:application_t-1} and off-peak in Figure~\ref{fig:sensitivity_20_to_8}. 
The magnitudes of the (log-log) elasticity estimates are comparable across subsets, but larger than the estimate from the joint sample. The (linear) on-peak response is roughly twice the off-peak response.
The pattern of the estimates is the same across both splits: we reject Models~I and~II, but not Model~III. In the linear off-peak specification, Model~II is rejected by the disagreement between $\#1$ and $\#5$ rather than by their disagreement with $\#2$, $\#4$ and $\#8$.

\clearpage
{\centerline{
SUPPLEMENTARY MATERIAL}\vspace{-0.4cm}} 
\section{Comparing Graph-based CIV Identification with SVAR Approaches}
\label{sec:SVAR}
The identification of causal effects in dynamic equilibrium systems has long been a central concern in applied econometrics. Structural vector autoregressions (SVARs), introduced in macroeconometrics by \citet{sims_macroeconomics_1980}, have become a standard framework for analyzing the propagation of structural shocks through impulse response functions (IRFs) \citep{kilian_structural_2017, ramey_macroeconomic_2016}. In contrast, applied microeconometrics has increasingly adopted the potential outcomes framework \citep{angrist_interpretation_2000, imbens_causal_2015} and
related graphical approaches based on directed acyclic graphs (DAGs) \citep[e.g.,][]{pearl_causality_2009, hernan_causal_2020, cunningham_causal_2021} to discuss and formalize identification strategies; the framework proposed in our work falls into the latter category.
 
Although these two traditions employ different representations and identification strategies, both seek to recover causal information from observational data generated by underlying economic processes, which may admit multiple causal representations.
In this paper, we use graphical methods to analyze a dynamic equilibrium system that admits an SVAR representation.
This appendix compares 
our framework with the SVAR framework
and makes the following three points.
 
First, the two frameworks target different causal estimands. 
Our conditional instrumental variable (CIV) estimators target $\beta^P$, a structural parameter of the demand equation (Section~\ref{sec:SVAR_DGP}).
SVAR analysis targets
structural impulse response functions (IRFs),
that is,
the responses of the observables to structural shocks; these include the effects of market clearing (Section~\ref{sec:SVAR_representation}).
 
Second, 
the identifiability conditions differ.
Both estimands are linked in that
a fully identified SVAR recovers the structural parameters; conversely, only the complete set of structural parameters reconstructs the IRFs.
In Section~\ref{sec:SVAR_RR}, we provide systems in which the structural IRFs are not identified while $\beta^P$ is still recovered via SVAR analysis (e.g., Model~III), and one system in which SVAR 
identifies neither IRFs nor $\beta^P$
but a CIV estimator identifies $\beta^P$.
 
Third,
in our view, the closest connection between the two frameworks lies in
the Proxy-VAR literature \citep{stock_identification_2018}. There, similarly to our framework,
identification
relies
on exogenous variation supplied by an instrument and
targets a single structural shock. With a supply-side instrument, it recovers an impact response that equals $\beta^P$.
However, the two sets of assumptions are not equivalent. Section~\ref{sec:SVAR-IV} introduces the Proxy-VAR estimator and gives a system in which no proxy constructed from $W_t$ is admissible while a CIV (nuisance) estimator remains valid.

 Finally, Section~\ref{sec:SVAR_numerical} compares the two approaches%
in simulation
and Section~\ref{sec:SVAR_discussion} adds a discussion.
 
\subsection{Mechanism-augmented SCM and its Estimand}\label{sec:SVAR_DGP}
We take Model~I of the main text as a data-generating process that either framework can analyze, now written with the equilibrium quantity $Q_t$ made explicit:
\begin{align}
    D_t &\coloneqq M_t^D(P_t, D_{t-1}) \coloneqq D^0 + \beta^{D1} D_{t-1} + \beta^P P_t + U^D_t, \label{eq:svar_demand}\\
    S_t &\coloneqq  M_t^S(P_t, W_t) \coloneqq S^0 + \gamma^P P_t + \gamma^W W_t + U^S_t, \label{eq:svar_supply}\\
    W_t &\coloneqq \beta^{W} W_{t-1} + U^W_t, \label{eq:svar_wind}\\
    P_t & \text{ is such that }    D_t = S_t =:Q_t, \label{eq:svar_eq}
\end{align}
with $U^D_t$, $U^S_t$, and $U^W_t$ mutually independent i.i.d.\ shocks. We call
\eqref{eq:svar_demand}--\eqref{eq:svar_wind} the structural equations of the economy and their coefficients the structural parameters.
The market-clearing condition~\eqref{eq:svar_eq} is a constraint rather than a structural assignment, and, as in Section~\ref{sec:demand_autocorrelated}, its solution defines the assignment for $P_t$,
\begin{align}
    P_t &\coloneqq \Phi(M_t^D, M_t^S; D_{t-1}, W_t) \notag\\
        &= \frac{1}{\beta^P - \gamma^P} \left( S^0 - D^0 + \gamma^W W_t - \beta^{D1} D_{t-1} + U_t^S - U_t^D \right), \label{eq:svar_eq_price}
\end{align}
where $\Phi$ is the solution operator; assuming $\beta^P \neq \gamma^P$, the solution exists and is unique. Equations~\eqref{eq:svar_demand}--\eqref{eq:svar_eq_price} are a mechanism-augmented SCM in the sense of Appendix~\ref{sec:constrained_graph}.
Our estimand is $\beta^P$, the causal effect of the equilibrium price $P_t$ on demand $D_t$. Section~\ref{sec:CIV-estimation} discusses graphical tools for identifying $\beta^P$.

\subsection{SVAR Representation, Estimand and Identification}\label{sec:SVAR_representation}
We now 
present the same example in the SVAR formalism\footnote{
We follow the notation of \citet{gambetti_structural_2020} with a few modifications. 
We use: $B(L)$ for 
the structural impulse response functions, $B_0$ for the structural impact matrix; and $C(L)$ for the reduced-form IRFs.
To make the discussion of \citet{rubio-ramirez_structural_2010} easier to follow, we write the structural vector autoregression form with a contemporaneous matrix $A_0 = B_0^{-1}$ and the lag matrix $A_1$; and then write the reduced-form VAR using matrices $\Pi_k$. We also depart from \citet{gambetti_structural_2020} 
in that we do not normalize the covariance matrix of the shocks, $\Sigma_V=I$.
Instead, we set one coefficient 
per row 
of $A_0$ to one 
(see~\eqref{eq:svar_structural_form}) 
and leave the shock variances free.}.
Let $X_t = (W_t, P_t, Q_t)'$ collect the observables.
Equations~\eqref{eq:svar_demand}--\eqref{eq:svar_eq} can be written as a \emph{structural vector autoregression} (SVAR),
\begin{equation}\label{eq:svar_structural_form}
A_0 X_t = \mu_0 + A_1 X_{t-1} + V_t,
\quad \text{where, here} \quad
A_0 :=
\begin{bmatrix}
1 & 0 & 0\\
0 & -\beta^P & 1\\
-\gamma^W & -\gamma^P & 1
\end{bmatrix}
\quad \text{and}  \quad V_t :=
\begin{bmatrix}
U_t^W \\ U_t^D \\ U_t^S
\end{bmatrix},
\end{equation}
$V_t = (V^W_t, V^D_t, V^S_t)'$ collects the structural shocks,
 $\mu_0 \coloneqq (0, D^0, S^0)'$, $(A_1)_{1,1} \coloneqq \beta^W$, $(A_1)_{2,3} \coloneqq \beta^{D1}$, and all other entries of $A_1$ are zero.
The shock variances $\sigma_W^2$, $\sigma_D^2$, and $\sigma_S^2$ are left unrestricted.
 
Under certain conditions, 
systems of the form~\eqref{eq:svar_structural_form}
admit two further representations: the structural moving-average form~\eqref{eq:svar_moving_average}, which contains the SVAR estimand, and the reduced-form VAR~\eqref{eq:svar_reduced_Form}, 
whose form motivates several widely used estimators.
Precisely,
if\footnote{
In our system, both conditions are satisfied: for~\eqref{eq:svar_structural_form}, the roots are $1/\beta^{W}$ and $(\gamma^P - \beta^P)/(\beta^{D1}\gamma^P)$, so stability amounts to $|\beta^{W}| < 1$ and $|\beta^{D1}\gamma^P| < |\gamma^P - \beta^P|$ and 
$A_0$ is invertible because $\gamma^P - \beta^P\neq0$.
}
the system is stable, that is, the roots of $z \mapsto\det(A_0 - A_1 z)$ lie outside the unit circle,
and $A_0$ is invertible, 
which ensures the structural shocks are recoverable from current and past observables,
then it admits the \emph{structural moving-average} representation
\begin{equation}\label{eq:svar_moving_average}
X_t  = \mu + \sum_{k=0}^{\infty} B_k V_{t-k} = \mu + B(L) V_t,
\end{equation}
where $\mu\coloneqq (A_0-A_1)^{-1}\mu_0$ is the mean of $X_t$, $L$ is the lag operator, $L^h X_t = X_{t-h}$, $B(L) \coloneqq \sum_{k=0}^{\infty} B_k L^k$, and $B_k := (A_0^{-1}A_1)^k A_0^{-1}$; in particular, $B_0 = A_0^{-1}$. The SVAR estimand is $B(L)$: the structural impulse response functions given by the sequence of $B_k$ matrices. 
Each matrix $B_k$ 
collects the responses of the observables at time $t$ to a unit structural shock at time $t-k$; $B_0$ is the matrix of
\emph{impact responses}, called the \emph{impact matrix}.

Given only that $A_0$ is invertible, 
the process \eqref{eq:svar_structural_form} also allows for%
the \emph{reduced-form VAR} representation
\begin{equation}\label{eq:svar_reduced_Form}
X_t = \nu + \sum_{k=1}^{p} \Pi_k X_{t-k} + \varepsilon_t,
\end{equation}
where $\nu\coloneqq A_0^{-1}\mu_0$, $p$ is the lag order and $\Pi_k = A_0^{-1}A_k$ are the reduced-form lag matrices, with $\Pi(L) \coloneqq I - \Pi_1 L - \cdots - \Pi_p L^p$; for~\eqref{eq:svar_structural_form}, $p = 1$.
The reduced-form innovations $\varepsilon_t = (\varepsilon_{t,W}, \varepsilon_{t,P}, \varepsilon_{t,Q})'$ satisfy
\[
\varepsilon_t = B_0 V_t, 
\qquad 
\Sigma_{\varepsilon} \coloneqq \mathbb{E}[\varepsilon_t\varepsilon_t'] = B_0 \Sigma_V B_0', 
\qquad 
\Sigma_V \coloneqq \mathbb{E}[V_tV_t'] = \mathrm{diag}(\sigma_W^2, \sigma_D^2, \sigma_S^2).
\]

We now describe the target of inference of the SVAR approach and the impact matrix 
in our example system~\eqref{eq:svar_demand}--\eqref{eq:svar_wind} 
in more detail and relate it to $\beta^P$.
Substituting~\eqref{eq:svar_eq_price} into the supply equation~\eqref{eq:svar_supply} gives the equilibrium quantity as a function of the contemporaneous shocks and the past,
\begin{equation}\label{eq:svar_eq_structural_shocks}
Q_t = S^0 + \gamma^W W_t + U_t^S
+ \frac{\gamma^P}{\beta^P - \gamma^P}
\left( S^0 - D^0 + \gamma^W W_t - \beta^{D1} D_{t-1} + U_t^S - U_t^D \right).
\end{equation}
Inverting $A_0$, or equivalently differentiating~\eqref{eq:svar_wind},~\eqref{eq:svar_eq_price}, and~\eqref{eq:svar_eq_structural_shocks} with respect to the shocks, gives the impact matrix
\begin{equation}\label{eq:impact_matrix}
B_0 =
\begin{bmatrix}
1 & 0 & 0 \\
\frac{\gamma^W}{\beta^P - \gamma^P} & \frac{-1}{\beta^P - \gamma^P} & \frac{1}{\beta^P - \gamma^P} \\
\frac{\gamma^W \beta^P}{\beta^P - \gamma^P} & \frac{-\gamma^P}{\beta^P - \gamma^P} & \frac{\beta^P}{\beta^P - \gamma^P}
\end{bmatrix},
\end{equation}
whose entries
are the impact responses of the observables $(W_t, P_t, Q_t)$ (rows) to shocks $(U_t^W, U_t^D, U_t^S)$ (columns).
To illustrate how the estimands
from Sections~\ref{sec:SVAR_DGP} and~\ref{sec:SVAR_representation}
differ, consider a positive supply shock $U^S_t$ that shifts supply outward and moves price and quantity simultaneously. 
The impact response of $Q_t$ to
the supply shock
equals
$\beta^P/(\beta^P - \gamma^P)$, the total equilibrium effect after the market adjusts, which is in general different from $\beta^P$. 
The right panel of Figure~\ref{fig:demand-supply-interventions} (Section~\ref{sec:application}) illustrates%
this distinction.
Thus, 
the 
two frameworks target different causal estimands even when applied to the same equilibrium data.
However,
$\beta^P$ can be computed from $B_0$: e.g.,
\[
\beta^P =
\frac{\text{impact of } U_t^S \text{ on } Q_t}{\text{impact of } U_t^S \text{ on } P_t}
=
\frac{(B_0)_{QS}}{(B_0)_{PS}}.
\]
Conversely, $\beta^P$, $\gamma^P$ and $\gamma^W$ reconstruct $B_0$, and together with $\beta^W$ and $\beta^{D1}$ (the coefficients in $A_1$), they reconstruct $B(L)$.

We now turn to the question of identifiability. 
The reduced form~\eqref{eq:svar_reduced_Form} can be 
obtained from the observational distribution by population
least squares, but, in general, $\Pi_1,\ldots,\Pi_p$ and $\Sigma_\varepsilon$ do not uniquely determine $B(L)$ without further restrictions.
The structural system in~\eqref{eq:svar_structural_form} carries three zero restrictions on $A_0$: (i)~price and (ii)~quantity do not enter the wind row, and (iii)~wind does not enter the demand row.
For restrictions of this form (zeros on the rows of $A_0$), 
\citet{rubio-ramirez_structural_2010} give rank conditions that are necessary (in just-identified systems such as ours) and sufficient for $A_0$ to be 
identifiable.%
\footnote{\citet{rubio-ramirez_structural_2010} write the system as $y_t' A_0 = \cdots$, so that an equation corresponds to a column of their $A_0$; we write $A_0 X_t = \cdots$, so an equation is a row. They also normalize $\Sigma_V = I$, whereas we normalize one coefficient per row of $A_0$ and leave the variances free. The two parametrizations differ by a rescaling of the rows of $A_0$, which leaves the zero pattern, and hence the rank condition, unchanged.}
Here, the rank condition reduces to $\beta^P \neq \gamma^P$ and $\gamma^W \neq 0$: the first 
condition appeared when defining the equilibrium, (see~\eqref{eq:svar_eq_price}), the second condition states instrument relevance.
Together with the shocks being mutually uncorrelated,
restrictions (i)--(iii) uniquely determine $A_0$, $\Sigma_V$, and because
$A_1 = A_0\Pi_1$, also $B(L)$.

Identification of $B(L)$ therefore rests on five premises, which we label for later reference: (P1)~the system is stable, so that~\eqref{eq:svar_moving_average} exists; (P2)~the reduced form~\eqref{eq:svar_reduced_Form} has finite order $p$; (P3)~its innovations satisfy $\varepsilon_t = B_0 V_t$; (P4)~the structural shocks are mutually uncorrelated, so that $\Sigma_V$ is diagonal; and (P5)~the zero restrictions~(i)--(iii) hold and the rank condition is satisfied.

\subsection{Comparing Classical SVAR and CIV Identification}
\label{sec:SVAR_RR}
\label{sec:SVAR_structural_params}
Identifying only $\beta^P$ from an SVAR  
is possible under strictly weaker conditions than 
(P1)--(P5).
We 
give 
three systems in which at least
one of (P2)--(P4) (i.e., 
a premise of the SVAR framework and not (P5), the identification conditions of \citet{rubio-ramirez_structural_2010}) fails: in the first two, \eqref{eq:svar_beta_ratio} still identifies $\beta^P$; in the third, it does not (but the nuisance IV estimator~$\#8$ remains valid). 

We first turn to the identifiability of $\beta^P$.
Since $\varepsilon_t = B_0V_t$, the rows of $A_0$ in~\eqref{eq:svar_structural_form} express the shocks in terms of the reduced-form innovations. Restrictions~(i) and~(ii) leave $V^W_t = \varepsilon_{t,W}$, and restriction~(iii) leaves $V_t^D = \varepsilon_{t,Q} - \beta^P \varepsilon_{t,P}$.
Writing $\sigma_{ij}$, $i,j \in \{W,P,Q\}$, for the entries of $\Sigma_\varepsilon$, 
the condition $\mathbb{E}[V_t^D V_t^W] = 0$ then gives
\begin{equation}\label{eq:svar_beta_ratio}
\beta^P = \frac{\sigma_{QW}}{\sigma_{PW}}.
\end{equation}
Recovering $\beta^P$ this way 
does not require all of
(P1)--(P5): 
it neither requires (P2), nor (P3) beyond its wind row (restrictions~(i) and~(ii), which give $\varepsilon_{t,W} = V^W_t$), nor (P4) beyond the single condition $\mathbb{E}[V^D_t V^W_t] = 0$. This is an internal instrument setting \citep{plagborg-moller_local_2021}.
The condition $\mathbb{E}[V_t^D V_t^W] = 0$ is the moment restriction that (CIV1) certifies: 
in our system,
it holds%
because
$W_t$ is d-separated from $D_t$ in $\mathcal{G}_{P_t \not\to D_t}$ given $\{W_{t-s}, P_{t-s}, D_{t-s}\}_{s=1}^{p}$, the conditioning set supplied by the projection on the past in~\eqref{eq:svar_reduced_Form}. If, instead, there was an unblocked confounding path between wind and demand 
the condition would generally fail
and~\eqref{eq:svar_beta_ratio} would generally be inconsistent for $\beta^P$.
Neither framework identifies $\beta^P$ in that case.

We now discuss 
the three systems.
In the first 
system, (P4) fails and $\beta^P$ is still identified. 
To define the system,
suppose the data are generated by a variant of~\eqref{eq:svar_demand}--\eqref{eq:svar_eq} in which the demand and supply mechanisms share an unobserved contemporaneous factor, so that the shocks are no longer mutually independent and $c \coloneqq \mathrm{Cov}(V^D_t, V^S_t) \neq 0$. The structural form~\eqref{eq:svar_structural_form},
and with it $\varepsilon_t = B_0V_t$, is unchanged, but $\Sigma_V$ is no longer diagonal. 
Imposing $\mathbb{E}[V^S_t V^D_t] = 0$ when it is false returns $(\gamma^P \sigma_D^2 - c\,\beta^P)/(\sigma_D^2 - c)$ in place of $\gamma^P$, so the supply row of $A_0$, and with it $B(L)$, is not recovered. $\beta^P$ is identified because $\mathbb{E}[V^D_t V^W_t] = 0$ still holds in this system.

In the second system, defined by Model~III \eqref{eq:demand_inventory}--\eqref{eq:price_inventory},
(P2) fails and $\beta^P$ is still identified. Model~III makes the inventory $K_{t-1}$ a geometric sum of past demand by~\eqref{eq:inventory}; eliminating it, by applying $(1 - (1-\delta^K)L)$ to~\eqref{eq:demand_inventory}, leaves a finite-lag equation in $D$ and $P$ whose error is the moving-average $U^D_t - (1-\delta^K)U^D_{t-1}$.
The system is therefore VARMA, no finite-order~\eqref{eq:svar_reduced_Form} represents it exactly, and (P3) fails at any finite $p$. Since $0 < 1-\delta^K < 1$, the moving-average polynomial $1-(1-\delta^K)L$ is invertible, so a reduced form of infinite order does exist, and its innovation covariance is the $\Sigma_\varepsilon$ that~\eqref{eq:svar_beta_ratio} refers to here.
Wind nevertheless follows an exogenous AR(1), so residualizing on lags leaves the wind innovation whatever the truncation, and estimator~$\#4$ (shown in Section~\ref{sec:SVAR-IV} to coincide with~\eqref{eq:svar_beta_ratio}) is valid under Model~III. 
 
In the third system, (P3) fails in its wind row and the SVAR approach identifies neither $\beta^P$ nor $B(L)$.
To define the system,
consider the confounded system described in Section~\ref{sec:consistent_estimators}, where, starting from Model~I, hidden confounding links $W_t$ with $D_{t-1}$ and $D_{t-1}$ with $D_t$. This makes $D_{t-1}$ a non-collider on $W_t \leftrightarrow D_{t-1} \to D_t$ and a collider on $W_t \leftrightarrow D_{t-1} \leftrightarrow D_t$: not conditioning on it leaves the first path open, which invalidates estimator~$\#2$, while conditioning on it, which is what estimator~$\#4$ and the projection in~\eqref{eq:svar_reduced_Form} do, opens the second.
More precisely, the 
contemporaneous condition $\mathbb{E}[V^D_t V^W_t] = 0$ remains intact, but 
the innovation $\varepsilon_{t,W}$ is no longer $V^W_t$, because~\eqref{eq:svar_reduced_Form} conditions on the collider $D_{t-1}$, and this correlates $\varepsilon_{t,W}$ with the constructed demand shock $\varepsilon_{t,Q} - \beta^P\varepsilon_{t,P}$. Applied to $\varepsilon_t$, the rows of $A_0$ therefore do not return the structural shocks, and~\eqref{eq:svar_beta_ratio} does not identify $\beta^P$ either. In the second system, the projection on the past still delivered the wind innovation; here it does not, and no admissible control set restores it (see Section~\ref{sec:SVAR-IV}).
The nuisance IV estimator~$\#8$ remains valid (Section~\ref{sec:consistent_estimators}).
 
Recovering $\beta^P$ when $B(L)$ is not identified is a different question from the one \citet{rubio-ramirez_structural_2010} ask, and closer to the Proxy-VAR approach \citep{stock_identification_2018} we turn to next.

\subsection{Comparing Proxy-VAR and CIV Identification}\label{sec:SVAR-IV}
\citet{stock_identification_2018} translated the IV identification strategy from microeconometrics to SVAR and local projection models.
Rather than recovering the full impact matrix from covariance restrictions, a Proxy-VAR uses an external variable $z_t$, correlated with one target shock and uncorrelated with the others, to trace that shock alone. The procedure has two steps. A reduced-form VAR~\eqref{eq:svar_reduced_Form} is estimated by least squares and its residuals $\hat\varepsilon_t$ retained, which removes the influence of the past; the impact vector is then obtained, up to scale, as the covariances of the proxy $z_t$ with those residuals, normalized so that the price response is one, and propagated through the estimated reduced-form moving-average matrices $\hat C_h$,
\begin{equation}\label{eq:proxy_var_estimator}
\hat b
=
\frac{\widehat{\mathbb{E}}\big[z_t\,\hat\varepsilon_{t}\big]}
     {\widehat{\mathbb{E}}\big[z_t\,\hat\varepsilon_{t,P}\big]},
\qquad
\hat\theta_{h} = \hat C_{h}\,\hat b,
\qquad h = 0, 1, 2, \ldots
\end{equation}
Here $\widehat{\mathbb{E}}$ denotes the sample average, $\hat\varepsilon_{t,j}$ the $j$th entry of $\hat\varepsilon_t$, and $C(L) = \Pi(L)^{-1}$ collects the \emph{reduced-form moving-average} matrices, with $C_0 = I$.
The first step is the projection on lags used in~\eqref{eq:svar_reduced_Form}; the second is what makes this an IV procedure, since the proxy supplies the exogenous variation and~\eqref{eq:proxy_var_estimator} is a ratio of reduced-form covariances.
 
Take the wind innovation as the proxy, $z_t = \hat\varepsilon_{t,W}$, and set $h = 0$. Then $\hat\varepsilon_{t,Q}$ and $\hat\varepsilon_{t,P}$ are $Q_t$ and $P_t$ residualized on $\{W_{t-s}, P_{t-s}, Q_{t-s}\}_{s=1}^{p}$, and $z_t$ is $W_t$ residualized on the same set, so by the Frisch--Waugh--Lovell theorem
\[
\hat\theta_{Q,0} = \mathrm{CIV}\big(W_t \mid P_t \to D_t \mid \{W_{t-s}, P_{t-s}, D_{t-s}\}_{s=1}^{p}\big),
\]
where $\hat\theta_{Q,0}$ is the entry of $\hat\theta_{0}$ that corresponds to quantity. This expression is estimator~$\#4$ of Table~\ref{tab:estimators} (with $L=p$ and given that in equilibrium $Q_t=D_t$), the 
\emph{lag-augmented local projection IV estimator} at $h=0$ \citep{montiel_olea_local_2021, plagborg-moller_local_2021}. It is also~\eqref{eq:svar_beta_ratio}, since both are $\sigma_{QW}/\sigma_{PW}$. The fully identified SVAR, the Proxy-VAR, and CIV estimator $\#4$ thus yield the same 
estimate
for $\beta^P$ in this system. The equivalence is exact at finite $p$ here because Model~I's reduced form is a VAR(1) \citep{plagborg-moller_local_2021}.

The assumptions
required by Proxy-VAR and CIV are closely related. Both cover relevance and exogeneity, but they are stated differently and are not equivalent. For a target shock $V^1_{t}$ and proxy $z_t$, \citet[Section~1.3]{stock_identification_2018} require
\begin{equation}\label{eq:proxy_var_conditions}
\underbrace{\mathbb{E}[z_t V^1_{t}] \neq 0}_{\text{relevance}},
\qquad
\underbrace{\mathbb{E}[z_t V^j_{t}] = 0 \ \ \forall j \neq 1}_{\text{contemporaneous exogeneity}},
\qquad
\underbrace{\mathbb{E}[z_t V_{t+k}] = 0 \ \ \forall k \neq 0}_{\text{lead--lag exogeneity}},
\end{equation}
where $V^j_{t}$ is the $j$th entry of $V_t$ and $V^1_{t}$ the targeted shock, together with regularity conditions ensuring well-defined covariances. They also allow these conditions to hold only after residualizing on control variables \citep[Condition LP-IV$^{\perp}$ of][Section~1.4]{stock_identification_2018}. The CIV criteria (CIV1)--(CIV3) are given in Section~\ref{sec:conditionsIV}. 
From here, we draw three comparisons.

First, our CIV framework and \citet{stock_identification_2018} are closely related. 
Consider wind in levels as the proxy. Strict lead--lag exogeneity
fails: $\mathbb{E}[W_t U^W_{t-s}] = (\beta^W)^s\sigma_W^2 \neq 0$ for $s \geq 1$, which is why the naive~IV estimator $\#1$ is inconsistent 
(Section~\ref{sec:consistent_estimators}).
Both frameworks agree on the remedy: conditioning on $\{W_{t-s}\}_{s=1}^{L}$, as estimator~$\#2$ does, residualizes wind on its own past and leaves the wind innovation, for which lead--lag exogeneity holds.
This works because wind's serial correlation comes from its own innovation alone. In general, \citet[Theorem~1, Section~2.2.3]{stock_identification_2018} show that adding lagged observables as generic controls restores lead--lag exogeneity if and only if the corresponding VAR is invertible.

Second, the CIV criteria admit designs that neither~\eqref{eq:proxy_var_conditions} nor its control-variable version admits.
The two frameworks depart when lead--lag exogeneity cannot be restored by residualization, that is, when no admissible control set exists. 
In the confounded system of Section~\ref{sec:consistent_estimators}, $D_{t-1}$ is a collider on one path from $W_t$ to $D_t$ and a non-collider on another (Section~\ref{sec:SVAR_RR}), so neither conditioning on it nor omitting it leaves $\hat\varepsilon_{t,W}$ a valid proxy. The nuisance IV estimator~$\#8$ is nevertheless valid.

Third, using a graph 
makes the estimand explicit.
A supply-side instrument moves the equilibrium price along the demand function, so the ratio of its impact effects on quantity and on price is $\beta^P$; conversely, a demand-side instrument moves it along the supply function, so that ratio is $\gamma^P$. 
More precisely, with $z_t = \hat\varepsilon_{t,W}$ the vector $\hat b$ estimates the wind column of $B_0$ normalized to a unit price impact. At $h = 0$ the rescaling divides out $\gamma^W/(\beta^P - \gamma^P)$ and leaves $\beta^P$; at $h > 0$ it leaves the propagation of the instrument's own shock through the system, not the response of $Q_t$ to a shock in prices. In the graph, by contrast, the estimator names the edge $P_t \to D_t$ (not $P_t \to Q_t$), so which effect is being estimated is part of its specification. Under CIV we claim identification of the structural coefficient $\beta^P$ only, not of a dynamic response.

\subsection{Numerical Illustration}\label{sec:SVAR_numerical}
\begin{figure}[t]
\centering
\includegraphics[width=\linewidth]{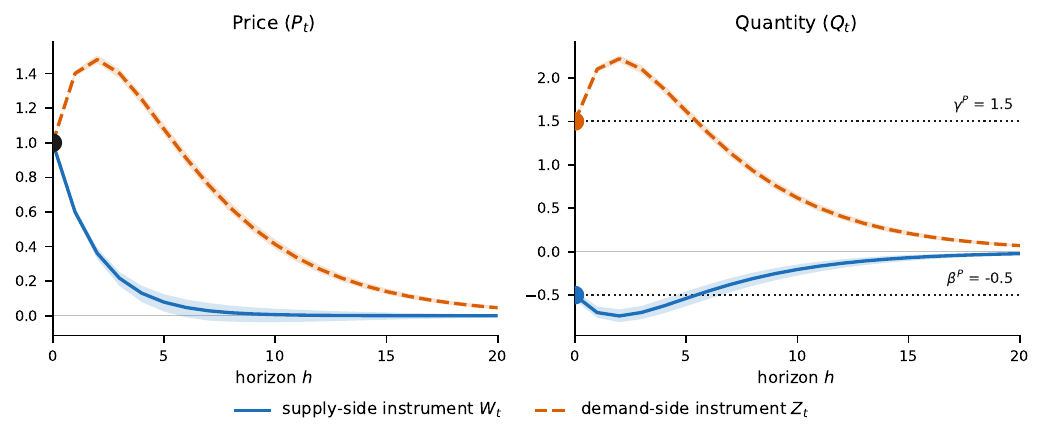}
\caption{Proxy-VAR estimates $\hat\theta_h$ of~\eqref{eq:proxy_var_estimator} 
over horizons $h = 0,\ldots, 20$. 
The left panel shows the price entry of $\hat\theta_h$, the right panel the quantity entry.
Solid blue: using the supply-side instrument $W_t$, whose normalized impact response of quantity
(right panel, $h=0$)
estimates the structural demand parameter $\beta^P$. Dashed orange: using the demand-side instrument $Z_t$, whose normalized impact response of quantity (right panel, $h=0$) estimates the structural supply parameter $\gamma^P$.
In the left panel, both curves equal one at $h=0$ by the normalization in~\eqref{eq:proxy_var_estimator}, which divides by the price entry;
at $h>0$ each curve traces the propagation of its own instrument's shock. Shaded bands are pointwise $2.5$--$97.5$ percentiles over $N = 200$ replications of $T = 40{,}000$ observations.
The figure illustrates the third
point 
of Section~\ref{sec:SVAR-IV}: which structural parameter a proxy recovers is determined by which side of the market it shifts. Dynamic effects for $h\geq1$ also depend on which proxy is used.
}
\label{fig:svar_proxy_var}
\end{figure}

We illustrate the two comparisons (to classical SVAR and Proxy-VAR) on simulated data.
Figure~\ref{fig:svar_proxy_var} shows the Proxy-VAR implemented with two different proxies: we extend Model~I with an observed demand-side instrument, $Z_t \coloneqq \rho^Z Z_{t-1} + U^Z_t$, entering demand~\eqref{eq:svar_demand} with coefficient $\beta^Z$ and excluded from supply, and estimate~\eqref{eq:proxy_var_estimator} on $X_t = (W_t, Z_t, P_t, Q_t)'$ using in turn $\hat\varepsilon_{t,W}$ and $\hat\varepsilon_{t,Z}$ as the proxy. On impact the two estimators return $\beta^P$ and $\gamma^P$, respectively; from $h = 1$ on they trace the propagation of their own instrument's shock. Throughout,
\begin{gather*}
\beta^P = -0.5, \quad \beta^{D1} = 0.8, \quad \gamma^P = 1.5, \quad \gamma^W = 1.0, \quad \beta^{W} = 0.8,\\
D^0 = 30, \quad S^0 = 10,  \quad \sigma_W = 1, \quad \sigma_D = 2, \quad \sigma_S = 0.5,\\
\text{and, where } Z_t \text{ is used,} \quad \beta^Z = 1.7, \quad \rho^{Z} = 0.8, \quad \sigma_Z = 1,
\end{gather*}
and all reduced forms are estimated with $p = 1$ lag.

Table~\ref{tab:svar_comparison} reports what each estimator returns under Model~I and under the confounded system of Section~\ref{sec:consistent_estimators}. Under Model~I, all four are centered on $\beta^P$. Under the confounded system the classical SVAR and the Proxy-VAR return the same incorrect value
(the two coincide here, see Section~\ref{sec:SVAR-IV}); estimator~$\#2$ is further off; and only the nuisance IV estimator~$\#8$ recovers 
the correct value of
$\beta^P$.
\begin{table}[ht]
\centering
\begin{tabular}{lcc}
\toprule
 & Model~I & Section~\ref{sec:consistent_estimators} \\
\midrule
Classical SVAR, \eqref{eq:svar_beta_ratio}  & $-0.5018$ $(0.0181)$ & $-0.4213$ $(0.0178)$ \\
Proxy-VAR, $h=0$ (estimator $\#4$)          & $-0.5018$ $(0.0181)$ & $-0.4213$ $(0.0178)$ \\
Estimator $\#2$                             & $-0.5011$ $(0.0239)$ & $-1.3549$ $(0.0542)$ \\
Estimator $\#8$ (nuisance IV)               & $-0.5019$ $(0.0181)$ & $-0.4999$ $(0.0643)$ \\
\bottomrule
\end{tabular}
\caption{Estimates of $\beta^P = -0.5$ under Model~I and under the confounded system of Section~\ref{sec:consistent_estimators}, in which $W_t$ is confounded with $D_{t-1}$ and $D_{t-1}$ with $D_t$. Mean over $N = 200$ replications of $T = 40{,}000$ observations; standard deviation in parentheses. The classical SVAR estimate is~\eqref{eq:svar_beta_ratio} and the Proxy-VAR estimate is~\eqref{eq:proxy_var_estimator} at $h = 0$; the two are numerically identical by the argument of Section~\ref{sec:SVAR-IV}.
Estimators $\#2$, $\#4$, and $\#8$ are those of Table~\ref{tab:estimators} with $L=1$.
}
\label{tab:svar_comparison}
\end{table}

 \subsection{Discussion}\label{sec:SVAR_discussion}
We have established that the two frameworks target different estimands, that $\beta^P$ can be identified under strictly weaker conditions than those that identify $B(L)$,
and that the CIV criteria admit designs that the Proxy-VAR conditions do not.

The CIV criteria characterize the set of valid $(\mathcal{I}, \mathcal{X}, \mathcal{B})$ configurations for a given graph, and typically several can be valid, of which Table~\ref{tab:estimators} lists a few. 
Disagreement among estimators that are valid under the same graph is evidence against the assumptions encoded in the model (Section~\ref{sec:consistent_estimators}), and in the application this allows us to reject Models~I and~II (Section~\ref{sec:application}).
A just-identified SVAR scheme (a set of zero restrictions or a single proxy) only yields a single estimator of $\beta^P$ and cannot directly be used in a similar way. 

Still, we do not think that either 
approach strictly dominates the other.
The same equilibrium system admits multiple causal representations, each suited to a different question and to the methods a given setting allows. Researchers interested in equilibrium propagation may prefer SVAR methods; those interested in recovering the primitive structural relationships 
and in obtaining several different estimators
may benefit from graphical identification. 

\section{Extension to Nonlinear Causal Effects}
\label{sec:non-linear-ext}
The main text assumes that all structural equations are linear. 
In the classical (i.e., non-equilibrium) nonlinear IV literature, 
identifiability 
has been studied \citep[e.g.,][]{amemiya, neweypowell, imbens_identification_2009, newey_nonparametric_2013,christiansen_causal_2021}; from a more practical perspective, 
several estimators have been proposed \citep{neweypowell, Hartford2017,NPREGIV-CRAN,chen2018optimal, singh_kernel_2019, christiansen_causal_2021}, many of which have not yet been extended to the conditional case.
This appendix relaxes the linearity assumption for equilibrium systems: we restate the
system in nonlinear form and define the estimand
(Appendix~\ref{sec:nonlinear-equilibrium}), ask what an IV analysis identifies and
connect 
our
graphical formulation 
to the average-derivative interpretation of
\citet{angrist_interpretation_2000}
(Appendix~\ref{sec:nonlinear-identification}), develop the tractable case of structural
equations that are linear after a known feature transformation
(Appendix~\ref{sec:feature-extension}), validate 
our results
in simulation
(Appendix~\ref{sec:nonlinear-simulation}), and 
discuss how the graphical view is helpful in the nonlinear case, too
(Appendix~\ref{sec:nonlinear-graphical}).

\subsection{Nonlinear Equilibrium Systems and the Causal Estimand}
\label{sec:nonlinear-equilibrium}

We return to the equilibrium system of Section~\ref{sec:demand_autocorrelated},
now without the linearity assumption. Taking Model~I as the running example,
the structural equations for demand, supply, and the instrument are
\begin{align}
    D_t &\coloneqq f^D(P_t, D_{t-1}, U^D_t), \label{eq:nl-demand}\\
    S_t &\coloneqq f^S(P_t, W_t, U^S_t), \label{eq:nl-supply}\\
    W_t &\coloneqq f^W(W_{t-1}, U^W_t), \label{eq:nl-wind}
\end{align}
with mutually independent i.i.d.\  shocks $U^D_t, U^S_t, U^W_t$, each with mean
zero. The alternative demand structures (Models~II and~III) are obtained by
changing the arguments of $f^D$ exactly as in
Section~\ref{sec:demand_alternatives}; the discussion below covers all three.

We assume that $f^D$ is continuously differentiable and strictly decreasing in
its price argument, that $f^S$ is continuously differentiable and strictly
increasing in its price argument, and that $(D_t, P_t, W_t)$ is strictly
stationary and ergodic with finite second moments. Under these conditions the
market-clearing condition
\begin{equation}\label{eq:nl-equilibrium}
    f^D(P_t, D_{t-1}, U^D_t) = f^S(P_t, W_t, U^S_t)
\end{equation}
admits, for every realization of $(D_{t-1}, W_t, U^D_t, U^S_t)$, a
solution for the equilibrium price, which we write as
$P_t = f^P(D_{t-1}, W_t, U^D_t, U^S_t)$. The monotonicity of $f^D$ and $f^S$ in
price guarantees the uniqueness of this solution; 
it is
the nonlinear counterpart of the requirement $\beta^P \neq \gamma^P$ in the
linear model, and coincides with the structural monotonicity conditions of
\citet[Lemma~1]{angrist_interpretation_2000}. 
Under the further assumption that the curves cross at one point, this 
ensures the existence of the solution.
As in the main text the equations
are structural in $(P_t, D_{t-1})$ and $(P_t, W_t)$; in potential-outcome
notation we write $D_t(p) \coloneqq f^D(p, D_{t-1}, U^D_t)$ and
$S_t(p,w) \coloneqq f^S(p, w, U^S_t)$ for the demand and supply that would be
realized at price $p$ (and instrument $w$).

In the linear model the causal effect of interest is the slope $\beta^P$. For a
general $f^D$ the natural causal object is not a single number but a function --
the \emph{average structural function} (ASF) \citep[e.g.,][]{imbens_identification_2009}, i.e., the average dose--response of
demand to an intervention on price,
\begin{equation}\label{eq:asf}
    m^*(p)
    \coloneqq
    \mathbb{E}\!\left[D_t \mid \mathrm{do}(P_t = p)\right]
    =
    \mathbb{E}\!\left[f^D(p, D_{t-1}, U^D_t)\right].
\end{equation}
Because the intervention is on $P_t$ alone, the expectation in~\eqref{eq:asf} is
over the stationary (observational) distribution of $(D_{t-1}, U^D_t)$, which the
intervention leaves unchanged. When $f^D$ is linear, $m^*(p) = \tilde{D}^0 + \beta^P p$
with $\tilde{D}^0 \coloneqq D^0 + \beta^{D1}\mathbb{E}[D_{t-1}] + \mathbb{E}[U^D_t]$, so
the main-text parameter $\beta^P$ 
is exactly
the (constant) slope $m^{*\prime}$.
When $f^D$ is nonlinear, $m^*$ is curved and is no longer captured by a single
slope.

The shape of $m^*$ also pins down what the edge $P_t \to D_t$ represents when the
shock does not enter additively. Suppose, for example, that $U^D_t$ moderates the
price response,
\begin{equation}\label{eq:moderator}
    f^D(P_t, D_{t-1}, U^D_t)
    = D^0 + \beta_0^P P_t + \beta^{D1} D_{t-1}
      + \theta^D\, P_t U^D_t + U^D_t .
\end{equation}
Then $m^*(p) = \tilde{D}^0 + (\beta_0^P + \theta^D\,\mathbb{E}[U^D_t])\,p$, so the slope of
interest is the \emph{total} price effect $\beta_0^P + \theta^D\,\mathbb{E}[U^D_t]$  
-- the average response of demand to an intervention on price -- rather than the
conditional slope $\beta_0^P$ that holds at a fixed value of $U^D_t$ (though in the case where $\mathbb{E}[U^D_t]=0$ this is the same). We take this
total effect (equivalently, the slope of $m^*$) to be the estimand throughout this appendix.

\subsection{Identifiability and Identification under Nonlinear Structural Equations}
\label{sec:nonlinear-identification}

We now ask what an instrument that shifts supply identifies about the average
structural function~\eqref{eq:asf}. To separate the role of nonlinearity from
the autocorrelation studied in the main text, we suppress the lagged arguments
and write $f^D(P_t, U^D_t)$; lagged parents are handled exactly as in
Section~\ref{sec:CIV-estimation}, by including them in the conditioning set
$\mathcal{B}$.

Let $m^*(p) \coloneqq \mathbb{E}[D_t \mid \mathrm{do}(P_t = p)] = \mathbb{E}[D_t(p)]$
denote the average structural function (ASF) of
Appendix~\ref{sec:nonlinear-equilibrium}, and define the residual
\begin{equation}\label{eq:residual}
    e_t \coloneqq D_t - m^*(P_t),
    \qquad\text{so that}\qquad
    D_t = m^*(P_t) + e_t .
\end{equation}
The decomposition~\eqref{eq:residual} holds by construction.
If, in addition,
\begin{equation}\label{eq:moment}
    \mathbb{E}[e_t \mid W_t] = 0,
\end{equation}
then $m^*$ solves the conditional moment restriction
$\mathbb{E}[D_t - m(P_t)\mid W_t]=0$ and is identified under a standard
completeness condition \citep{neweypowell}.
We now discuss whether
the
ASF satisfies the instrument moment condition~\eqref{eq:moment}.

Crucially, \eqref{eq:moment} is, in general, \emph{not} implied by the graphical validity of $W_t$ as an
instrument for linear SCMs (conditions (CIV1) and (CIV2)). A graphically valid instrument 
implies
exogeneity,
$W_t \perp\!\!\!\perp U^D_t$, and the exclusion restriction that $W_t$ does not enter $f^D$;
these properties constrain how $W_t$ relates to the structural shock 
$U^D_t$
but say nothing about
how that shock enters $f^D$. In fact, \eqref{eq:moment} holds automatically if the shock is additively separable, 
that is,
\begin{equation}\label{eq:separable}
    D_t = m^*(P_t) + U^D_t .
\end{equation}
Under~\eqref{eq:separable} the residual coincides with the structural shock,
$e_t = U^D_t$, and \eqref{eq:moment} 
coincides with
instrument exogeneity
$\mathbb{E}[U^D_t \mid W_t]=0$. Note that \eqref{eq:separable} does not
require $m^*$ to be linear: the price response may be arbitrarily curved, as long
as the shock does not interact with price.

When the shock is not additively separable, $e_t$ ceases to equal $U^D_t$ and
inherits a price-dependent, confounded component. For the moderating
specification~\eqref{eq:moderator}, the ASF slope is
$\beta^P = \beta_0^P + \theta^D\mathbb{E}[U^D_t]$ and the residual contains the term
$\theta^D P_t (U^D_t-\mathbb{E}[U^D_t])$. Because the equilibrium
price~\eqref{eq:nl-equilibrium} depends on $U^D_t$, the price $P_t$ is correlated
with the shock, so $\mathbb{E}[e_t\mid W_t]$ does not vanish in general.

The decisive point is that additive separability is a restriction on the
functional form of $f^D$, not a property of the causal graph. Conditioning on a
set $\mathcal{B}$, as in the criteria (CIV1)--(CIV3), can restore a violated
exogeneity or exclusion restriction by blocking a path, but it cannot, in general, turn a
nonseparable structural equation into a separable one. Note that nonseparability does not preclude identification per se: \citet{imbens_identification_2009}, for example, identify the ASF in a nonseparable triangular model by exploiting a scalar-monotone first stage to construct a control variable, together with an appropriate common-support condition.

\subsubsection{The Estimator-side reading.}
\label{sec:esr}
The same phenomenon has a complementary description on the side of the estimator,
due to \citet{angrist_interpretation_2000}. Rather than asking whether the ASF
satisfies the moment condition, they fix the linear IV estimator -- the ratio of
the covariance of the outcome and the instrument to the covariance of price and
the instrument -- and ask what it identifies when demand is heterogeneous.
Casting the simultaneous system in terms of the potential outcomes $D_t(p)$,
$S_t(p,w)$ rather than functional-form-specific residuals, they show, under four
assumptions,
\begin{itemize}
    \item[(i)] \emph{independence}: $W_t \perp\!\!\!\perp \{D_t(p), S_t(p,w)\}_{p,w}$ (given
    covariates), the instrument is as good as randomly assigned;
    \item[(ii)] \emph{exclusion}: $W_t$ does not enter demand;
    \item[(iii)] \emph{relevance}: $W_t$ has a non-trivial effect on the
    equilibrium price;
    \item[(iv)] \emph{monotonicity}: the equilibrium price is monotone in $W_t$;
\end{itemize}
that the IV estimand equals a weighted average of the derivative of the demand
function,
\begin{equation}\label{eq:agi}
    \beta^{\mathrm{IV}}
    = \int_0^\infty
        \mathbb{E}\!\left[
            \frac{\partial D_t}{\partial p}(p) \,\middle|\, p \in \mathcal{P}_t
        \right]
        \omega(p)\, \mathrm{d}p,
    \qquad
    \omega \ge 0,\quad \int \omega(p)\,\mathrm{d}p = 1,
\end{equation}
where the weights $\omega(p)$ are proportional to the number of periods whose
price the instrument shifts across $p$, and $\mathcal{P}_t$ is the
corresponding price interval. Monotonicity~(iv) makes these
weights non-negative; in the linear constant-coefficient model it holds
automatically, 
and otherwise it has to be assumed. Related recent work by \citet{kolesar_dynamic_2025} establishes
weighted-average causal interpretations for linear local projections and VARs
based on observed shocks or proxies under nonlinear data-generating processes.

The two assumptions on the shock map onto the two regimes above. When demand is
identical across periods up to an additive shift, $D_t(p)=m^*(p)+U^D_t$ -- the
separable case~\eqref{eq:separable} -- the estimand~\eqref{eq:agi} is a weighted
average of the slope of the population-average demand curve, with no averaging
over heterogeneity \citep[Cor.~3]{angrist_interpretation_2000}; this is the
regime in which $e_t=U^D_t$ and IV recovers the slope of $m^*$ (exactly so
when that slope is constant). When instead the price response is heterogeneous --
say $D_t(p)=a_t-b_t\,p$ with random coefficients -- \eqref{eq:agi} becomes a
price-gap--weighted average of the period-specific slopes $b_t$, which in general
differs from the average slope $\mathbb{E}[b_t]$
\citep[Cor.~2]{angrist_interpretation_2000}. In this regime
$\mathbb{E}[e_t\mid W_t]\neq 0$, and the weighted average in~\eqref{eq:agi} is
precisely the object hidden inside the residual $e_t$.

These complications motivate the restriction to linear structural equations in
the main text. They also delimit a useful intermediate case: if the outcome
equation is linear and separable after a known, finite feature transformation of price, the
difficulties above disappear and the conditional-IV estimators of
Section~\ref{sec:CIV-estimation} apply with only a change of variables. We
develop this case next.

\subsection{Feature-Expansion without Interactions}
\label{sec:feature-extension}

Assume a structural causal model for time series as in
Appendix~\ref{sec:appendix_scm}, and suppose the structural equation for the
outcome $Y$ is linear in a known feature map of the endogenous variables,
\begin{equation}\label{eq:feature-structural}
    X_Y = \beta^{\top}\phi(X_\mathcal{X}) + \gamma^{\top} X_K + \epsilon^Y,
\end{equation}
where $\phi:\mathbb{R}^{|\mathcal{X}|}\to\mathbb{R}^{d_{\phi}}$ is a feature map
and the shock $\epsilon^Y$ enters additively. A feature map is a fixed,
known transformation of the endogenous variables -- for instance
$\phi(p)=(p,p^2)$ for a quadratic price response, or a spline or
piecewise-linear basis -- chosen by the analyst rather than estimated. Two points
about the assumptions in~\eqref{eq:feature-structural} are worth making
explicit. Treating $\phi$ as known is what keeps the problem inside the linear
identification theory of the main text: once the features are fixed,
$\phi(X_\mathcal{X})$ is an ordinary (vector-valued) regressor and no
non-parametric estimation of the functional form is required. The requirement
that the shock
$\epsilon^Y$ enters~\eqref{eq:feature-structural} additively is the separability
condition of Appendix~\ref{sec:nonlinear-identification}: it is what lets the
coefficient $\beta$ keep its causal reading as the (feature-transformed) average
derivative, rather than the heterogeneity-weighted average of~\eqref{eq:agi}.

We extend the conditional-IV condition to the feature-expanded setting by
replacing (CIV3) with the following relevance condition:
\begin{itemize}
    \item[(CIV3$'$)] there exists a known feature map $\psi$ such that
    \begin{equation*}
        \mathbb{E}\!\left[\mathrm{cov}\!\left(\phi(X_{\mathcal{X}}),\,
        \psi(X_\mathcal{I}) \mid X_\mathcal{B}\right)\right]
    \end{equation*}
    has full row rank, i.e., rank equal to $d_{\phi}$.
\end{itemize}
Assume that (CIV1), (CIV2), and (CIV3$'$) are satisfied, that the time series is
strictly stationary and ergodic, and that it has finite second moments. Then,
following the same arguments as in \citet{thams_identifying_2022}, $\beta$ is the
unique vector satisfying
\begin{equation*}
    \mathrm{cov}\!\left(X_Y - b^{\top}\phi(X_\mathcal{X}),\,
    \psi(X_\mathcal{I}) \mid X_\mathcal{B}\right) = 0,
\end{equation*}
and
\begin{equation*}
    \hat{\beta} \coloneqq \arg\min_{b}
    \left\|
    \widehat{\mathrm{cov}}\!\left(X_Y - b^{\top}\phi(X_\mathcal{X}),\,
    \psi(X_\mathcal{I}) \mid X_\mathcal{B}\right)
    \right\|_F
\end{equation*}
is a consistent estimator of $\beta$.
Operationally, $\hat\beta$ is the
conditional-IV estimator of Section~\ref{sec:CIV-estimation} applied to the
transformed regressors $\phi(X_\mathcal{X})$ and instruments
$\psi(X_\mathcal{I})$; consequently the whole validity analysis of the main text
-- the choice of conditioning set, the nuisance-effect construction, and the
estimator-discrepancy diagnostic -- carries over unchanged, now identifying the
feature-transformed effect $\beta$. The features that can be accommodated are
exactly those that enter the outcome equation additively and do not interact with
the shock; price-by-shock interactions of the form~\eqref{eq:moderator} fall
outside this class and return us to the general problem of
Appendix~\ref{sec:nonlinear-identification}.

\subsection{Simulation: A Quadratic Demand Curve}
\label{sec:nonlinear-simulation}

We illustrate the feature-expansion estimator of Appendix~\ref{sec:feature-extension}
on simulated electricity-market data, specializing the general demand
equation~\eqref{eq:nl-demand} to the feature-expanded linear
form (Equation~\ref{eq:feature-structural}) with the quadratic feature map
$\phi(P_t) = (P_t, P_t^2)^\top$.
We take the same assumptions as in Model~I but replace the demand equation by
\begin{equation}\label{eq:nlsim-demand}
    D_t \coloneqq D^0 + \beta^{PL} P_t + \beta^{PS} P_t^2 + \beta^{D1} D_{t-1} + U^D_t,
\end{equation}
where $\beta^{PL}$ and $\beta^{PS}$ are the linear and quadratic own-price
coefficients, $\beta = (\beta^{PL}, \beta^{PS})^\top$ in the notation
of~\eqref{eq:feature-structural}. 
The supply and wind equations are  
as in
Section~\ref{sec:Simulation}.
Equating demand and supply gives a quadratic equation in $P_t$,
\begin{equation}\label{eq:nlsim-equilibrium}
    \beta^{PS} P_t^2 + (\beta^{PL} - \gamma^P) P_t
    + \left(D^0 - S^0 + \beta^{D1}D_{t-1} + U^D_t - \gamma^W W_t - U^S_t\right) = 0,
\end{equation}
with two roots given by the quadratic formula. 
We use the values $\beta^{PL}=-100$ and $\beta^{PS}=-1$. The two roots lie on either side of the vertex $(\gamma^P-\beta^{PL})/(2\beta^{PS}) = -300$ EUR/MWh, at which supply is already negative for the wind levels we simulate, so the smaller root implies a negative equilibrium quantity and we select the larger one. On that branch demand is strictly decreasing, since $\beta^{PL}+2\beta^{PS}P_t<0$ for $P_t>-50$ EUR/MWh, which holds throughout the simulated price range. Equilibrium is therefore always well-defined.
The choice $\beta^{PS} < 0$ implies that the marginal price response
becomes more negative as $P_t$ rises: consumers curtail demand more 
aggressively per unit price increase at high price levels.

We apply the feature expansion to each of the eight conditional-IV estimators 
of Table~\ref{tab:estimators} by applying $\phi$ to every price regressor and
$\psi(W_t) = (W_t, W_t^2)^\top$ to every instrument, as in the relevance
condition (CIV3$'$). 
We also apply $\omega(X_t) = (X_t, X_t^2)^\top$ to every
variable in the conditioning set. Blocking a path requires conditional independence, which residualizing linearly on the conditioning set delivers only when the dependence on it is linear; here the wind process is linear and the expansion is redundant, but in general it is not.
The path-blocking logic of (CIV1)--(CIV2) 
is unaffected by this change, since it depends only on the topology of the 
marginalized graph, which is the same as in the linear case: $P_t$ enters 
demand through a single edge $P_t \to D_t$, whether that edge corresponds to
a linear or a nonlinear (feature-expanded) causal effect. We therefore 
expect the same validity pattern across Models~I--III as in 
Table~\ref{tab:estimators}, now applied jointly to $\hat\beta^{PL}$ 
and $\hat\beta^{PS}$.

\begin{figure}[t]
    \centering
    \includegraphics[width=0.85\linewidth]{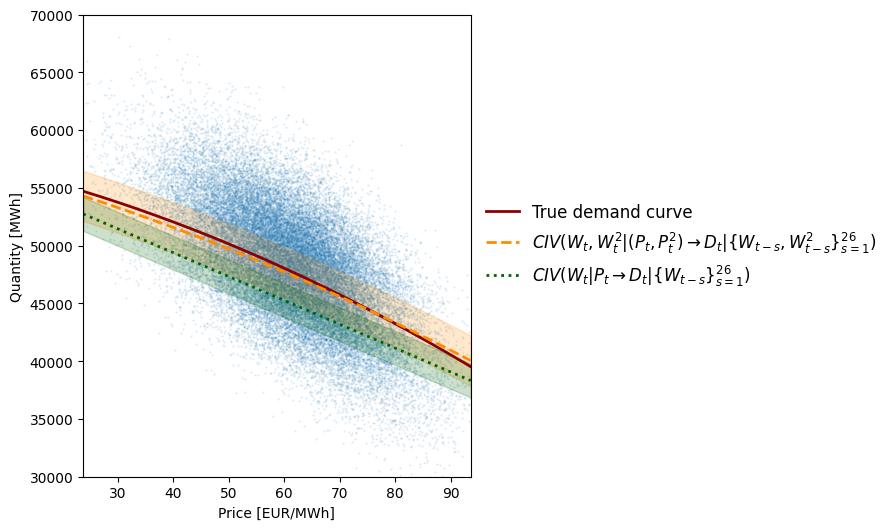}
    \caption{Simulated equilibrium data $(P_t, D_t)$ and recovered demand curves under the quadratic expansion of Model~I. The solid red line is the average structural function $m^*(p) = \mathbb{E}[D_t \mid \mathrm{do}(P_t = p)] 
    = D^0 + \beta^{D1}\mathbb{E}[D_{t-1}] + \beta^{PL}p + \beta^{PS}p^2$, 
    evaluated at the sample mean of $D_{t-1}$. The dashed
    line is estimator~$\#2$ implemented with the quadratic feature maps
    $\phi(P_t)=(P_t,P_t^2)^\top$, $\psi(W_t)=(W_t,W_t^2)^\top$ 
    and $\omega(X_t) = (X_t, X_t^2)^\top$,
    which satisfies (CIV1)--(CIV3$'$) and traces the curvature of $m^*$
    closely. The dotted line is the same estimator implemented with only the
    linear terms $P_t$ and $W_t$: although it still satisfies
    (CIV1)--(CIV2), it targets a single weighted-average slope rather than
    the curve itself, and deviates from $m^*$. 
    }
    \label{fig:nlsim-curve}
\end{figure}

\begin{figure}[t]
    \centering
    \includegraphics[width=1\linewidth]{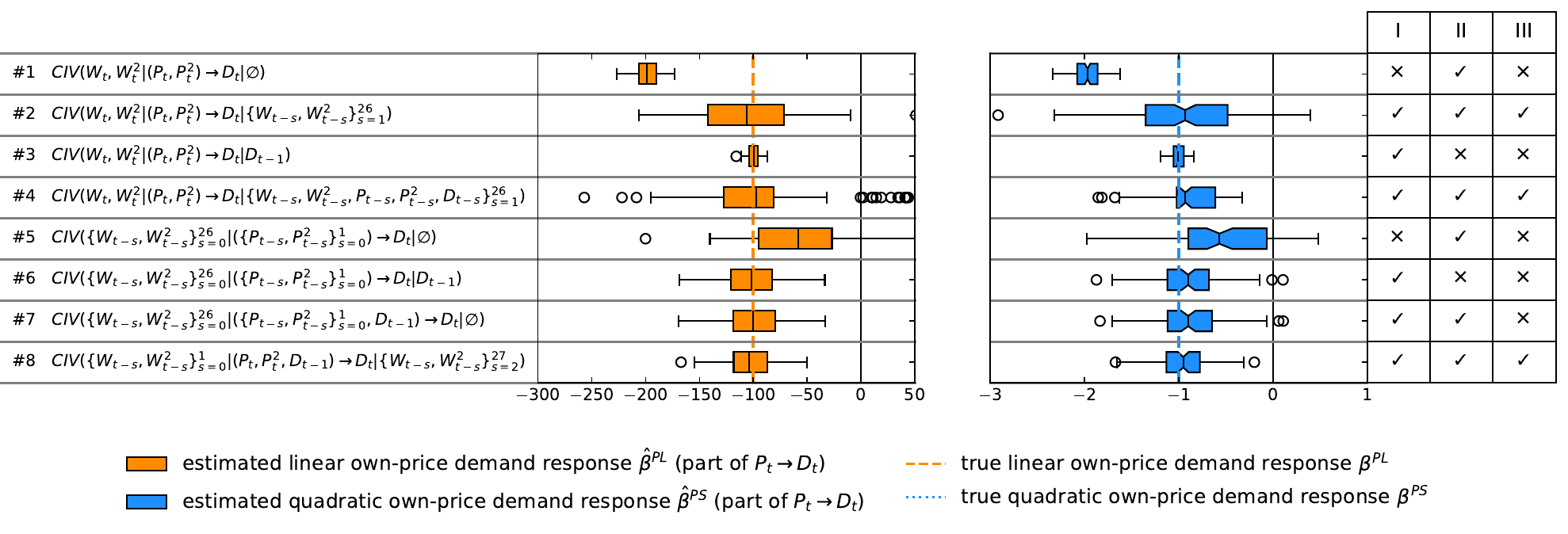}
    \caption{Recovered linear and quadratic demand-response coefficients
    across the eight feature-expanded CIV estimators and $N=100$
    simulated datasets, for the quadratic expansion of Model~I. Left:
    estimated $\hat\beta^{PL}$ (orange) against the true value $\beta^{PL}$
    (dashed line). Right: estimated $\hat\beta^{PS}$ (blue) against the true
    value $\beta^{PS}$ (dotted line). 
    The validity columns on the right give the (CIV1)--(CIV2) classification of Table~\ref{tab:estimators}. The variable sets are as printed in the figure; some include additional lags of the instrument, which does not change the classification.
    Estimators valid for a given model recover both coefficients.
    }
    \label{fig:nlsim-validity}
\end{figure}

\begin{figure}[t]
    \centering
    \includegraphics[width=1\linewidth]{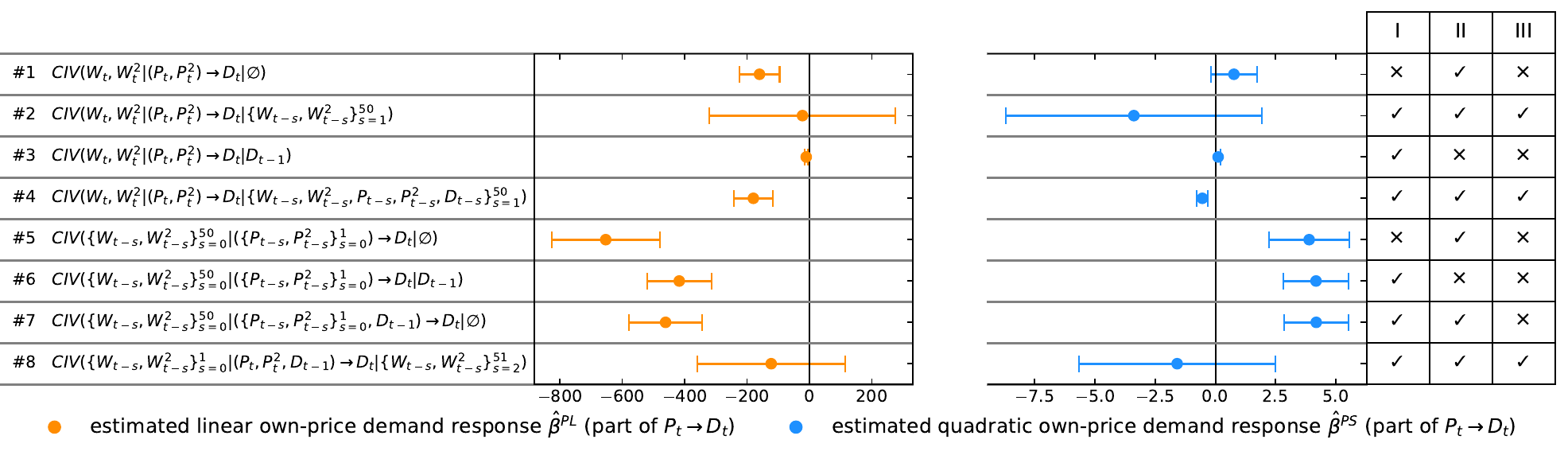}
    \caption{The own-price elasticity of aggregate German electricity demand, obtained by CIV estimators under a linear-quadratic feature expansion. Left:
    estimated $\hat\beta^{PL}$ (orange). Right: estimated $\hat\beta^{PS}$ (blue). The validity columns on the right give the (CIV1)--(CIV2) classification of Table~\ref{tab:estimators}. The variable sets are the feature-expanded counterparts of those in Table~\ref{tab:estimators}, as printed in the figure.}
    \label{fig:nlsim-application}
\end{figure}

Figure~\ref{fig:nlsim-curve} shows simulated equilibrium data together with the
average structural function $m^*(p)$
and two implementations of estimator~$\#2$: one using the
quadratic feature map for both the regressor and the instrument, and one using
only the linear terms $P_t$ and $W_t$. Both implementations satisfy
(CIV1)--(CIV2) for Model~I, since neither changes the conditioning set; only
the feature-expanded implementation also satisfies the relevance condition
(CIV3$'$) for the quadratic component. The linear-only implementation
illustrates the point of Appendix~\ref{sec:nonlinear-identification}
empirically: rather than recovering the curve $m^*(p)$, it recovers a single
slope, an instrument-weighted average of the local derivative of $m^*$ over the 
simulated price support (Equation~\eqref{eq:agi}). Figure~\ref{fig:nlsim-validity}
confirms this systematically: across simulated datasets and all eight
estimators, applied to Model~I, only estimators satisfying
(CIV1)--(CIV3$'$) for a given model recover both $\beta^{PL}$ and $\beta^{PS}$
without bias, reproducing the validity pattern of Table~\ref{tab:estimators}. Figure~\ref{fig:nlsim-application} shows results for the real-world data of Section~\ref{sec:application}. Across both settings, estimator $\#4$ yields a rank-deficient covariate matrix under the feature expansion; we therefore drop a (randomly selected) subset of the time series in the conditioning set to restore full column rank.

\subsection{The Graphical View Remains Useful under Nonlinearity}
\label{sec:nonlinear-graphical}

Graphical IV validity alone does not in general
identify the causal effect in nonlinear equilibrium systems. Additional functional,
distributional, or support assumptions may be required
\citep[e.g.,][]{neweypowell, chesher2003identification, matzkin2008identification, imbens_identification_2009,
matzkin2015estimation,
torgovitsky2015identification}. Nevertheless, the graphical part of the
identification argument remains useful:
Consider the
standard IV setting (instrument $Z$, endogenous regressor $X$, outcome $Y$) with
an additional \emph{observed} confounder $V$ ($V \to X$, $V \to Y$) and a
homogeneous first stage. Whether $V$ must enter the conditioning set is governed
entirely by $f^Y$ and the graph. If $Y$ is linear in $X$ -- even with $V$
modifying the slope, $Y = \beta(V)X + \dots$ -- the instrument induces variation
in $X$ that is orthogonal to $V$ and so averages over $V$ automatically, and IV
identifies the ASF without conditioning on $V$. If instead $Y$ is nonlinear in
$X$ with a $V$-dependent shape, e.g.,\ $Y = \beta X + \theta\, V X^2 + \dots$, the
term $V X^2$ couples the confounder to the regressor; ignoring $V$ biases the
estimated ASF, and conditioning on $V$ restores it. The contrast depends only on
$f^Y$ and the edges $V \to X$, $V \to Y$, not on the structural equations of $X$,
$Z$, or $V$. The graph together with the form of $f^Y$ thus still determines the
conditioning set that targets the ASF; what nonlinearity removes is the
automatic equivalence between a valid instrument and a single scalar effect, not
the graphical reasoning itself.

\section{Additional Details on the Simulation and the Application}\label{app:simulation}

\subsection{Details of the Two-stage Least-squares Estimator}
\label{app:tsls}
We include a constant in the regressions below, bringing the models of Section~\ref{sec:graphical_representation}, which have non-zero intercepts, into the zero-mean setting of Section~\ref{sec:conditionsIV}.
Let $T \in \mathbb{N}$ and $\mathcal{B}, \mathcal{I}, \mathcal{X}$ and $Y$ satisfy (CIV1)--(CIV3). Assume  we have observations at time points $\{1, \dots, T\}$. Let $\{(t_1,s_1), \dots, (t_n,s_n)\} = \mathcal{B}$, where $(t_i,s_i)$ are in lexicographic order. Without loss of generality, assume $T \in \{t_1, \dots, t_n\}$ and let $k \coloneqq \max \{|t_i - t_j| \ \vert \ i,j \in \{1,\dots, n\}\}$. Define
\begin{equation*}
    \mathbf{B} \coloneqq
    \begin{bmatrix}
        X^{s_1}_{t_1} & \dots & X^{s_n}_{t_n}\\
        X^{s_1}_{t_1-1} & \dots & X^{s_n}_{t_n-1}\\
        \vdots & & \vdots \\
        X^{s_1}_{t_1-T+k+1} & \dots & X^{s_n}_{t_n-T+k+1}
    \end{bmatrix}
\end{equation*}
and define $\mathbf{X}, \mathbf{Y}$ and $ \mathbf{I}$ analogously.
Define $r_{\mathbf{I}}$ as the residuals of regressing $\mathbf{I}$ on $\mathbf{B}$ with ordinary least squares and define $r_{\mathbf{X}}$ and $ r_{\mathbf{Y}}$ analogously. Defining $P_{\mathbf{I}} \coloneqq r_{\mathbf{I}}({r_{\mathbf{I}}}^{\top}r_{\mathbf{I}})^{-1} r_{\mathbf{I}}^{\top}$,
\begin{equation*}
    \hat{\beta} \coloneqq \left(r_{\mathbf{X}}^{\top} P_{\mathbf{I}} r_{\mathbf{X}} \right)^{-1} r_{\mathbf{X}}^{\top} P_{\mathbf{I}} r_{Y}
\end{equation*}
solves \eqref{eq:civ}. By \citet[Theorem 5]{thams_identifying_2022} $\hat{\beta}$ is consistent for $T \to \infty$.

\subsection{Magnitude and Diagnosis of the Naive~IV Bias}\label{app:autocorrelation_plots}
\begin{figure}[t]
    \centering
    \includegraphics[width=0.9\linewidth]{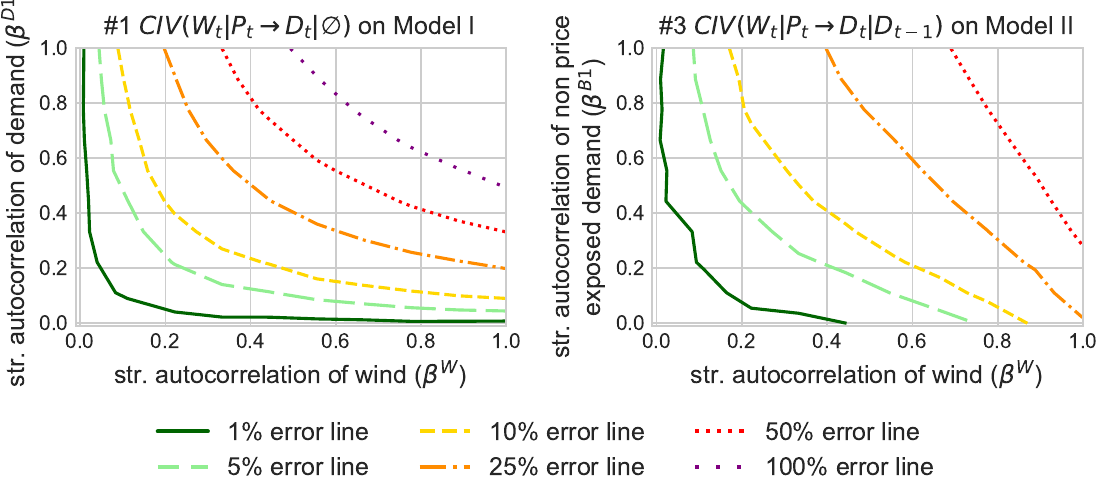}
    \caption{
    Estimated isolines showing the average absolute percentage error of the point estimate as a function of the structural autocorrelation of wind (instrument) and the structural autocorrelation of demand (dependent variable).
    Left: estimator $\#1$ on data from Model~I, where demand is autocorrelated with $\beta^{D1}$.
    Right: estimator $\#3$ on data from Model~II, where the non-price-exposed demand is autocorrelated with $\beta^{B1}$.
    For the estimation, we use a $10\times10$ grid of equidistant values in $[0, 1]$ and run $20$ simulations with five years of data $(T = 43{,}800)$ per point.
    }
    \label{fig:heatmap}
\end{figure}

This appendix shows the simulation evidence behind the ``Magnitude of bias'' discussion in Section~\ref{sec:Simulation}.
Figure~\ref{fig:heatmap} (left) depicts the average absolute percentage error of the naive~IV estimator $\text{CIV}(W_t|P_t \to D_t|\emptyset)$ (estimator $\#1$) for data generated from Model~I, as a function of the structural autocorrelation of the instrument $(\beta^W)$ and of demand $(\beta^{D1})$. For low instrument autocorrelation $(\beta^W<0.2)$ the bias is moderate $(<25\%)$. However, in our application the lag-one (observed) autocorrelation of wind generation exceeds $0.9$ (see Figure~\ref{fig:autocorrelation}). At those values, a small structural autocorrelation of demand of $\beta^{D1}\approx0.2$ already produces an error of $25\%$ and $\beta^{D1}\approx0.55$ an error of $100\%$. Figure~\ref{fig:heatmap} (right) shows the corresponding plot for Model~II, where the autocorrelation is induced by the autocorrelated error term with $\beta^{B1}$, and estimator $\text{CIV}(W_t| P_t \to D_t| D_{t-1})$ (estimator $\#3$).

\begin{figure}[t]
    \centering
    \includegraphics[width=0.95\linewidth]{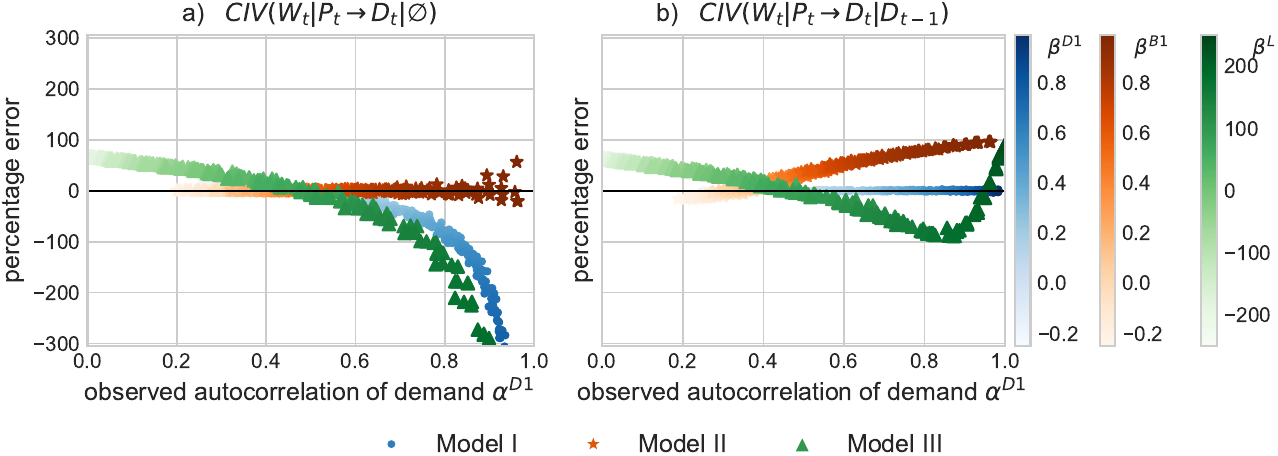}
    \caption{
   The percentage error is shown as a function of the observed autocorrelation of demand $(\alpha^{D1})$ for (a) the naive~IV estimator $\#1$ and (b) the CIV estimator $\#3$, which conditions on $D_{t-1}$; both panels share the vertical axis.
   In each simulation we vary the structural dependencies: In Models~I and II, the strengths of the structural autocorrelations, $\beta^{D1}$ and $\beta^{B1}$, are both varied between $-0.25$ and $0.99$.
   In Model~III, $\beta^{K}$ is varied between $-250$ and $+250$ (relaxing the sign restriction $\beta^K<0$). Color intensity increases with the parameter value. For each model and panel, we conduct $400$ simulations with a sample period of five years $(T = 43{,}800)$.
   }
    \label{fig:observed_ar}
\end{figure}

Knowing the observed autocorrelation of demand does not suffice to identify this bias. In Figure~\ref{fig:observed_ar}(a) we plot the percentage error of estimator $\#1$ against the observed autocorrelation of demand ($\alpha^{D1}$, to distinguish it from the structural coefficient $\beta^{D1}$); color intensity encodes the structural coefficients. The same observed autocorrelation at lag one arises from different structural dependencies and yields different errors: at $\alpha^{D1}\approx0.2$ the error is about $+50\%$ under Model~III, whereas a value as high as $\alpha^{D1}\approx0.9$ can accompany no bias at all under Model~II. Consequently, the observed autocorrelation is insufficient to derive a structural model or predict a bias. Figure~\ref{fig:observed_ar}(b) extends the analysis to estimator $\#3$, which conditions on past demand: the same observed level ($\alpha^{D1}=0.8$) maps to errors between $-100\%$ and $+100\%$, or to no error.

\subsection{Additional Indicator Plots for Model~II and Model~III}\label{app:indicator_plots}
\begin{figure}[t]
    \centering
    \includegraphics[width=1\linewidth]{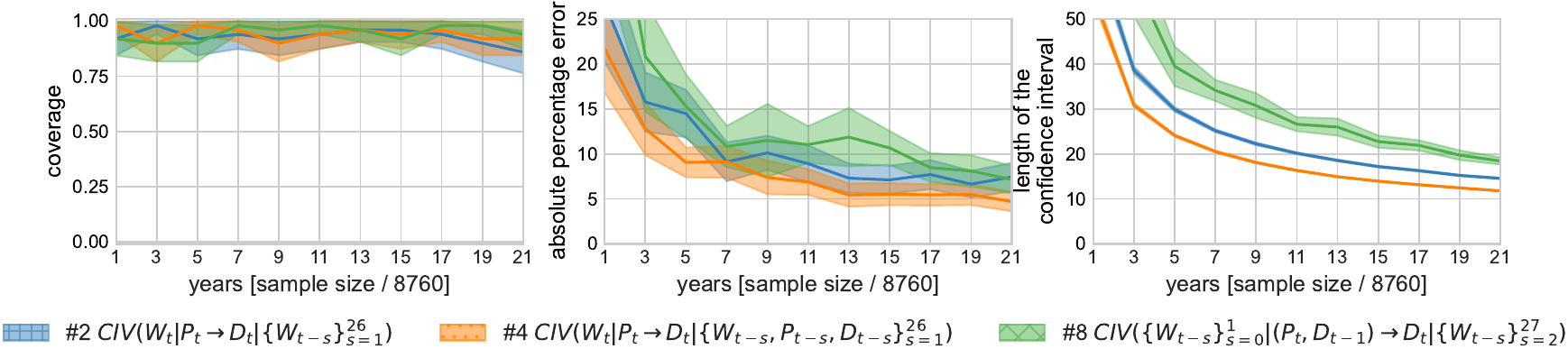}
    \caption{The performance of valid CIV estimators based on the indicators (a) coverage, (b) percentage error, and (c) length of the confidence interval for Model~II at different sample sizes. The shaded areas represent the $95\%$ confidence interval for the indicators. For each sample size, $ 50$ simulations were run. For data generation, the parameters of the demand equation~\eqref{eq:quantities_partially_unexposed} are set to $\beta^{P} = -100$ MW/(EUR/MWh) and $\beta^{B1} = 0.9$.}
    \label{fig:indicators_model_2}
\end{figure}

\begin{figure}[t]
    \centering
    \includegraphics[width=1\linewidth]{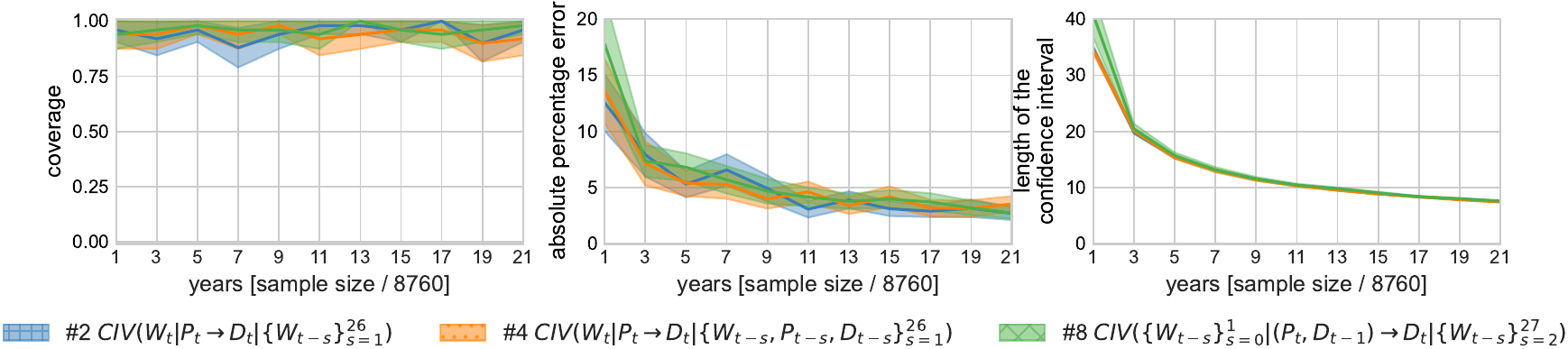}
    \caption{The performance of valid CIV estimators based on the indicators (a) coverage, (b) percentage error, and (c) length of the confidence interval for Model~III at different sample sizes. The shaded areas represent the 95\% confidence interval for the indicators. For each sample size, $50$ simulations were run. For data generation, the parameters of the demand equation~\eqref{eq:demand_inventory} are set to $\beta^{P} = -100$ MW/(EUR/MWh), $\beta^{K} = - 100$, $\delta^K = 0.1$, $\delta^D=0.001$ and $K^0 = 10$.}
    \label{fig:indicators_model_3}
\end{figure}

In Section~\ref{sec:Simulation}, we have compared the statistical performance of estimators $\#2$,~$\#8$,~and~$\#4$ on data generated from Model~I.
Figure~\ref{fig:indicators_model_2} and Figure~\ref{fig:indicators_model_3} compare the performance on data generated by Models~II and~III. As expected from theory, all estimators maintain nominal coverage regardless of sample size. Under Model~II, $\#8$ shows a markedly higher absolute percentage error and longer confidence intervals than $\#2$ and $\#4$; under Model~III the three estimators are practically indistinguishable on all three indicators.


\subsection{Disagreement between CIV Estimates for Simulated Data}\label{app:divergence}
\begin{figure}[t]
    \centering
    \includegraphics[width=.9\linewidth]{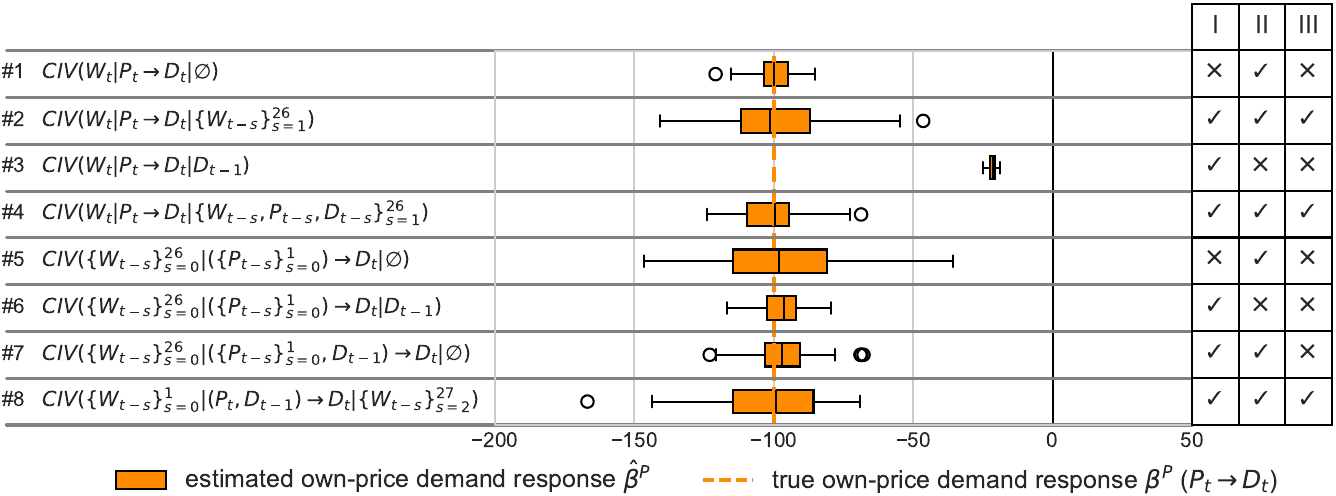}
    \caption{
    CIV estimates of 
    the demand response 
    $\beta^P$ when the simulated data do not exhibit structural autocorrelation of price-exposed demand (Model~II, see Figure~\ref{fig:DAG_demand_variants} (left)). The slope of the 
    price-exposed demand $\beta^P$
    is set to $-100$ MW/(EUR/MWh), and the autocorrelation coefficient of the non-price-exposed demand is set to $\beta^{B1}=0.9$.}
    \label{fig:simulation_two_demand_types}
\end{figure}

\begin{figure}[t]
    \centering
    \includegraphics[width=.9\linewidth]{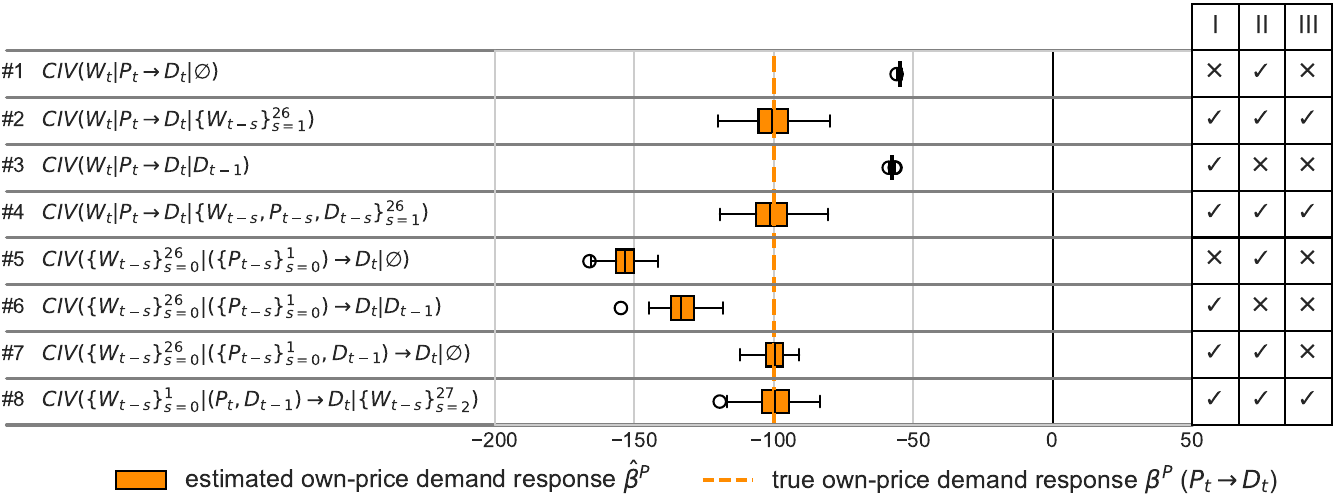}
    \caption{
    CIV estimates of the demand response $\beta^P$ when the data-generating process includes an inventory (Model~III, see Figure~\ref{fig:DAG_demand_variants} (right)). The true effect is $\beta^P = -100$ MW/(EUR/MWh) and the coefficient on the inventory level is $\beta^{K} = - 100$. According to theory (see Table~\ref{tab:estimators}), the estimators in column~III should be consistent, and the simulation results support this. 
    }
    \label{fig:simulation_cross}
\end{figure}

In Section~\ref{sec:CIV-estimation}, we have argued that non-overlapping confidence intervals of presumably valid estimators should lead to model rejection (see also Figure~\ref{fig:simulation_autocorrelated} for data from  Model~I).
Figures~\ref{fig:simulation_two_demand_types}~and~\ref{fig:simulation_cross} show the corresponding patterns of CIV estimators for Model~II and Model~III, respectively. The tables to the right of the figures show which estimators are valid according to the CIV criteria. 

For Model~II (Figure~\ref{fig:simulation_two_demand_types}), the estimators with a checkmark in column II are valid, and their estimates overlap. 
As for estimator $\#3$, conditioning on lagged demand generally induces a bias because the estimator conditions on a collider. For Model~II, estimator $\#6$ is a special case: based on the CIV criteria alone, the estimator is invalid, yet the estimated effect $P_t \to D_t$ is close to the true effect. The same happens with estimator $\#7$ for Model~III. The criteria are sufficient for validity, not necessary.
The pattern of estimators we observed in Figure~\ref{fig:simulation_cross} is consistent with what Model~III predicts. A researcher assuming Model~III would find little evidence against it; a researcher who incorrectly assumed Model~I would find evidence against that assumption (see, e.g., estimators~$\#2$ and~$\#3$). 

\section{Microfoundations for the Dynamic Demand Models}
\label{sec:microfoundations}

In the main text we introduce the demand equations as single-equation dynamic models in the econometric tradition \citep[chapter~7]{hendry_dynamic_1995}: the relevant intertemporal dependence is captured directly in one structural equation, and the equation is motivated by a behavioral mechanism.
This appendix makes the connection to optimizing behavior explicit by stating the assumptions under which each demand equation coincides with the decision rule of a utility-maximizing consumer.
Our analysis in the main text does not depend on these microfoundations: the identification results in Section~\ref{sec:CIV-estimation} require only the structural equations as introduced in Section~\ref{sec:graphical_representation}. 
Yet the derivation serves three purposes. First, it delivers sign restrictions on the coefficients of the structural equations. Second, it shows that consumers' price expectations are an identification assumption: consumers must not use the instrument to forecast future prices. 
Third, it separates the short-run price response we estimate from the response to a change in the price process itself.

\subsection{Models I and II}

We take Model~I as the leading case. The structural autocorrelation is due to an adjustment cost: changing consumption
between adjacent periods is costly, reflecting physical inertia as, for example, in multi-hour industrial processes.
Consider a consumer with per-period payoff
\begin{equation}\label{eq:payoff_modelI}
    \pi_t(d_t, d_{t-1})
    \coloneqq
    (a + u_t)\,d_t
    - \tfrac{b}{2}\,d_t^2
    - p_t\,d_t
    - \tfrac{\varphi}{2}\,(d_t - d_{t-1})^2,
\end{equation}
where $u_t$ is an i.i.d.\ taste shock with $\mathbb{E}[u_t]=0$, $b>0$, and $\varphi \geq 0$ is the adjustment cost\footnote{Lower-case letters ($d_t$, $p_t$, $u_t$) denote the choice variables, prices and shocks of an individual consumer, upper-case letters the market variables of Section~\ref{sec:graphical_representation}. We work with a representative consumer, so $d_t = D_t$ and $p_t = P_t$; aggregation is discussed for Model~II, where it matters.}. 
Two observations follow. 
First, the adjustment cost makes past consumption a state variable: if $\varphi = 0$, the problem is static and demand is not (structurally) autocorrelated.
Second, the adjustment cost does not 
determine whether the consumer is forward-looking (that is a separate assumption about behavior), but it makes the distinction consequential: if $\varphi = 0$ 
the problem 
separates across periods and the two behaviors coincide.

Suppose that the consumer is forward-looking, maximizing

\begin{equation}\label{eq:expected_utility_modelI}
    \mathbb{E}_t \left[ \sum_{s \geq t} \eta^{\,s-t}\, \pi_s(d_s, d_{s-1}) \right]
\end{equation}
with discount factor $\eta \in [0,1)$. The Euler equation is
\begin{equation}\label{eq:euler_modelI}
    a + u_t - b\,d_t - p_t - \varphi\,(d_t - d_{t-1})
    + \eta\varphi\,\mathbb{E}_t[\,d_{t+1} - d_t\,] = 0.
\end{equation}
Its stable solution is the decision rule\footnote{We refer to the expression that gives the demand of the consumer as the decision rule to distinguish it from the structural demand equations introduced in the main text.}
\begin{equation}\label{eq:policy_forward}
    d_t
    = \lambda\, d_{t-1}
    + \frac{\lambda}{\varphi} \sum_{j \geq 0} (\eta\lambda)^j\,
      \mathbb{E}_t\!\left[\, a + u_{t+j} - p_{t+j} \,\right],
\end{equation}
where $\lambda \in (0,1)$ is the stable root of
$\eta\varphi z^2 - \bigl(b + \varphi(1+\eta)\bigr) z + \varphi = 0$.\footnote{
For $\varphi>0$, $b>0$ and $\eta\in[0,1)$, exactly one root lies in $(0,1)$. The root is increasing in $\varphi$ and decreasing in $b$, and covers $(0,1)$ as $\varphi$ runs from zero to infinity.
}
The decision rule therefore depends on the entire forecast path $\mathbb{E}_t[p_{t+j}]$, $j \geq 1$.

In our equilibrium system, the instrument carries information about current and future prices: 
$P_t \leftarrow W_t \rightarrow W_{t+1} \rightarrow P_{t+1}$. 
The following three conditions holding true at the same time
creates a problem:
(i)~consumers are forward-looking ($\eta>0$); 
(ii)~consumers observe $W_t$ at time $t$ (and not only at $t+1$ or later); and
(iii)~$W_t$ has predictive content for $P_{t+1}$ beyond the other variables consumers observe (here it does, via $W_t \to W_{t+1} \to P_{t+1}$).
Under (i)--(iii), $W_t$ enters the forecast in~\eqref{eq:policy_forward}. The demand equation then contains $W_t$, which requires us to draw a direct edge $W_t \to D_t$, failing (CIV1), that is, the exclusion restriction that makes $W_t$ a valid instrument. 

We therefore assume that consumers do not include $W_t$ in their price forecasts. Three narratives support this: consumers may be myopic ($\eta = 0$), so that no forecast enters the problem; $W_t$ may not be in their information set at the time of consumption, since most electricity consumers do not observe real-time wind generation (whereas they do observe the current price); or they may perceive prices as a first-order Markov process and so not track the variables that drive price formation. We adopt the third, as \citet[Assumption~2]{hendel_measuring_2006} do,\footnote{
They assume prices follow an exogenous first-order Markov process, conceding that it is hard to specify a supply model that generates one, but arguing that such a process is ``a reasonable assumption about consumers' memory and formation of expectations''. We adopt the same restriction on beliefs. It is less plausible at the hourly frequency we study, but identification requires only that consumers not use $W_t$ to forecast prices. We do not adopt the exogeneity: here the price is determined in equilibrium and correlated with the demand shock by construction, which is what the instrument is for.
} which gives
\begin{equation}\label{eq:belief_restriction}
    \mathbb{E}_t[p_{t+j}] = h_j(p_t).
\end{equation}
This restriction is stronger than identification requires. For instance, beliefs may also depend on the state variable already included in the demand equation ($d_{t-1}$ in Model~I, $B_{t-1}$ in Model~II, $K_{t-1}$ in Model~III) without adding any edge to the graph. A forecast using further variables (e.g., past prices) adds the corresponding edges. Which estimators are valid is then read off the graph in the usual way. However, all estimators in Table~\ref{tab:estimators} would be invalid if consumers used $W_t$ to forecast prices. We adopt~\eqref{eq:belief_restriction} because it makes the coefficient mapping below exact.

Under~\eqref{eq:belief_restriction}, the forecast sum in~\eqref{eq:policy_forward}
is a function of $p_t$ alone, and with $u_t$ i.i.d.\ the decision rule
reduces to
\begin{equation}\label{eq:policy_collapsed}
    d_t = g(p_t) + \lambda\, d_{t-1} + \frac{\lambda}{\varphi}\, u_t,
\end{equation}
for the function
\begin{equation*}
    g(p_t) = \frac{\lambda}{\varphi}\Bigl[\,\frac{a}{1-\eta\lambda}
    - p_t - \sum_{j \geq 1} (\eta\lambda)^j h_j(p_t)\Bigr],
\end{equation*}
which combines preferences, the adjustment cost (both through $\lambda$), discounting (through $\eta$), and beliefs (through $h_j$).
The decision rule thus depends on $(p_t, d_{t-1}, u_t)$ only: it has the causal structure of Model~I, with $\beta^{D1} = \lambda$. 
If, in addition, beliefs are affine,
\begin{equation}\label{eq:belief_linear}
    h_j(p_t) = (1-\rho^{\,j})\,\bar p + \rho^{\,j} p_t, \qquad \rho \in [0,1),
\end{equation}
the $j$-step-ahead forecast of an AR(1) price process with long-run mean $\bar p$, then $g$ is linear and the decision rule is exactly Equation~\eqref{eq:demand_autocorrelated}. We obtain
\begin{equation}\label{eq:mapping_forward}
\begin{aligned}
    D^0 &= \frac{\lambda}{\varphi}\,\frac{a(1-\eta\lambda\rho) - \eta\lambda(1-\rho)\,\bar p}{(1-\eta\lambda)(1-\eta\lambda\rho)},
    &\qquad
    \beta^P &= -\frac{\lambda}{\varphi\,(1-\eta\lambda\rho)},
    \\
    \beta^{D1} &= \lambda,
    &\qquad
    U^D_t &= \frac{\lambda}{\varphi}\,u_t .
\end{aligned}
\end{equation}
The coefficient $\beta^P$ combines the contemporaneous effect with the anticipation of future prices, and $|\beta^P|$ is increasing in $\rho$: with adjustment costs, a price change believed to persist is met with a larger response.
The primitives deliver the sign and range restrictions $\beta^P < 0$ and $\beta^{D1} = \lambda \in (0,1)$.
If beliefs are nonlinear, demand is a nonlinear function of the current price with the same causal structure; this case is 
discussed in Appendix~\ref{sec:non-linear-ext}.

This derivation clarifies our estimand by separating three objects. The parameter $\beta^P$ is the instantaneous own-price response: the effect of $\mathrm{do}(P_t = p)$ on $D_t$, holding the state variable fixed and leaving beliefs and the price process unchanged. 
The long-run multiplier $\beta^P/(1-\beta^{D1})$ is the response to a price held at $p$ for many periods, still at unchanged beliefs; it simply adds the propagation through the autocorrelation.%
The third object 
is the response to a change in the price-generating process itself. For instance, a change in market design alters beliefs and with them $\beta^P$: by~\eqref{eq:mapping_forward} a consumer who came to regard price changes as permanent ($\rho \to 1$) would respond more strongly, by the factor $(1-\eta\lambda\rho)/(1-\eta\lambda)$, which exceeds one whenever $\eta\lambda > 0$, since $\rho < 1$.
No single-equation estimate identifies that response \citep{lucas_econometric_1976}. Our estimand is the first object.

Model~II follows from the same per-period payoff, applied to each component with its own primitives. The price-exposed component is the $\varphi_A \to 0$ limit of~\eqref{eq:mapping_forward}, in which $\lambda \to 0$ and the belief parameters drop out: $A_t = A^0 + \beta^P P_t + U^A_t$ with $A^0 = a_A/b_A$, $\beta^P = -1/b_A$ and $U^A_t = u^A_t/b_A$, so $\beta^P$ is a pure preference parameter regardless of $\eta$. The non-exposed component $B_t$ faces a flat retail tariff, so $P_t$ does not enter its problem (the tariff is absorbed into the constant), while an adjustment cost $\varphi_B > 0$ gives $B_t = B^0 + \beta^{B1} B_{t-1} + U^B_t$ with $\beta^{B1} = \lambda_B$. Aggregate demand is the sum, as in Equation~\eqref{eq:quantities_partially_unexposed}. 
Summing over type-A consumers is immediate, since their decision rules are static and linear. Summing over type-B consumers requires that they share the root $\lambda_B$; heterogeneous adjustment speeds would generate higher-order aggregate dynamics.

\subsection{Model III}

We now turn to Model~III, the stockpiling model inspired by \citet{hendel_measuring_2006}.
We maintain the belief restriction~\eqref{eq:belief_restriction} in the affine form~\eqref{eq:belief_linear} with $\bar p > 0$, since the level to which prices are expected to revert is what determines whether the consumer stockpiles. Only the one-step-ahead forecast will enter below, unlike in Model~I. Purchases are unrestricted, so the consumer reaches her preferred inventory in every period. This keeps the problem in closed form but leaves the inventory with no dynamics of its own, as we discuss at the end of the subsection.

Let the consumer value consumption $s_t$ drawn from the inventory of energy services $K_t$. The consumer chooses purchases $d_t$ and consumption $s_t$:
\begin{equation}\label{eq:payoff_modelIII}
    \max_{\{d_{t+j}, s_{t+j}\}_{j\ge0}} \quad
    \mathbb{E}_t \sum_{j\ge0}  
        \left[
            \eta^j
            \left(
             (\alpha + u_{t+j}) s_{t+j}
             - \frac{c}{2} s_{t+j}^2
             - p_{t+j} d_{t+j} 
             - \frac{\kappa}{2}K_{t+j}^2
            \right)
        \right]
\end{equation}
\begin{equation}\label{eq:law_of_motion_stock}
    \text{s.t.} \quad
    K_t = (1-\delta^K)\, K_{t-1} + \delta^D d_t - s_t,
\end{equation}
with $\alpha, c, \kappa > 0$; $\delta^K \in (0,1)$ is the depreciation rate of the inventory, and $\delta^D > 0$  the conversion rate of electricity to energy services. 
We do not impose $K_t \geq 0$: in the linear-quadratic formulation~\eqref{eq:payoff_modelIII}, $K_t$ measures how far the inventory sits above or below its usual level, around which holding costs are symmetric; the constant $K^0$ in~\eqref{eq:inventory} is what centers it. Here, a negative $K_t$ is, for instance, an electric vehicle charged less than usual. This is standard in linear-quadratic inventory models \citep[see][]{ramey_inventories_1999}. 

To solve this problem we introduce the value function $V_t(K)$, the best expected discounted payoff achievable from period $t$ onward given that the consumer enters $t$ with inventory $K$; the time subscript carries the dependence on $p_t$ and $u_t$, which the consumer takes as given. It satisfies
\begin{equation}\label{eq:modelIII_value_function}
    V_t(K_{t-1}) = \max_{\{d_t, s_t\}} \quad
        \left\{
        \underbrace{
            (\alpha + u_{t}) s_{t}
            - \tfrac{c}{2} s_t^2
            - p_{t} d_{t}
            - \tfrac{\kappa}{2}K_{t}^2
        }_{\text{payoff at $t$}}
        +
        \underbrace{
            \eta\mathbb{E}_t [V_{t+1}(K_t)]
        }_{\text{value of future payoffs}}
        \right\}.
\end{equation}

Substituting the law of motion into~\eqref{eq:modelIII_value_function} and writing $\mu_t \coloneqq \eta \mathbb{E}_t[V_{t+1}'(K_t)]$ for the discounted marginal value of the inventory carried from $t$ to $t+1$, the first-order conditions for $d_t$ and $s_t$ are
\begin{equation}\label{eq:modelIII_focs}
    -p_t -\kappa K_t \delta^D + \mu_t \delta^D = 0
    \qquad\text{and}\qquad
    (\alpha + u_{t}) - c s_t + \kappa K_t - \mu_t = 0 .
\end{equation}
Both conditions equate the net marginal value of the inventory, $\mu_t - \kappa K_t$, to a margin: the consumer buys until the marginal cost of adding to the inventory, $p_t/\delta^D$, equals it, and withdraws until the marginal utility of energy services, $\alpha + u_{t} - c\,s_t$, equals it. Substituting the first condition into the second gives optimal consumption
\begin{equation}\label{eq:modelIII_optimal_consumption}
    s_t^*=\tfrac{1}{c}\bigl(\alpha+u_t-p_t/\delta^D\bigr).
\end{equation}
Differentiating~\eqref{eq:modelIII_value_function} with respect to the state $K_{t-1}$ and applying the envelope theorem, $V_t'(K_{t-1}) = (1-\delta^K)(\mu_t - \kappa K_t) = (1-\delta^K)p_t/\delta^D$, using the first condition in~\eqref{eq:modelIII_focs}. The same holds one period later, so $V_{t+1}'(K_t) = (1-\delta^K)p_{t+1}/\delta^D$ and
\begin{equation}\label{eq:modelIII_mu}
    \mu_t = \eta (1 - \delta^K)\frac{\mathbb{E}_t[p_{t+1}]}{\delta ^D}.
\end{equation}
Note that $V'_t$ does not depend on $K_{t-1}$: the value function is affine in the inventory.
Substituting~\eqref{eq:modelIII_mu}, the law of motion, the belief rule~\eqref{eq:belief_linear} and $s_t^*$ into the first condition in~\eqref{eq:modelIII_focs} and rearranging gives the consumer's decision rule for demand,
\begin{equation}\label{eq:modelIII_final_mapping}
\begin{split}
    d_t &=
    \frac{-1}{\kappa (\delta^D)^2}
    \Bigl(
    p_t
    + \kappa \delta^D(1-\delta^K)\, K_{t-1}
    - \tfrac{\kappa \delta^D}{c}\bigl(\alpha+u_t-p_t/\delta^D\bigr)
    \\
    &\hspace{7.2em}
    - \eta(1 - \delta^K)\bigl((1-\rho)\,\bar p + \rho \, p_t\bigr)
    \Bigr)
    \\
    &= \frac{\alpha}{c\delta^D} + \frac{\eta(1-\delta^K)(1-\rho)\,\bar p}{\kappa (\delta^D)^2}
    - \frac{1+\kappa/c - \eta(1-\delta^K)\rho}{\kappa (\delta^D)^2}\, p_t
    \\
    &\hspace{7.2em}
    - \frac{1 - \delta^K}{\delta^D}\, K_{t-1}
    + \frac{u_t}{c\delta^D} ,
\end{split}
\end{equation}
which maps to~\eqref{eq:demand_inventory} as
\begin{equation*}
\begin{aligned}
    D^0 &= \frac{\alpha}{c\delta^D} + \frac{\eta(1-\delta^K)(1-\rho)\,\bar p}{\kappa (\delta^D)^2},
    &\qquad
    \beta^P &= - \frac{1 + \kappa/c - \eta(1-\delta^K)\rho}{\kappa (\delta^D)^2},
    \\
    \beta^K &= - \frac{1 - \delta^K}{\delta^D},
    &\qquad
    U_t^D &= \frac{u_t}{c\delta^D}.
\end{aligned}
\end{equation*}
The mapping delivers the sign restrictions the main text assumes and decomposes each coefficient into primitives:
\begin{itemize}
    \item 
    Sign of the price response.
    Since $\eta(1-\delta^K)\rho < 1$, the numerator of $\beta^P$ exceeds $\kappa/c > 0$, so $\beta^P<0$. In this model $\beta^P$ bundles preferences ($c$), technology ($\kappa, \delta^D, \delta^K$), beliefs ($\rho$), and discounting ($\eta$), but it still has the interpretation of the short-run demand response to prices.
    \item Sign of the inventory coefficient. The parameter $\beta^K < 0$ is purely technological, driven by the conversion and depreciation rates of the inventory.
    \item Anticipation. The term $\eta(1-\delta^K)\rho$ makes $|\beta^P|$ smaller when prices are persistent: a high price today signals a high price tomorrow, weakening the incentive to respond to prices (e.g., to postpone purchases). The sign is opposite to that of Model~I, where anticipation raises $|\beta^P|$ by~\eqref{eq:mapping_forward}.
    \item Stockpiling. Combining~\eqref{eq:modelIII_focs} with~\eqref{eq:modelIII_mu} gives the inventory the consumer chooses to carry,
    \begin{equation}\label{eq:modelIII_target_stock}
        K_t = \frac{\eta(1-\delta^K)(1-\rho)\,\bar p - \bigl(1-\eta(1-\delta^K)\rho\bigr)p_t}{\kappa\,\delta^D},
    \end{equation}
    which is positive if and only if $p_t < \eta(1-\delta^K)(1-\rho)\bar p / \bigl(1-\eta(1-\delta^K)\rho\bigr)$, a threshold that approaches $\bar p$ as $\eta \to 1$ and $\delta^K \to 0$. The consumer builds up the inventory when the current price is low relative to what she expects and runs it down when it is high. More precisely, carrying an inventory forward is worthwhile only when the discounted, depreciated price she expects to pay next hour exceeds the price she pays now. At $\eta = 0$ the threshold is zero, so a myopic consumer never stockpiles; but given the hourly setup, $\eta$ is close to one. $K_t$ does not depend on $K_{t-1}$: the inherited inventory changes how much she must buy, not the level she wants to hold.
    \item Consumption from the inventory. Optimal consumption $s_t^*$ is not constant but varies with $p_t$ and $u_t$; the law of motion in the main text~\eqref{eq:inventory} replaces it by its mean $K^0=\mathbb{E}[s_t^*]$.

    \item Short- and long-run responses. The same three objects separate here. Under the main text's law of motion~\eqref{eq:inventory} the long-run response is $\delta^K\beta^P$; under~\eqref{eq:law_of_motion_stock}, which keeps $s^*_t$, it is larger in absolute value by $(1-\delta^K)/(c(\delta^D)^2)$, since a permanently higher price also lowers consumption drawn from the inventory. A change in the price process itself could change $\rho$, and with it $\beta^P$, or $\bar p$, and with it $D^0$.

    \item Inventory dynamics. The linear-quadratic problem reproduces the demand equation~\eqref{eq:demand_inventory}, its sign restrictions and its decomposition into primitives, but not the inventory dynamics of Model~III. Because the value function is affine, the inventory the consumer carries~\eqref{eq:modelIII_target_stock} is affine and strictly decreasing in the current price, so along the optimal path $K_{t-1}$ is redundant given $p_{t-1}$ and~\eqref{eq:modelIII_final_mapping} reduces to a distributed lag in current and lagged prices. A cost of adjusting purchases, or the non-negativity constraint that leads \citet{hendel_measuring_2006} to solve their model numerically, would make the value function quadratic and $K_{t-1}$ a genuine state, at the cost of coefficients that no longer have closed form.
\end{itemize}

\end{appendix}

\end{document}